\documentclass{appolb}
\usepackage{epsfig}
\usepackage{latexsym}
\usepackage{amsmath}
\usepackage{amssymb}

\begin{document}

\title{Large N and confining flux tubes as strings \\
-- a view from the lattice
\thanks{Lectures at the $49^{th}$ Cracow School of Theoretical Physics}%
}
\author{Michael Teper
\address{ Rudolf Peierls Centre for Theoretical Physics, University of Oxford,\\
1 Keble Road, Oxford OX1 3NP, UK}
}
\maketitle
\begin{abstract}
I begin these three lectures by describing some of the useful 
things that we have learned about large-$N$ gauge theories 
using lattice simulations. For example that
the theory is confining in that limit, that for many quantities
SU(3) $\simeq$ SU($\infty$), and that this includes the strongly 
coupled gluon plasma just above $T_c$, thus providing some of the
justification needed to make use of gauge-gravity duality in analysing
QCD at RHIC/LHC temperatures. 
I then turn, in a more detailed discussion, to recent progress
on the problem of what effective string theory describes 
confining flux tubes. I describe lattice calculations of the 
energy spectrum of closed loops of confining flux, and some dramatic 
analytic progress in  extending the `universal Luscher correction' 
to terms that are of higher order in $1/l^2$, where $l$ is the length of 
the string. Both approaches point increasingly to the Nambu-Goto
free string theory as being the appropriate starting point for
describing string-like degrees of freedom in SU($N$) gauge theories.  
\end{abstract}
\PACS{11.15.-q, 11.15.-Pg, 11.15.Ha, 11.25.Pm}
  
\section{Introduction}
Over the last ten years lattice simulations have helped us 
learn a great deal about 't Hooft's large-$N$ limit of gauge theories 
and QCD. This resurgence of interest on the lattice side
\cite{MT98d3}
coincided (co-incidentally) with the culmination of the `second
superstring revolution' in Maldacena's AdS/CFT correspondence
\cite{maldacena}
and the gauge-gravity dualities that have flowed from it.
These dualities between weakly coupled string theories and
strongly coupled gauge theories at large $N$, have led to
a common interest in what is the physics of large $N$ gauge 
theories.

At the same time, Maldacena's work has provided a new and
unexpected twist to the very old question of what, if any, string
theory describes the strong interactions and hence gauge theories.
A narrower version of this question is to ask what
effective string theory describes the dynamics of confining
flux tubes in gauge theories. Certainly at large $N$ this 
latter question, when applied to long flux tubes and to their
low-lying excitations, should become entirely well-defined.
Any answer promises to provide essential insights into what might
be the answer to the first and much more speculative question.

I begin these lectures with some cursory remarks about the 
large $N$ limit (with which I assume you are all familiar) and
I then give you an overview of how one does lattice calculations,
describing how one can obtain predictions for interesting physical 
quantities, such as ratios of masses, in the continuum gauge theory.
In Section~\ref{section_largeN} I select a few topics about large
$N$ gauge theories that have been addressed (and largely resolved)
by lattice calculations in recent years. Is SU($\infty$) linearly
confining? Is SU(3) close to SU($\infty$)? Is QCD close to 
QCD$_{\infty}$? How does the coupling run, and should we keep
$g^2N$ fixed for a smooth large-$N$ limit? I finish this Section
by discussing how the computational cost increases with $N$, and
show that large-$N$ calculations are surprisingly inexpensive
and accessible. I then devote Section 4 to discussing large $N$
gauge theories at finite temperature; in particular above but not
too far from the deconfining transition. This is of particular
interest since it has become the focus of a large AdS/CFT effort
in recent years. Section 5 is very brief and just lists some
topics on which there has been interesting work, but which I have
no time to discuss in these lectures. The remainder of my lectures
is devoted to my second topic: what is the effective string theory
that describes confining flux tubes. Section 6 summarises
the analytic work that has accompanied and motivated (and been
motivated by) the large amount of numerical work on this question
that has been carried out over the last three decades. After
some general background, and a detailed description of both
the Gaussian approximation and the Nambu-Goto free string theory,
I describe the dramatic progress that has been achieved in the
last five years (with some startling papers appearing even as I write 
these lectures). I finish the Section with a potted and 
inadequate history of numerical calculations during this period.
I then move on to the numerical calculations of the energy spectrum
of closed flux tubes that I have been involved in for the last 
4 or 5 years. Section 7 discusses our calculations in 
SU($N$) gauge theories in $2+1$ dimensions. We obtain very
accurate energy estimates for quite a large number of low-lying
eigenmodes, and I display how remarkably well Nambu-Goto describes
these even when the flux tube is so short that it is hardly longer 
than it is wide. In Section 8 I give a brief preview of our
unpublished work on SU($N$) gauge theories in $3+1$ dimensions.
Here there is an interesting distinction between the majority
of states, which adhere closely to a Nambu-Goto-like spectrum,
much as in $D=2+1$, and a significant minority of states that
behave quite differently. I will finish, in Section 9, with some
concluding remarks, although the bulk of my conclusions are
embedded, at appropriate points, in these lectures.

\section{Preliminaries}

\subsection{Large $N$}

In $D=3+1$ gauge theories one has a dimensionless coupling $g^2$ and 
one might hope to be able to use it as a general expansion 
parameter for the theory. However because the scale invariance is 
anomalous, setting $g^2$ to some particular value  $g_0^2$ means 
that we can only hope to use it as a useful expansion parameter 
for physics close to the scale $l_0$ where the running 
coupling takes that value, i.e. where $g^2(l=l_0) \simeq  g_0^2$.
In $D=2+1$ gauge theories  $g^2$ has dimensions of mass so that the
dimensionless expansion parameter for physics on the scale $l$ is $g^2 l$
and one immediately sees that the coupling cannot serve as a useful 
expansion parameter for the theory as a whole. 

\begin{figure}[htb]
\begin	{center}
\leavevmode
\epsfig{figure=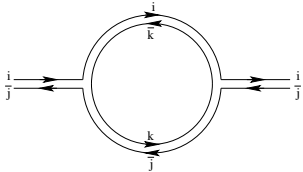, angle=360, width=4cm} 
\end	{center}
\caption{Gluon loop insertion using 't Hooft's double line notation.}
\label{fig_planar}
\end{figure}

Faced with this, 't Hooft suggested, back in 1974
\cite{GtH74},
that an alternative, less obvious but more general expansion 
parameter might be
provided by $1/N$. That is to say one thinks of expanding SU($N$)
gauge theories in powers of $1/N^2$ around  SU($\infty$). Pictorially:
\begin{equation}
SU(N) = SU(\infty) + O\left(\frac{1}{N^2}\right).
\label{eqn_largeN}
\end{equation}
That the expansion parameter is $1/N^2$, follows from 't Hooft's analysis 
of all-order perturbation theory using his clever double-line notation 
for gluon propagators and vertices, in which the gluon is represented
by a fundamental line and its conjugate. (For simplicity I will here
ignore the difference between U($N$) and SU($N$).) It also follows 
that a smooth large-$N$ limit can only be achieved if one keeps  
$g^2N$ fixed. We can begin to see why this should be so 
by considering a gluon loop insertion in 
the gluon propagator using the double-line notation, as shown in 
Fig.~\ref{fig_planar}. The two vertices give a factor of $g^2$
and the sum over the colour index  in the closed  loop
gives a factor of $N$. So such an insertion produces a factor
of $g^2N$ in the amplitude. If we want smooth physics as $N\to\infty$,
then at the very least we require that the number of such insertions, 
in the dominant diagrams contributing to the physics of interest,
should be roughly fixed as we vary $N$. This requires that we 
keep $g^2N$ fixed in that limit.
One can readily generalise the argument to all diagrams.

This argument looks perturbative, but it is nonetheless convincing
because in demanding that there is a smooth large-$N$ limit, we are 
demanding {\it{inter alia}} that SU($\infty$) should be asymptotically 
free. If instead of keeping $g^2N$ fixed we vary 
$g^2 \propto 1/N^{1+\epsilon}$ with $\epsilon < 0$ it is clear
that at $N=\infty$ infinite order diagrams will dominate at any 
scale where we attempt to apply perturbation theory, i.e. there is no
asymptotic freedom. If on the other hand we take $\epsilon > 0$,
one would be driven to a free theory on any scale
where one attempts to apply perturbation theory. Neither result is
what we want, so the only limit that can work is to keep
$g^2N$ fixed as $N\to\infty$. Note however that while we can easily
argue this condition to be necessary, there is no guarantee
that it is sufficient.

In the gauge theory the coupling runs, so a little care is needed
in defining what we mean by keeping $g^2N$ fixed. So first let us
define the length scale for the running coupling in units of
say the mass gap, i.e. $lm_G$. Then define the 't Hooft coupling
$\lambda_N = g^2N$ for the SU($N$) gauge theory. Then the appropriate 
way to encode $g^2N = \mathrm{const}$ is simply as follows: 
\begin{equation}
\lambda_N(lm_G) \longrightarrow \lambda_\infty(lm_G) 
\end{equation}
i.e. the 't Hooft coupling on a scale $l$, where the
scale $l$ itself is fixed in physical units, tends to a non-trivial
limit as we sweep through the various SU($N$) gauge theories. 

The large-$N$ expansion will only be useful if the theory we are
expanding about, SU($\infty$), is sufficiently simple. Although it 
is, in fact, not so simple that we can solve it analytically, some 
things are very simple in a large-$N$ confining phase
\cite{Nreviews}. 
This essentially
arises because the probability of any quarks or gluons forming 
a colour singlet (rather than some other representation) goes to 
zero as the  number of colours increases. This, together with
the fact that $g^2 \propto 1/N \to 0$ is
the reason that we have zero decay widths, zero mixing, a perfect OZI
rule and no colour singlet scattering in that limit.

Of course it would be nice to show that there is in fact a large $N$
confining phase, and that a smooth physics limit does in fact exist.
And we would like to determine what precisely that physics is.
And we would like to check whether, for at least some reasonably wide 
class of important physical quantities, we can say that SU(3) (and $QCD$) 
are indeed close to SU($\infty$) (and $QCD_{N=\infty}$). This is after 
all the reason many of us might be interested in the large-$N$ limit.
At present these questions can only be addressed numerically.

Before moving on to these numerical calculations, let me emphasise
that there is no expectation that all the physics of SU(3) is 
close to that of SU($\infty$). Indeed it is clear that this cannot
be so and in that sense eqn(\ref{eqn_largeN}) is misleading.
Indeed it is entirely possible that other large-$N$ limits
may be more appropriate for some physics. For example, in QCD we have
2 or 3 light flavours, so $N_f/N \sim 1$. It might appear plausible
that the limit $N\to\infty$ with  $N_f/N$ fixed might be more
appropriate for some physical quantities
\cite{GVNcNf}.
(Or a limit with fermions in the bi-fundamental representation 
\cite{MSGV}
which has the virtue that
some quantities become calculable at $N=\infty$ because the
theory effectively has some supersymmetry in that limit.) 
However it is worth noting that the 't Hooft limit has turned out to
be phenomenologically useful even in some cases where one would
naively have not expected it to be. A good example is the case of 
baryons, which consist of $N\to\infty$ quarks, but where
one has inferred a useful SU(6)-like symmetry to describe
the multiplet structure
\cite{DMsu6}.
\subsection{Lattice calculations \cite{lat_books}}
\label{subsection_lattice}

We want to calculate correlation functions numerically. These
can be expressed as Feynman path integrals. Generically:
\begin{equation}
Z = \int \prod_{x\in M_4} d\phi(x) e^{iS[\phi]}
\end{equation}
After we explicitly integrate over any Grassmanian quark fields, the 
integrand depends just on ordinary numbers (which may be grouped into
SU($N$) matrices) so one can imagine doing the integral numerically.

To be able to do this we first need to get rid of the phase
factor $\exp\{iS\}$ since numerical integrations involving phase
factors are notoriously ill conditioned (the sign problem).
This we achieve (in the cases of interest in these lectures)
by going to Euclidean space-time, so that $ i\int dt \to  \int dt$.
(How that affects what we can calculate will be addressed in a moment.)
The next problem is that we have an infinite number of integrations.
For a numerical calculation this must be made finite: so the
space-time volume must be made finite and discrete. It is usually 
convenient to use a hypertorus and a hypercubic lattice for these
purposes. (Sometimes other choices are useful.) So these two steps 
look like:
\begin{equation}
\int \prod_{x\in M_4} d\phi(x) e^{iS[\phi]}
\to
\int \prod_{x\in R^4} d\phi(x) e^{-S_E[\phi]}
\to
\int \prod_{n\in T^4} d\phi_L(n) e^{-S_L[\phi_L]}
\end{equation}
where we express the discrete space-time points as $x=na$ where $n$ is a 
$D$-tuple of integers, $a$ is the lattice spacing, and $\phi_L$ is a 
dimensionless lattice field variable, with action $S_L$, chosen so that
\begin{eqnarray}
\phi_L(x) & \stackrel{a\to 0}{\longrightarrow} &  a^{-dim(\phi)}\phi(x) 
\nonumber \\
S_L  & \stackrel{a\to 0}{\longrightarrow} & S_E 
\end{eqnarray}
where $dim(\phi)$ is the length dimension of the field $\phi$.
Obviously this schematic outline of the continuum limit misses many
essential details. 

Two asides. Firstly, putting the theory in a finite box of size $l$ 
can be harmless if, as here, the theory has a finite mass gap, $m_G$,
because in such a case we expect finite-size effects to be
$O(e^{-m_G l})$ which can easily be made very small by choosing
$l \gg 1/m_G$. Secondly, since the theory is renormalisable, 
what we do at short distances, e.g. introducing a lattice cut-off,
can be simply absorbed into the renormalisation of parameters such 
as the coupling, if we choose $a \ll 1/\Lambda$, where $\Lambda$ 
is the typical dynamical energy scale. 
Moreover since the theory is asymptotically free, we can analyse 
the lattice spacing corrections in a perturbation expansion in the 
small coupling $g^2(a)$.

The continuum degrees of freedom are the gauge potentials
$A_\mu(x)$ which belong to the SU($N$) Lie algebra. They tell us
how to compare colour at infintesimally neighbouring points. If
we want to compare colour between points with a finite separation,
we use the path ordered exponential of the gauge potential along
some path $c$ joining the points: $P\exp\{i\int_c A.ds\}$. 
This is an SU($N$) group element
which `lives on' the particular path chosen. So the natural
gauge degrees of freedom on the lattice, where all points are a 
finite distance apart, are group elements, $U_l$, that live on the 
links, $l$, of the lattice. More specifically, if the link goes
from the site $x$ to the site $x+a \hat{\mu}$, we label this group
element by $U_\mu(x)$. Under a gauge transformation $V(x)$,
it is defined to transform as
\begin{equation}
U_\mu(x) \longrightarrow V(x)U_\mu(x)V^\dagger(x+a \hat{\mu})
\end{equation}
as would the corresponding path ordered exponential in the
continuum theory. To the same link but taken in the reverse 
direction, i.e.  $x+a \hat{\mu}$ to  $x$, we assign 
$U^\dagger_\mu(x)$. Thus if we go out along a link and then 
return by the same link to the same point, the colour comparison 
matrix is the unit matrix, as it should be.

We choose for our integration measure the standard Haar measure
which has the nice property that it invariant under left
or right multiplication and hence gauge invariant:
\begin{equation}
d U_\mu(x) = d \{V(x)U_\mu(x)V^\dagger(x+a \hat{\mu})\}.
\end{equation}
Now all we need is a gauge-invariant action. It is easy to see
that the trace of the product of link-matrices around any closed 
path $c$, $ \mathrm{Tr} \prod_{l\in c} U_l$, is gauge 
invariant. (To any backward going link $l$ we
assign $U^\dagger_l$.) Now, the continuum action compares fields
at (infinitesimally separated) neighbouring points. We can 
compare neighbouring link matrices on a lattice by taking
their product. To be gauge-invariant this product should appear
within a product of matrices being taken around some closed
path. The simplest and most common choice is to use the path
that is an elementary square on the lattice, called
a plaquette. So for the action we take
\begin{equation}
S = \sum_p \left\{1 - \frac{1}{N} \mathrm{ReTr}U_p\right\}
\label{eqn_Splaq}
\end{equation}
where $U_p$ is the path-ordered product of link matrices around  
the plaquette $p$. The $\sum_p$ ensures that the action is
translation and rotation invariant. Taking the real part of
the trace ensures that it has $C=P=+$. So our lattice path
integral is  
\begin{equation}
Z = \int \prod_{l} dU_l e^{-\beta S}
\end{equation}
where $\beta$ is any constant. It is the only free parameter in $Z$,
so it is by varying $\beta$ that we will be able to vary the
lattice spacing $a$. Since our lattice theory has the important
continuum symmetries (albeit a subgroup for rotations and translations)
we expect that in the continuum limit, barring an unnatural choice
of lattice action (and here I mean unnatural in the technical sense
as applied, for example, to a light Higgs scalar) we must obtain
the continuum action, i.e.
\begin{equation}
\int \prod_{l} dU_l e^{-\beta S} 
\stackrel{a\to 0}{\longrightarrow}
\int \prod_{x,\mu} dA_\mu(x) 
e^{-\frac{4}{g^2} \int d^4 x Tr F_{\mu\nu} F_{\mu\nu} } 
\end{equation}
(up to a possibly infinite constant).
This tells us that in the continuum limit $\beta = c/g^2$.
A more careful analysis tells us what $c$ is:
\begin{equation}
\beta = \frac{2N}{g^2_L(a)} \qquad ; \qquad
g^2_L(a) \stackrel{a\to 0}{\longrightarrow} g^2(a)
\label{eqn_betaD4}
\end{equation}
where $g^2_L(a)$ is a running coupling on the scale $a$ in 
(this particular) lattice coupling scheme, and $g^2(a)$
is a running coupling in a(ny) continuum scheme. (In this
limit any difference between schemes is $O(g^4)$.) 
Since $g^2(a) \to 0$ as $a\to 0$ we know how to find the continuum 
limit; one simply takes $\beta \to \infty$.

Although the above has been for $D=3+1$ dimensions, we can follow
the same steps in $D=2+1$. In this case $g^2$ has dimensions $[m]$,
so the dimensionless bare lattice coupling is $ag^2$ and, not 
surprisingly, we find
\begin{equation}
\beta = \frac{2N}{a g^2}  \stackrel{a\to 0}{\longrightarrow} \infty
\qquad : D=2+1
\label{eqn_betaD3}
\end{equation}
where this $g^2$ becomes the $g^2$ of the continuum theory
when $a\to 0$.

Suppose we want to calculate the expectation value of some
functional $\Phi[U]$ of the gauge fields. We generate a set of
$n$ gauge fields $\{U^I\}; I=1,...,n$ distributed not just with
the measure $\prod_l dU_l$, but with the Boltzmann-like action
factor included, i.e. as  $\prod_l dU_l \exp\{-\beta S[U]\}$.
I won't go into the details -- there are standard heat bath,
Metropolis and HMC algorithms available. We thus obtain:
\begin{equation}
\langle \Phi \rangle
=
\frac{1}{Z}
\int \prod_{l} dU_l  \Phi[U] e^{-\beta S}
=
\frac{1}{n} \sum^n_{I=1} \Phi[U^I] + O\left( \frac{1}{\surd n}\right)
\label{eqn_avPhi}
\end{equation}
where the last term is the statistical error.

How, for example, should we calculate the mass gap? 
Recall the standard decomposition of a Euclidean correlator 
of some operator $\phi(t)$ in terms of the energy eigenstates:
\begin{eqnarray}
\langle \phi^\dagger(t=an_t)\phi(0) \rangle
& = &
\langle \phi^\dagger e^{-Han_t} \phi \rangle
=
\sum_i |c_i|^2 e^{-aE_in_t} \nonumber \\
& \stackrel{t\to \infty}{=} & 
|c_0|^2 e^{-am_0n_t}
\label{eqn_cortomass}
\end{eqnarray}
where the lightest mass is $m_0$ and its exponential falls slowest
with $t$ and hence will dominate at large $t$, as shown. Note that
the only states that can contribute are those that have
$c_j = \langle vac | \phi^\dagger | j \rangle \neq 0$, so we should
match the quantum numbers of the operator $\phi$ to those of
the state we are interested in. So typically we construct a $\phi$
with the desired $J^{PC}$ quantum numbers, and if we are interested 
in masses we also make $\phi$ have $\vec{p} = 0$.
Note also that because what we know is the value of $n_t$, we will 
always obtain the mass in lattice units, i.e as $am_0$, when we fit 
our numerical `data' with an exponential in $n_t$. 

So, having decided on a suitable operator $\phi$, we calculate it 
as in eqn(\ref{eqn_avPhi}) with $\Phi=\phi^\dagger(t)\phi(0)$. This 
will produce an estimate of $\langle \phi^\dagger(t)\phi(0) \rangle$ 
with a finite statistical error. To extract a value of $am_0$,
using eqn(\ref{eqn_cortomass}), we
clearly need to have significant evidence for the exponential 
behaviour $\propto e^{-a m_0 n_t}$, over some range of $n_t$, and 
this range needs to be at small enough $n_t$ that the 
exponential is still clearly visible above the statistical
errors. This is obviously harder to achieve for larger $m_0$,
so the systematic error will be larger for heavier states.
However, even for the lighter states we need the (normalised) 
$|c_0|^2 \simeq 1$: our operator needs to be a good wavefunctional 
for the state whose energy we are interested in, so that its
correlator is dominated by this state even at small $n_t$.

Let us now have an explicit example of how to calculate the lightest
$0^{++}$ glueball mass in the $D=3+1$ SU(3) gauge theory. Our
space-time is a $32^4$ hypercubic lattice with periodic boundary
conditions (i.e. a hypertorus). We use a Monte Carlo to generate
typical lattice gauge fields at $\beta = 6/g^2 = 6.515$. Once we
calculate the string tension we will find that this corresponds
to $a\surd\sigma \simeq 0.101$ i.e. $a \simeq 0.05 \mathrm{fm}$
if we use the conventional value of $\surd\sigma \sim 0.5 \mathrm{fm}$.
So the lattice spacing is very small and we are close to the continuum 
limit. 

\begin{figure}[htb]
\begin	{center}
\leavevmode
\input	{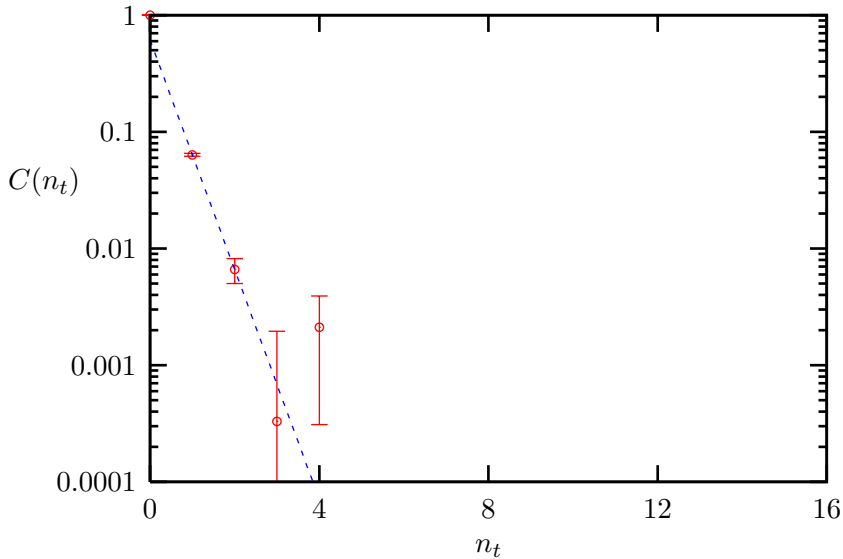}
\end	{center}
\caption{Glueball correlation function with a simple plaquette operator.}
\label{fig_corrpl}
\end{figure}

As a first attempt, we shall try the simplest operator we can think of,
i.e. one that is based on the plaquette that appears in our lattice action:
\begin{equation}
\phi_p(t)
=
\sum_{\vec{x}} \sum^3_{\mu,\nu=1} \mathrm{ReTr}U_{\mu\nu}(\vec{x},t)
\label{eqn_plaqop}
\end{equation}
where $U_{\mu\nu}(x)$ is the product of link-matrices around the
boundary of an elementary square $p$ in the ${\mu\nu}$ plane emanating
out of the site $x$ (in the directions that have been chosen as positive). 
This operator clearly has, as desired, $\vec{p}=0$ and $J^{PC}=0^{++}$, 
since it is explicitly translation and rotation invariant.
However it also has a non-zero overlap onto the vacuum, so we shall
use the shifted operator
$\phi_p(t) \longrightarrow \phi_p(t) - \langle \phi_p \rangle $ 
so as to prevent the vacuum from appearing in the sum over energy
eigenstates in eqn(\ref{eqn_cortomass}). The numerical result for the
correlation function (based on about 100,000 Monte-Carlo generated
gauge fields) is shown in Fig.~\ref{fig_corrpl}. This result
is clearly disappointing. Only for $n_t = 0,1,2$ are the errors
small enough for $C(n_t)$ to be useful. And we cannot put a simple
exponential through these 3 points, although we can do so trivially
for the $n_t=1,2$ points as shown in Fig.~\ref{fig_corrpl}. That is 
to say we have no evidence that the lightest state dominates this
correlation function and we are unable to extract a mass for the
lightest state.

The problem is that the plaquette is so local that it does not see 
the structure of a wave-function and will therefore have a roughly 
equal overlap onto all the eigenstates. Since the number of excited states 
increases rapidly with decreasing $a$, the  (normalised) overlap onto 
the groundstate will decrease rapidly. At the small value of 
$a$ at which we are here working, this overlap is presumably very small 
and we would have to go to quite large $t=an_t$ to suppress the excited 
states and reveal the ground state. Looking at the errors in 
Fig.~\ref{fig_corrpl}, and taking into account that this is based on 
$O(10^5)$ lattice fields, this is clearly unrealistic. So what we need 
are operators that are  `smooth' on a scale $\sim 1 \mathrm{fm}$
and which will therefore have very little overlap onto the 
`oscillating' wavefunctions of excited states. Note that simple
Wilson loops that are merely larger than the plaquette will in general
not be sufficient. We need Wilson loops that are densely packed within
a volume $\sim 1 \mathrm{fm}$ (like a `ball of wool').   
We can efficiently construct such operators that are 
`smooth' on physical length scales, by the iterative spatial
`smearing'
\cite{smear}
and `blocking' 
\cite{block}
of the lattice gauge fields. (This essentially consists of summing
over several paths between two sites, projecting the sum back into
the group, and calling this a smeared link. And then iterating the
procedure.) We can then use these `blocked' link matrices to construct 
appropriate Wilson loops that will be summed as in eqn(\ref{eqn_plaqop}) 
to produce corresponding operators. Using different Wilson loops
and different iteration levels of the blocking will thus produce some
set $\{\phi_i; i=1,...,n\}$ of operators of the desired quantum numbers.
Linear combinations of these operators form a vector space  $V_\phi$,
and we can perform a variational calculation to obtain our best 
estimate, $\psi_0$, of the ground state operator 
\cite{variational,veryoldG,uwmlvar}:
\begin{equation}
\langle {\psi_0}^\dagger(t_0)\psi_0(0)\rangle 
=
\max_{\phi\in V_\phi}
\langle \phi^\dagger(t_0) \phi(0)\rangle 
=
\max_{\phi\in V_\phi}
\langle \phi^\dagger e^{-Ht_0} \phi\rangle 
\end{equation}
where $t_0$ is some convenient value of $t$. Then $\psi_0$ is our 
best variational estimate for the true eigenfunctional of the ground 
state (with these quantum numbers) and we can now use the correlator
$\langle {\psi_0}^\dagger(t)\psi_0(0)\rangle$ to obtain our 
best estimate of the ground state mass. This generalises in an obvious 
way to calculating excited state energies. One constructs from $V_\phi$
the vector space orthogonal to $\psi_0$, repeats the above within this
reduced vector space, and obtains  $\psi_1$ which  is our 
best variational estimate for the true eigenfunctional of the first
excited state. And so on.

\begin{figure}[htb]
\begin	{center}
\leavevmode
\input	{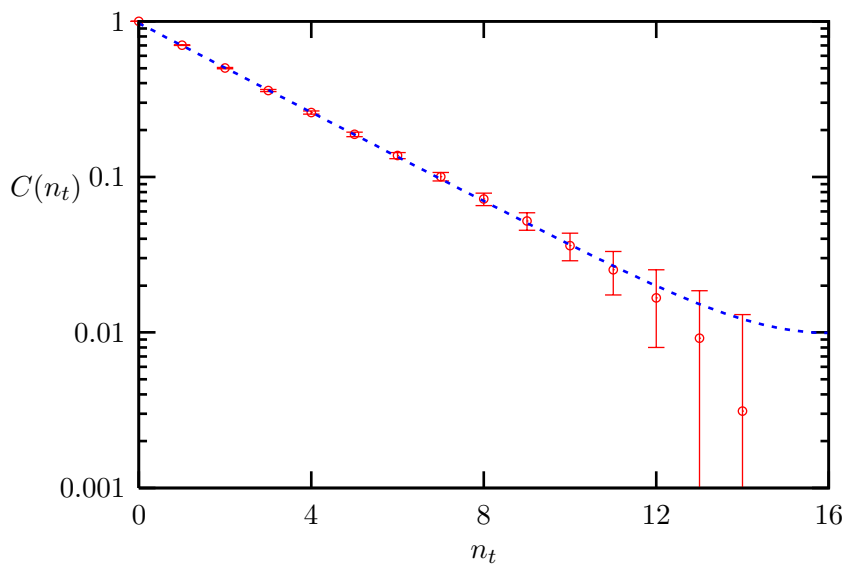}
\end	{center}
\caption{Glueball correlation function with best blocked operator
and cosh fit.}
\label{fig_corr}
\end{figure}

If we do this in our present example we get the 
ground state correlation function
shown in Fig.~\ref{fig_corr}. This can be fitted with a single 
exponential (or rather a cosh because of the periodicity in $t$) 
over a large range of $n_t$ values where $C(n_t)$ is accurately
determined. (We cannot include $n_t=0$ with a good $\chi^2$, and
in fact the overlap factor is $|c_0^2| \simeq 0.97$.) From the
fit we obtain an estimate of $am_{0^{++}} = 0.330(7)$ for the
lightest scalar glueball mass. It turns out that this is the lightest
mass {\it tout court} -- it is the mass gap of the SU(3) gauge theory.
The reassuring point is that we are able to calculate masses 
numerically with errors that are at the percent level. This means
that we will be able to perform meaningful continuum extrapolations. 

\begin{figure}[htb]
\begin	{center}
\leavevmode
\input	{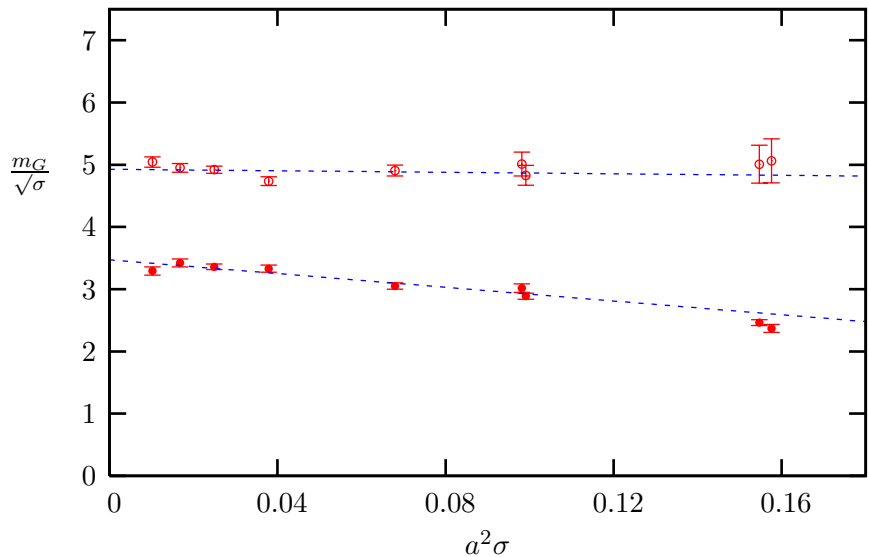}
\end	{center}
\caption{Scalar, $\bullet$, and tensor, $\circ$, glueball masses
with $O(a^2)$ extrapolations to the continuum limit.}
\label{fig_gkN3}
\end{figure}

To obtain the continuum limit from masses that are in lattice units 
and are distorted by the finite lattice cutoff, we take dimensionless 
ratios of masses, to get rid of lattice units, and then extrapolate 
these ratios to $a=0$ with an $O(a^2)$ lattice correction, which is 
what is appropriate for a pure-gauge plaquette action 
\cite{Symanzik},
e.g.
\begin{equation}
\frac{am(a)}{a\surd\sigma(a)}
=
\frac{m(a)}{\surd\sigma(a)}
=
\frac{m(0)}{\surd\sigma(0)}
+ c_0 a^2 \sigma + O(a^4).
\end{equation}
Here we choose to use the square root of the string tension $a^2\sigma$ 
as one of the masses, but we could just as well have chosen some other
glueball mass. (Aside: here $c_0$ is in fact a power series in the bare 
coupling, but the logarithmic variation with $a$ is weak and can usually 
be ignored within the errors of current calculations. This will not remain 
so for ever.) To do such an extrapolation we need to perform the mass
calculations at several values of $\beta$. Doing so for the lightest 
$0^{++}$ and $2^{++}$ glueball masses, we obtain the results displayed in
Fig.~\ref{fig_gkN3}. I show there the linear $O(a^2)$ extrapolations
to the continuum limit, $a^2\sigma = 0$. Clearly they are well-determined 
by these numerically determined mass ratios. We obtain 
\begin{equation}
\frac{m_{0^{++}}}{\surd\sigma} = 3.47(4) - 5.52(75) a^2\sigma
\end{equation}
for the mass gap, and  $m_{2^{++}}/\surd\sigma=4.93(5)-0.61(1.36)a^2\sigma$
for the lightest tensor. (Errors statistical only.) 
In the continuum limit we thus obtain
\begin{equation}
m_{0^{++}} \simeq 3.5 \surd\sigma  \simeq 1.6 \ {\mathrm{GeV}}.
\end{equation}
This now quite old lattice prediction 
\cite{mGold},
has helped to motivate
the now popular phenomenological interpretation (e.g. 
\cite{CloseG})
of the three observed $J^{PC}=0^{++}$ 
flavour 'singlet' states, the $f_0(1350), f_0(1500)$ and $f_0(1700)$, 
as arising from the mixing of nearby $u\overline{u}+d\overline{d}$, 
$s\overline{s}$ and glueball states. 

For our purposes here, the important point is that we are able to
calculate mass ratios in the continuum gauge theory at 
the percent level of accuracy. This means that we should have the
necessary accuracy to  compare physical mass ratios in different 
SU($N$) gauge theories, and to extrapolate to $N=\infty$.

\section{Large $N$ Physics : some basic results from the lattice}
\label{section_largeN}

If the large-$N$ expansion is to be useful for understanding the
strong interactions, the SU($N\to\infty$) gauge theory needs to 
be confining at low temperatures, $T$, and at least 
some of the physics needs to be very similar to that of
the SU(3) theory. This is what we will seek
to establish in the first part of this section.
I shall then describe some recent calculations that include quarks,
and which begin to address the question whether the mesonic spectrum
of   $QCD$ is `close to' $QCD_{N=\infty}$.
Finally I will return to the question of how one should take
a smooth large-$N$ limit : do our non-perturbative calculations
support the diagrammatic expectation that you hold $g^2 N$ fixed?

Our method is simple. We calculate physical mass 
ratios first for SU(2), then for SU(3), then for SU(4), ..., and
continue for larger groups until we have good evidence that our
results are indeed converging to a large-$N$ limit with the
expected leading $O(1/N^2)$ corrections. This method is pedestrian 
but effective.

I shall finish with some remarks about how the cost of the calculations
grow with $N$. In fact the growth is unexpectedly modest in many
situations -- which I hope will encourage some of you to get 
actively involved.

\subsection{Are large-$N$ gauge theories linearly confining?}

I will answer this question first by example and then, later
on, more quantitatively.

The example is taken from SU(6) in $D=3+1$
\cite{HMMT04}.
We calculate the mass of the lightest state in which one unit of 
fundamental flux closes upon itself by winding once around a spatial
torus. Suppose this torus is of length $l$. Then if we have linear
confinement the flux will organise itself into a flux tube (of the
same kind as would join two distant fundamental sources) and the
mass will grow linearly with $l$ for large $l$, 
$m(l) \stackrel{l\to\infty}{=} \sigma l$. How you actually do this
and what operators you need to use, will be discussed in some detail 
later on in these lectures.

\begin{figure}[htb]
\begin	{center}
\leavevmode
\input	{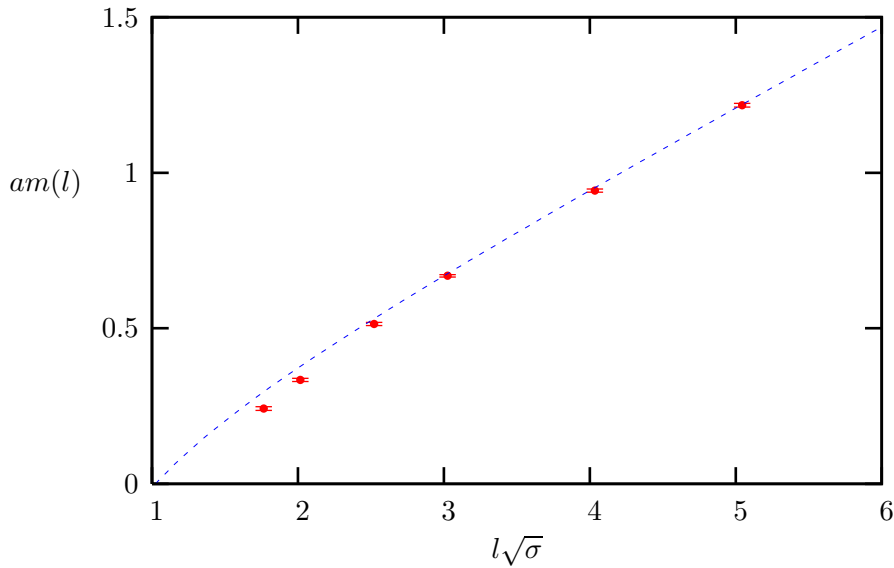}
\end	{center}
\caption{Ground state energy of a flux loop winding around a spatial 
torus of length $l$, in SU(6) and $D=3+1$.}
\label{fig_K1n6d4}
\end{figure}

The results of this calculation are shown in Fig.~\ref{fig_K1n6d4}.
It is immediately apparent from the plot that we do indeed have the 
(approximately) linear increase with $l$ that indicates linear 
confinement. So that you can judge what is the length $l$ in physical 
units, I have used the value of $a\surd\sigma$ from our fits to 
translate the lattice size $l=aL$ into physical units using
$l\surd\sigma = aL\surd\sigma=L \times a\surd\sigma$. Since we expect
the intrinsic width of a flux tube to be $O(1/\surd\sigma)$ we
can see that our largest values of $l$ are indeed large compared to
the flux tube width and it is reasonable to infer that what we are 
seeing is the onset of an asymptotic linear behaviour. 

The dashed line shown on the plot represents a linear piece modified 
by the Luscher correction term
\begin{equation}
m(l) = \sigma l - \frac{\pi}{3l}.
\end{equation}
This $O(1/l)$ correction is universal and the value used here 
corresponds to the universality class of a simple bosonic string
where the only massless modes are those of the transverse oscillations.
We can see from  Fig.~\ref{fig_K1n6d4} that this correction captures 
the bulk of the observed deviation from linearity. (One of course 
expects further corrections that are higher powers of $1/l$.)
So we have good evidence not only that linear confinement persists
at large $N$, but that it remains in the same universality class
as has been established by previous work for  SU(2) and SU(3).

\subsection{Is SU(3) close to SU($\infty$)?}

\begin{figure}[htb]
\begin	{center}
\leavevmode
\input	{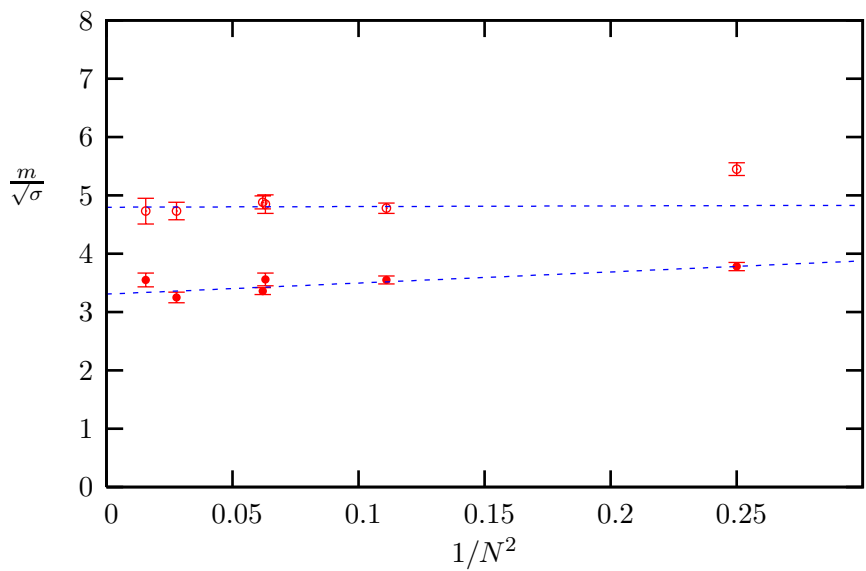}
\end	{center}
\caption{Scalar, $\bullet$, and tensor, $\circ$, glueball masses
in various continuum SU($N$) gauge theories, with $O(1/N^2)$
extrapolations to $N=\infty$ limit.}
\label{fig_gkNwa}
\end{figure}

In Fig.~\ref{fig_corr} we showed how to calculate a mass on the 
lattice, and in Fig.~\ref{fig_gkN3} how to obtain ratios of masses
in the physical continuum limit of the theory.
That example was for SU(3) but one can do the same for other values
of $N$. In Fig.~\ref{fig_gkNwa} I plot the resulting continuum ratios
for $N=2,3,4,6,8$
\cite{blmtuwG}.
Since the leading correction is expected to
be $O(1/N^2)$, I plot the ratios against $1/N^2$. In such a plot
the large-$N$ extrapolation should be a simple straight line
for large enough $N$. In practice, within our errors, large
enough $N$ turns out to mean $N\geq 3$ (indeed, $N\geq 2$ for
the scalar glueball) as we see from the linear fits on the
plot. It is also evident that the coefficient of this $1/N^2$ 
term is typically quite modest (compared to the $N=\infty$ value 
of the ratio). Thus this provides an example of the fact that
for many basic physical quantities  
\begin{equation}
SU(3) \simeq SU(\infty).
\end{equation}

There is something else that we can infer from  Fig.~\ref{fig_gkNwa}.
We provided evidence earlier on that linear confinement persists
at large $N$. However it could still be that it disappears as 
$N\to\infty$ through $\sigma$ vanishing in that limit. What 
Fig.~\ref{fig_gkNwa} demonstrates is that this is not the case.
In units of the physical glueball masses  $\sigma$ remains 
non-zero at $N = \infty$. (Logically, one should first show a
plot of the ratio of the scalar to tensor masses in order to 
establish the existence of a smooth large $N$ limit, but it is
clear from Fig.~\ref{fig_gkNwa} that this is the case.)  

\subsection{Is $QCD_3$ close to $QCD_\infty$?}

To estabish the phenomenological relevance of the large-$N$ limit
we need to consider mesons as well as glueballs. Once we have
fields in the fundamental representation, like quarks, we have
$O(1/N)$ corrections. (A quark self-energy loop in the gluon 
propagator will look just like Fig.~\ref{fig_planar} except 
without the innermost closed loop and the accompanying factor 
of $N$ from the sum over colours.) Thus the fact that we see
$SU(3) \sim SU(\infty)$ in the pure gauge theory, where the
leading correction is $O(1/N^2)$, does not guarantee that
the meson spectrum will be so well-behaved. Whether it is  
needs to be checked, and there have been three recent calculations 
that have begun to do precisely that
\cite{gbfb,ldd,ahrn}.

As $N \to\infty$ quark loops are suppressed by a factor $1/N$
compared to gluon loops,
and so to leading order the vacuum of $QCD_{N=\infty}$ is the same 
as that of the SU($N$) gauge theory. (As long as the quarks are not
precisely massless, when subtle issues arise.) We can of course still
ask what happens to the spectrum of mesons at $N=\infty$, by explicitly
calculating their propagators. When doing so we are calculating them in 
in what one would usually call the relativistic valence quark  
approximation except that here it is not an approximation, because 
the quark loops are not being neglected but are dynamically suppressed.
(And in addition the gluonic vacuum in which the quarks propagate
is the complete non-perturbative vacuum and not some crude
approximation thereof.)

This suggests an efficient way to proceed. (A straightforward
calculation of full $QCD_N$ with light quarks being too 
expensive.) At various finite $N$ one performs the meson spectrum 
calculation without vacuum quark loops -- what is called the
`quenched approximation' in the lattice community. One then
extrapolates the quenched results to $N=\infty$. Since we have
no quark loops the leading correction should be $O(1/N^2)$.
The extrapolated values are the correct values for $QCD_{N=\infty}$ 
since that theory is dynamically quenched. We now compare the
spectrum at $N=\infty$ with the experimental spectrum (or that of
recent full lattice $QCD$ calculations, which indeed agree with 
experiment). 

This looks like a win-win approach except for the fact that
quenched QCD at finite $N$ is not unitary. However the pathologies 
are subtle and appear primarily at small quark masses, so if
one extrapolates to $N=\infty$ at fixed non-zero quark masses
and only then to small quark masses, one should be largely protected
from them. Current calculations are not so pedantic, but since
they are probably not accurate enough to be sensitive to such 
pathologies anyway, this does not really matter.

\begin{figure}[htb]
\begin	{center}
\leavevmode
\centerline{
\includegraphics[scale=.7,angle=360,width=5.5cm]{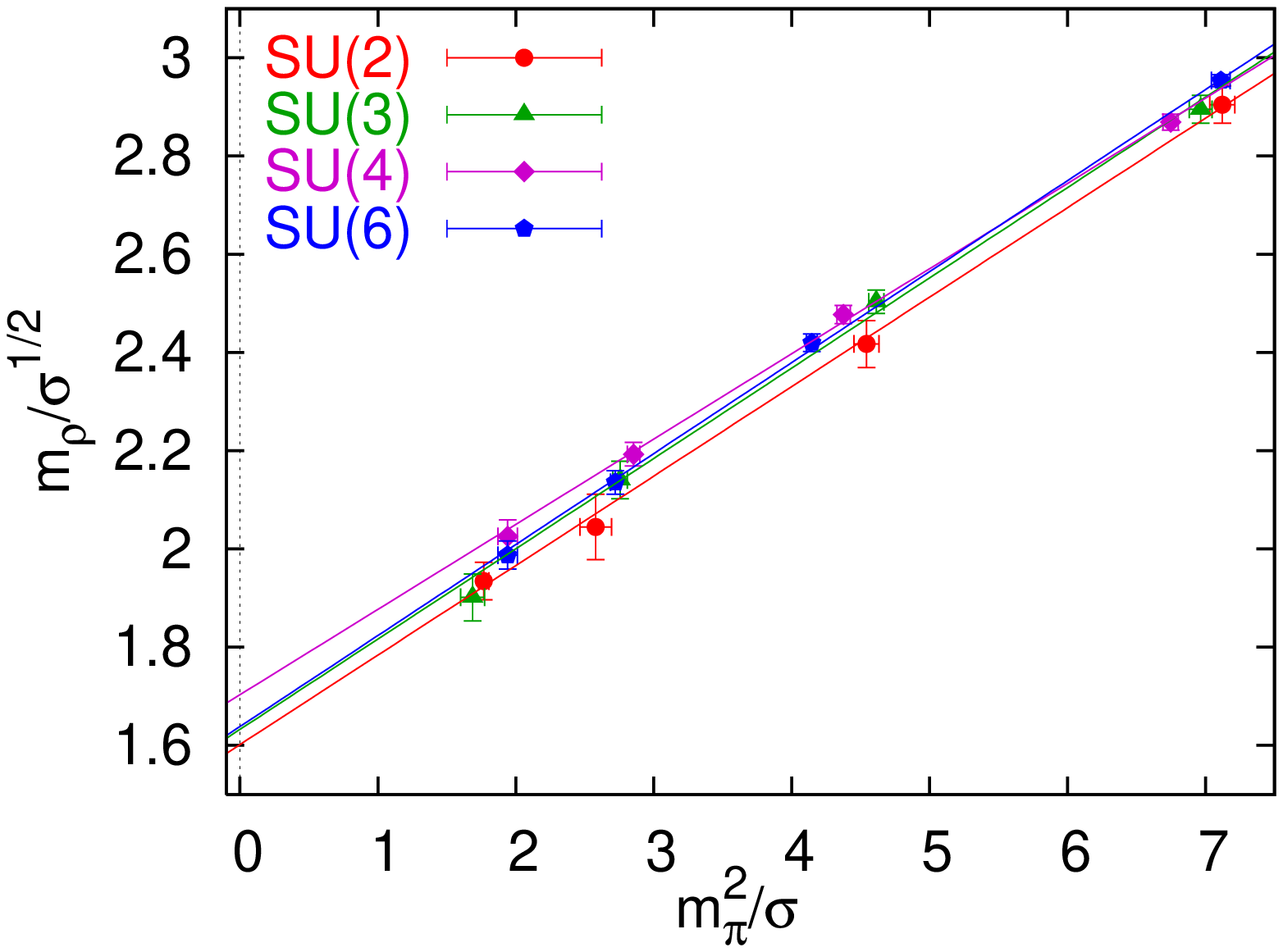} \hspace{10mm} \includegraphics[scale=.7,angle=360,width=5.5cm]{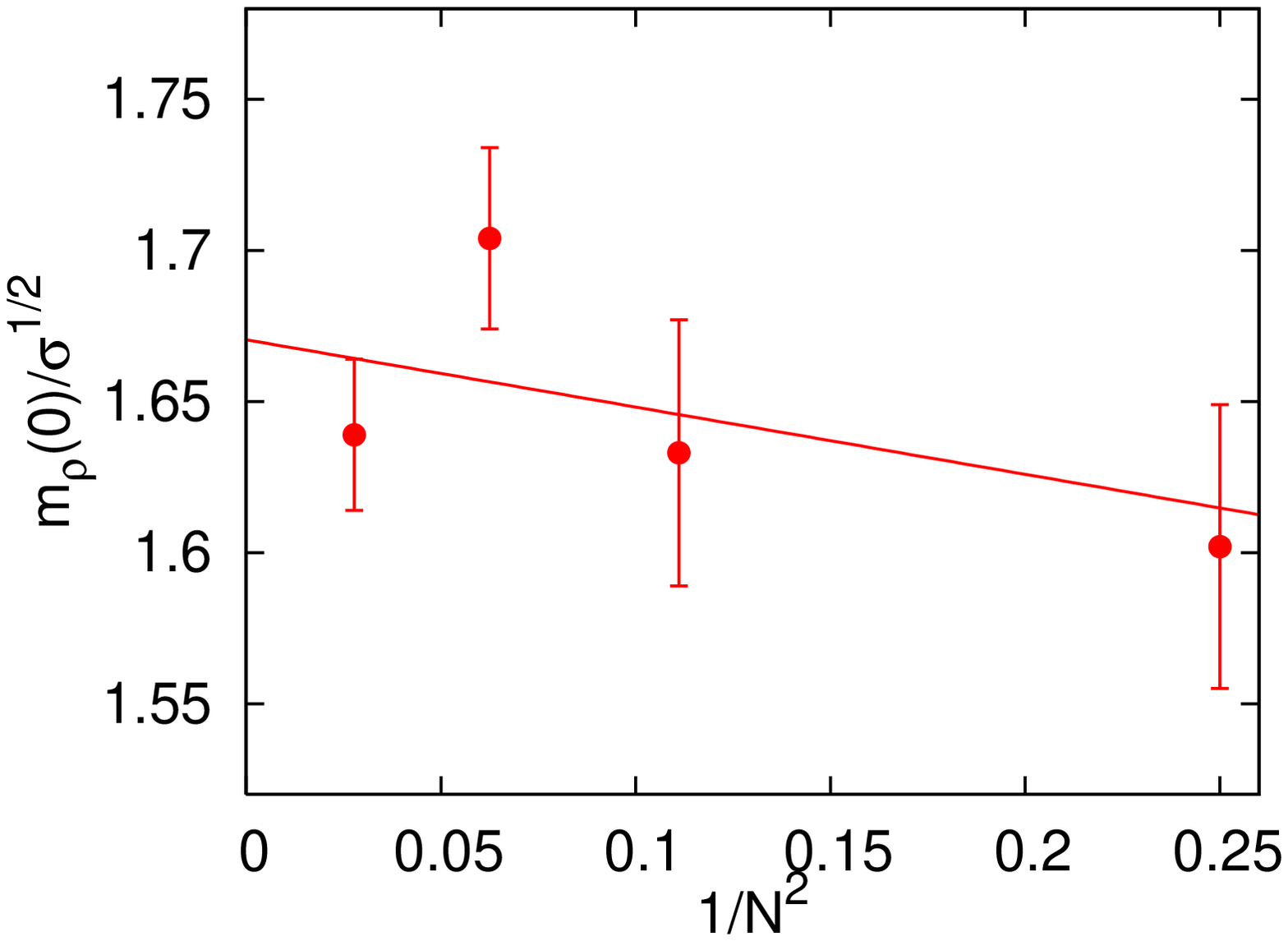} }
\end	{center}
\caption{Left: the $\rho$ mass against $m^2_{\pi}$ for various SU($N$).
Right: chiral extrapolation of $m_\rho$ for various $N$ and an $O(1/N^2)$
extrapolation to $N=\infty$.}
\label{fig_qcdN}
\end{figure}

In Fig.~\ref{fig_qcdN} I show some plots borrowed from
\cite{gbfb}.
On the left is a plot of the $\rho$-meson mass against $m^2_\pi$
for $N=2,3,4,6$ and obtained at one value of $a\surd\sigma$ (chosen 
to be very similar for all the SU($N$) groups). Such a linear behaviour 
is what one expects if $m_{\rho}(m_q) = m_{\rho}(0) + c m_q$ and 
$m^2_\pi \propto m_q$ (spontaneous chiral symmetry breaking).
We see that there is very little variation with $N$. On the right
is a plot of the chirally extrapolated $m_{\rho}(0)$  against
$1/N^2$, which also shows little variation.

So at this particular value of $a$, corresponding to $a\sqrt\sigma = 0.209$,
one obtains $\lim_{N\to\infty} {m_\rho}/{\sqrt\sigma}=1.688(25)$.
Now, fortunately there has been another calculation
\cite{ldd}, 
with exactly the same lattice action, but at a different lattice spacing,
$a\sqrt\sigma = 0.335$, obtaining 
$\lim_{N\to\infty} {m_\rho}/{\sqrt\sigma}=1.627(10)$. This allows us 
to make an $O(a)$ continuum extrapolation. (Although with two points 
it is of course not under good control.) This gives a value
\cite{gbfb}, 
\begin{equation}
\lim_{N\to\infty , a\to 0} \frac{m_\rho}{\sqrt\sigma}=1.79(5)
\end{equation}
to be compared with the real world value of
\begin{equation}
\frac{m_\rho}{\sqrt\sigma} 
\simeq \frac{770\mathrm{MeV}}{440\mathrm{MeV}}
\simeq 1.75
\end{equation}
So the unambiguous conclusion from these two numerical studies
\cite{gbfb,ldd}
is that as far as the $\rho$-mass is concerned, $QCD$ is indeed
close to $QCD_{\infty}$. 

Unfortunately things are not so clear-cut. There is a third and
very recent study
\cite{ahrn}
using very different methods that comes to a quite different
conclusion. These calculations are at much larger $N$,
SU(17) and SU(19), and on a small volume, using the fact that
as $N\to \infty$ finite volume effects vanish (for a broad class
of observables). The propagators are calculated in the pure gauge
theory with the same plaquette action, but with Neuberger 
(overlap) rather than Wilson fermions, and hence have good
chiral properties at $a\neq 0$. Moreover they are calculated in 
momentum rather than position space. And the value of $a\surd\sigma$ 
is very similar to that in
\cite{gbfb}.
However the conclusion is very different:
\begin{equation}
m_\rho \stackrel{N=19}{\simeq} 5.86 T_c \simeq 3.5 \sqrt\sigma 
\simeq 2 m^{QCD}_\rho.
\end{equation}
where $T_c$ is the deconfining temperature. 
(Here I have translated the value given in
\cite{ahrn}
into units of $\surd\sigma$ using the known values of $T_c/\surd\sigma$
\cite{blmtuwTc}
rather than normalising to $f_\pi$ as done in
\cite{ahrn}.
I prefer to set the scale this way because we know that the
mutual ratios of $T_c$, $\surd\sigma$ and $m_G$ have modest
$O(1/N^2)$ corrections.) 
That is to say, $QCD$ is far from $QCD_{\infty}$ for meson masses.

This clear-cut discrepancy needs to be sorted out. Since the value
of $a$ is the same, that only leaves the $N$-dependence. However
I think it is completely implausible that there should be very large
corrections between $N=6$ and $N=19$ if the variation is already 
very weak for $N \in [2,6]$ (as we have seen in Fig.~\ref{fig_qcdN}).
I would just remark that the calculations of
\cite{gbfb,ldd}
are completely standard in lattice QCD and all the systematic
errors are supposedly well-understood. By contrast the calculations
in 
\cite{ahrn}
are novel in several respects, being designed specifically
to make very large $N$ calculations possible. In particular
the momentum space propagator is evaluated for only a small
number of small momenta $p^2 \ll m^2_\rho$, and one might therefore
wonder if there might not be a large excited state contribution 
that cannot be readily resolved though a fit with more than 
one (Euclidean) pole term. 

So for the moment we must hold our breath. However once this
discrepancy is understood, it will be very interesting to 
address any number of other questions in $QCD_\infty$,
where all particles are stable and well-defined and do not mix.
In particular it would be very nice to see the scalar nonet
and scalar glueball within one calculation. Also the tensor
and pseudoscalar glueballs and the nearby mesons with those 
quantum numbers. (These mesons will presumably be radial
excitations.) All this could serve as a very useful guide for 
the  corresponding phenomenology in real QCD.

\subsection{$g^2\propto 1/N$ for a smooth large $N$ limit?}

In the calculations I have been describing, at each $N$ we 
calculate quantities such as  $a\surd\sigma$ at a number of
values of $\beta=2N/g^2_L(a)$. Using this we can compare how
the bare coupling runs with its scale at different $N$ and we can 
check whether our non-perturbative results confirm the
perturbative expectation that $g^2\propto 1/N$. This I will
describe in this Section.

The bare coupling necessarily has lattice spacing corrections,
and this presents some minor complications. However there are
now some very nice lattice calculations of the running coupling 
in the continuum theory, which I will also describe.

The same question can also be asked and answered in $D=2+1$.
Since the analysis is much more direct there (because $g^2$ has 
dimensions of mass), I will begin with that case.

\subsubsection{$D=2+1$}

\begin{figure}[htb]
\begin	{center}
\leavevmode
\input {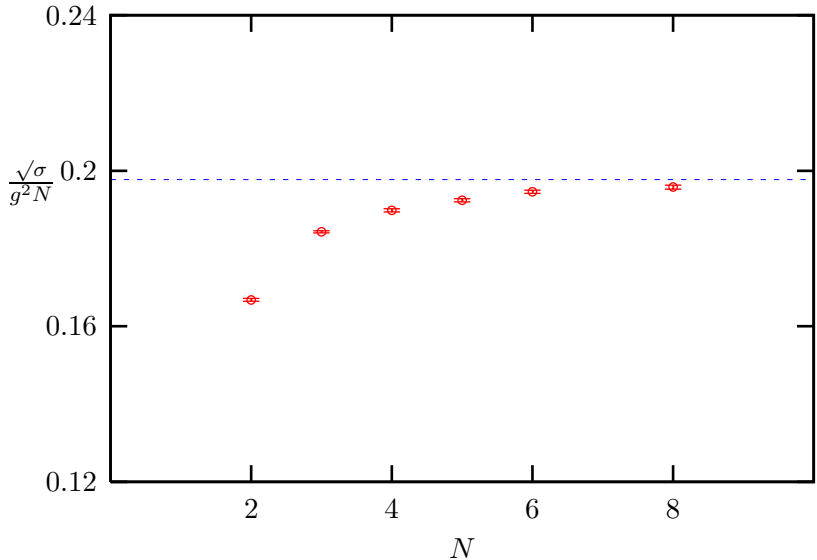}
\end	{center}
\caption{String tension in units of $g^2N$ for various continuum
 SU($N$) gauge theories in $D=2+1$.}
\label{fig_g2ND3}
\end{figure}

Suppose we calculate $a\surd\sigma$ at a number of values of $\beta$. 
Then we can calculate the continuum value of $\surd\sigma/g^2$ 
as follows
\cite{MT98d3}:
\begin{equation}
\beta a\surd\sigma(a)
=
\frac{2N}{ag^2}  a\surd\sigma(a)
=
2N \frac{\surd\sigma(a)}{g^2}
\stackrel{\beta \to \infty}{\longrightarrow}
2N\frac{\surd\sigma}{g^2}
\label{eqn_g2ND3}
\end{equation}
We can now examine how $\surd\sigma/g^2$ varies with $N$. The statement
that $g^2\propto 1/N$ is equivalent, in this context, to saying that 
\begin{equation}
\frac{\surd\sigma}{g^2 N}
\stackrel{N \to \infty}{\longrightarrow}
\mathrm{const}
\label{eqn_g2ND3b}
\end{equation}
In Fig.~\ref{fig_g2ND3} I display the continuum values of 
$\surd\sigma/g^2N$ for $N \in [2,8]$ as calculated in
\cite{BBMTk1d3}.
It is clear from this plot that we have very strong numerical evidence
for eqn(\ref{eqn_g2ND3b}) being correct.

The errors in  Fig.~\ref{fig_g2ND3} are so small (they are given
by the vertical spread of the horizontal error bars) that we
can hope to say something about the power of the leading correction
to the asymptotic behaviour in eqn(\ref{eqn_g2ND3b}).
Fitting with a correction $\propto 1/N^\gamma$ we find
\begin{equation}
\frac{\surd\sigma}{g^2 N}
=
c_0 + \frac{c_1}{N^\gamma}
\qquad \longrightarrow \qquad
\gamma = 2.01 \pm 0.20.
\label{eqn_g2ND3corr}
\end{equation}
So if we assume that $\gamma$ has to be integer, we can conclude 
that the leading correction is indeed $O(1/N^2)$ as predicted by
't Hooft's diagrammatic analysis.

So our numerical calculations of $\surd\sigma/g^2$ 
in $D=2+1$ SU($N$) gauge theories have confirmed that it is
$\surd\sigma/g^2N$ that has a smooth limit as $N\to\infty$,
and that the way this limit is approached is with a  $O(1/N^2)$
correction. Thus our fully non-perturbative calculation
confirms the conventional expectations based on 't Hooft's
diagrammatic analysis.

\subsubsection{$D=3+1$ : running bare coupling}

Once again, we have a calculation of  $a\surd\sigma$ at a number 
of values of $\beta$, for each value of $N$. However in 
$D=3+1$ the bare coupling is dimensionless so the analysis
will be less direct than the above. 

Recall that $\beta=2N/g^2_L(a)$ gives us a definition of
the running coupling on the distance scale $a$, in what we
can call the `lattice scheme' $L$. It is more 
useful to write it as  $g^2_L(a\surd\sigma)$ so that the argument 
is expressed in physical units in a way that is common for all $N$.
Now it has been known for a long time, in the lattice community, 
that $g^2_L$ is a `bad' scheme 
in the sense that higher order corrections are typically much larger 
than you would have with something like the $\overline{MS}$ scheme.
One of the earliest and simplest remedies for this was Parisi's 
mean-field improvement
\cite{Parisi_MFI}
(nowadays often known as tadpole improvement
\cite{Lepage_TI}).
This involves defining a new coupling, $g^2_I(a)$, 
\begin{equation}
g^2_I(a)
=
\frac{g^2_L(a)}{u_p}
=
\frac{2N}{\beta}\frac{1}{u_p}
\label{eqn_g2I}
\end{equation}
where  $u_p \equiv \langle \mathrm{Tr} U_p \rangle /N$ is the
average plaquette.
Since the plaquette is trivial to calculate, this is a convenient
improvement to apply.

\begin{figure}[htb]
\begin	{center}
\leavevmode
\input	{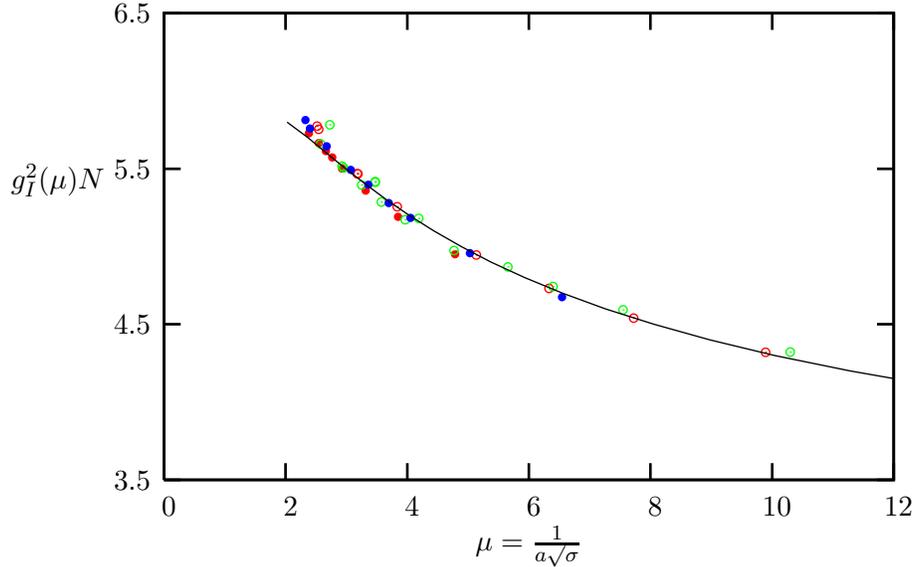}
\end	{center}
\caption{The running of the (improved) lattice 't Hooft coupling
for various $N$, from SU(2) (green open circles) to SU(8) 
(red solid points).}
\label{fig_g2IND4}
\end{figure}

Having calculated $g^2_I(a\surd\sigma)$ for various $a\surd\sigma$ 
and for various $N$, we plot the results for the product $g^2_IN$
in Fig.~\ref{fig_g2IND4}. We plot it against the corresponding 
energy scale, $\mu = 1/a\surd\sigma$ so that it looks more like 
the plots of the running coupling (against $Q$) that you will
normally encounter. (An earlier version of this kind of plot
appeared in
\cite{BLMTgkd4}
with this version being borrowed from
\cite{MTlat08}.)
This plot includes values for $N = 2,3,4,6,8$. Although the points 
are perhaps hard to distinguish, it is clear that there is a
common running 't Hooft coupling, $\lambda = g^2N$, for all these 
values of $N$ to a very good approximation. (One sees some 
dispersion at the coarsest lattice spacings
where we are around  the crossover to lattice strong coupling,
which becomes a phase transition for $N\geq 5$.) 

The solid line is a fit that incorporates 3-loop continuum running 
and $O(a^2)$ lattice spacing corrections
\cite{CAMTAT}:
\begin{eqnarray}
a \sqrt\sigma(a)
& = &
{\mathrm{lattice \ correction}\times \mathrm{3 \, loop \ continuum \ running}} 
\nonumber \\
& = &
\frac{\sqrt\sigma(0)}{\Lambda_I}
\left( 1 + c a^2\sigma \right)
e^{-\frac{1}{2\beta_0 g_I^2}}
\left(\frac{\beta_1}{\beta_0^2}+\frac{1}{\beta_0 g_I^2}
\right)^\frac{\beta_1}{2\beta_0^2} 
e^{-\frac{\beta^I_2}{2\beta_0^2}g_I^2} 
\label{eqn_gIrun}
\end{eqnarray}
where $\beta_0, \beta_1$ are the first two (and universal)
coefficients of the $\beta$-function, while  $\beta^I_2$ is the
third (scheme-dependent) coefficient. 
The fit shown is actually to the SU(3) running coupling, but
on this plot it fits other $N$ almost as well. If we perform 
such fits separately at each $N$, extract a value of $\Lambda_I$ 
in each case, and then convert it to $\Lambda_{\overline{MS}}$, 
we find that the latter can be well fitted by
\cite{CAMTAT}
\begin{equation}
\frac{\Lambda_{\overline{MS}}}{\surd\sigma}
=
0.503(2)(40) + \frac{0.33(3)(3)}{N^2}
\label{eqn_LambdaN}
\end{equation}

So it is clear that the diagrammatic prediction $g^2 \propto 1/N$ 
is confirmed at the non-perturbative level in $D=3+1$ as well
as in $D=2+1$ gauge theories.

\subsubsection{$D=3+1$ : a running continuum coupling}

The calculation in the previous section suffers both from the
complication of lattice spacing corrections, as in eqn(\ref{eqn_gIrun}),
and, more importantly, from a really small range of energy scales,
as we see in Fig~\ref{fig_g2IND4}. This is because with this method
we need to calculate the string tension at each value of $a$, and
this requires a lattice that grows as $1/a$ in lattice units
to avoid large finite volume corrections. So it is not practical
to go to extremely small value of $a$.

\begin{figure}[htb]
\hspace{0cm}
\vspace{-4.2cm}
\begin	{center}
\leavevmode
\epsfig{figure=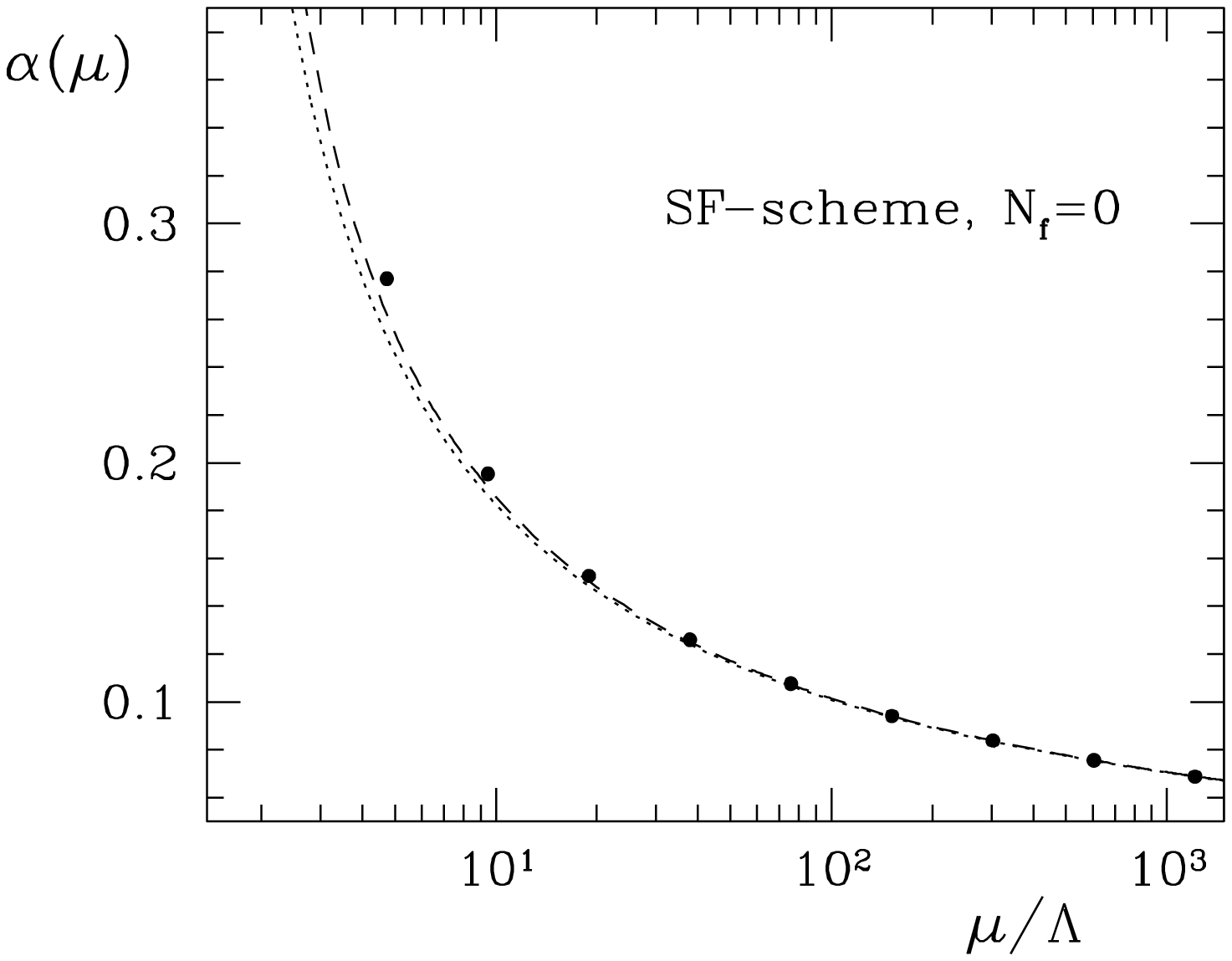, angle=360, width=14cm} 
\end	{center}
\vspace{-1.2cm}
\caption{Continuum running coupling in SU(3) in the SF scheme
 from \cite{SFsu3}.}
\label{fig_gSFN3}
\end{figure}

\begin{figure}[htb]
\begin	{center}
\leavevmode
\epsfig{figure=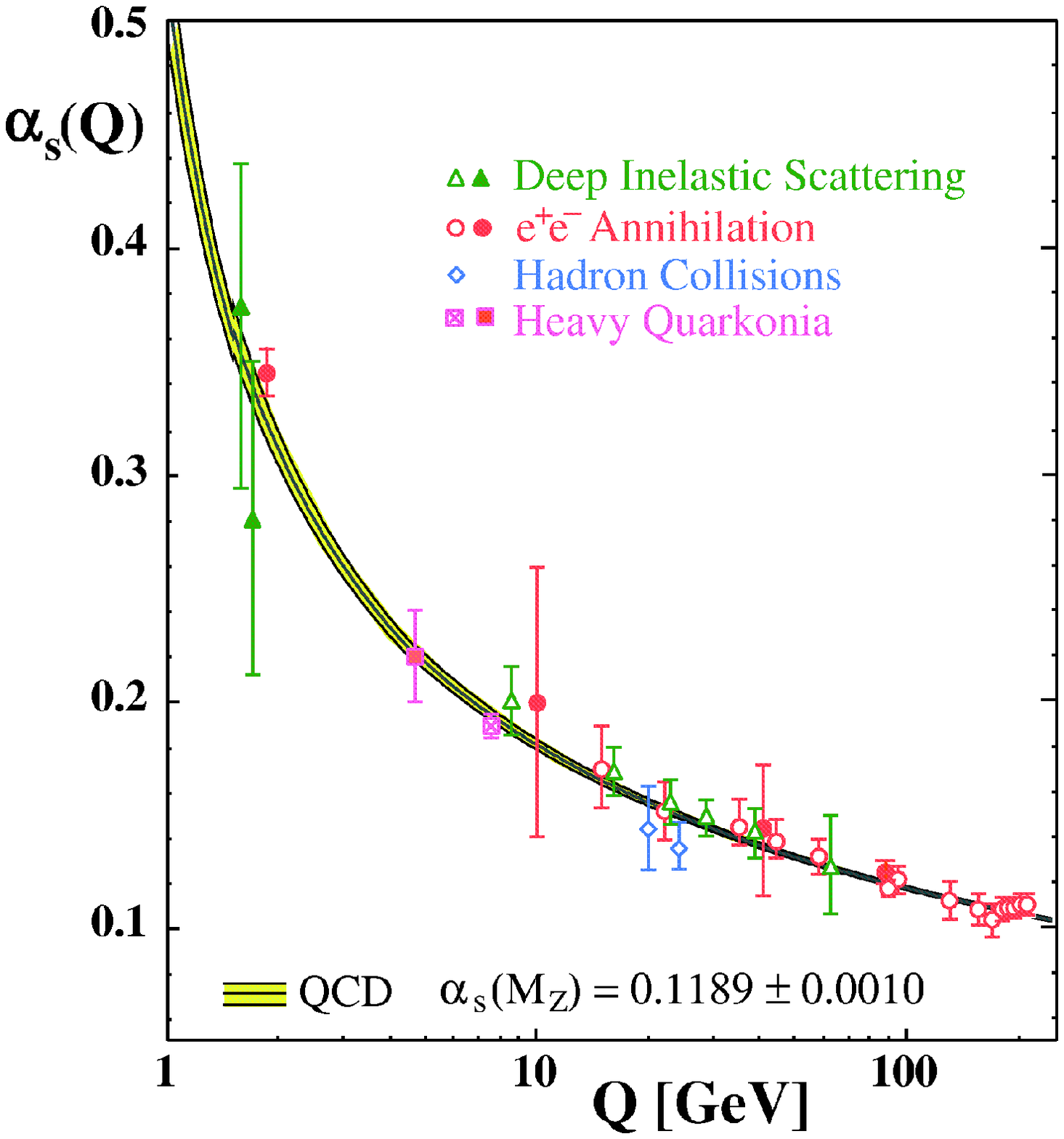, angle=360, width=10cm} 
\end	{center}
\caption{${\overline{MS}}$ running coupling: compilation of experimental
results from \cite{bethke}.}
\label{fig_gMSexp}
\end{figure}

Both of these issues are addressed in the step-scaling analysis
developed by the $Alpha$ collaboration
\cite{SFsu3}.
I do not have the time to discuss this very nice method, but
will just remark that it is designed to allow a continuum
extrapolation of the running coupling, over a very large range of 
energy scales. I have borrowed a plot of the SU(3) running
coupling from
\cite{SFsu3}
which I display in Fig.~\ref{fig_gSFN3}. For comparison I show 
in Fig.~\ref{fig_gMSexp} a quite recent compilation of experimental 
determinations of the running coupling that I have borrowed from
\cite{bethke}.
As you can see, the lattice calculation (which includes an extrapolation
to the continuum limit) is more accurate than the experimental one,
and extends over a range of scales that is at least as large.
We see a very impressive comparison with 3-loop continuum running,
beginning at very high energies where we can have confidence in the
applicability of perturbation theory.

\begin{figure}[htb]
\hspace{0cm}
\vspace{1.0cm}
\begin	{center}
\leavevmode
\epsfig{figure=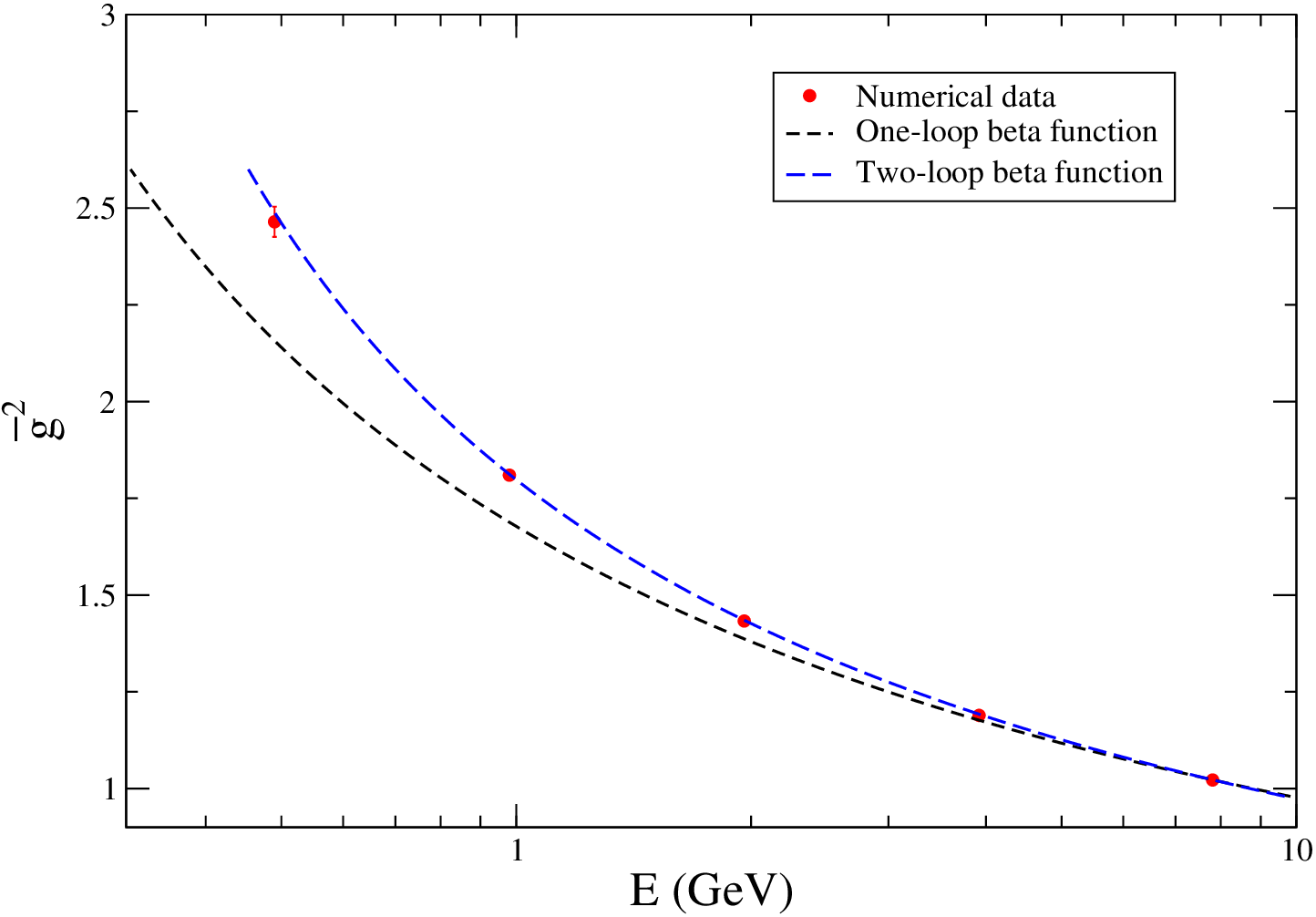, angle=360, width=10cm} 
\end	{center}
\caption{Continuum running coupling in SU(4) in the SF scheme
 from \cite{SFsu4}.}
\label{fig_gSFN4}
\end{figure}

However my purpose here is not to dwell upon these calculations
in any detail, but to point out that there have recently 
\cite{SFsu4}
been calculations of this kind in SU(4). I show the corresponding 
plot, borrowed from
\cite{SFsu4},
in Fig.~\ref{fig_gSFN4}. The range of energies is less impressive 
but is still non-trivial. Extracting the $\Lambda$ parameter
from the fits, and converting to the standard $\overline{MS}$
scheme, and using the results of earlier calculations
for $N=2$ and $N=3$, one finds
\cite{SFsu4}
\begin{equation}
\frac{\Lambda_{\overline{MS}}}{\surd\sigma}
=
0.528(40) + \frac{0.18(36)}{N^2} \qquad : N\geq 3
\label{eqn_LambdaSFN}
\end{equation}
Since this is a straight-line fit to just 2 points, $N=3$ and $N=4$,
it will require further confirmation, but it is reassuring that it is
consistent with the result in eqn(\ref{eqn_LambdaN}), obtained from 
the quite different approach of fitting the running of the bare 
lattice coupling.

\subsection{How hard are large $N$ lattice calculations?}

In the pure gauge theory, we are mostly calculating products
of $N\times N$ matrices, and the computational cost of that
clearly increases as $\propto N^3$. The Monte Carlo update of the 
matrices proceeds by updating the SU(2) sub-groups (using
standard Cabibbo-Marinari). This cost grows only as  $\propto N^2$
if one updates all the subgroups and so is relatively unimportant. 
The $O(N^3)$ increase in cost can be partially reduced
if one takes advantage of the fact that one can often work on a small
volume (as long as $l > 1/T_c$) at large $N$. As an extreme example,
the deconfining transition 
\cite{blmtuwTc,blmtuwTcprop}
can be calculated with just about the same precision on $10^3 5$ lattices 
in SU(8) as on $64^3 5$ lattices in SU(3). Here the volume gain 
more than outweighs the $N^3$ loss. And for a dramatic example of this, 
at very large $N$, see the string tension calculation in
\cite{jkrn}.

The above has to do with how the cost in generating a single lattice 
field grows with $N$. However the relevant question is what is
the cost of achieving a given error/signal ratio in the calculation
of some physical quantity like the mass gap.

Indeed, since we calculate masses from connected correlators of
traced operators, i.e. correlations between singlet fluctuations, 
and since we know that all such fluctuations vanish as $N\to\infty$, 
one might wonder whether this renders mass calculations impossible 
in that limit.

The answer is no: the errors on the fluctuations are themselves
determined by higher order correlators, which generically vanish at
the same rate in the pure gauge theory. For example consider
$C(t) = \langle \phi^\dagger(t) \phi(0) \rangle$ where $\phi$
is a trace of some gauge field operator, and $\langle\phi \rangle = 0$ 
so that there is only the disconnected piece to consider. Then in
a numerical estimate of  $C(t)$ its
fluctuation squared is proportional to the higher order 
correlator $\langle [\phi^\dagger(t) \phi(0) - C(t)]^2 \rangle$. An 
analysis of this using the usual large $N$ counting rules, shows
that both the average value of the correlator and the fluctuation 
around that average disappear with the same power of $N$.
That is to say, as $N\to\infty$ there are no extra hidden costs 
to extracting masses from correlation functions.

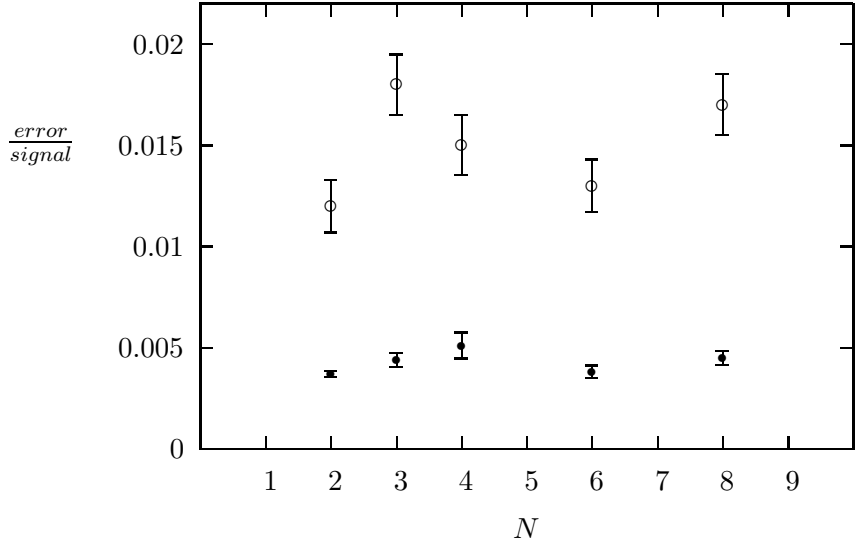
\begin{figure}[htb]
\begin	{center}
\leavevmode
\setlength{\unitlength}{0.240900pt}
\ifx\plotpoint\undefined\newsavebox{\plotpoint}\fi
\sbox{\plotpoint}{\rule[-0.200pt]{0.400pt}{0.400pt}}%
\begin{picture}(1500,900)(0,0)
\sbox{\plotpoint}{\rule[-0.200pt]{0.400pt}{0.400pt}}%
\put(400.0,150.0){\rule[-0.200pt]{4.818pt}{0.400pt}}
\put(375,150){\makebox(0,0)[r]{\ \ {$0$}}}
\put(1405.0,150.0){\rule[-0.200pt]{4.818pt}{0.400pt}}
\put(400.0,309.0){\rule[-0.200pt]{4.818pt}{0.400pt}}
\put(375,309){\makebox(0,0)[r]{\ \ {$0.005$}}}
\put(1405.0,309.0){\rule[-0.200pt]{4.818pt}{0.400pt}}
\put(400.0,468.0){\rule[-0.200pt]{4.818pt}{0.400pt}}
\put(375,468){\makebox(0,0)[r]{\ \ {$0.01$}}}
\put(1405.0,468.0){\rule[-0.200pt]{4.818pt}{0.400pt}}
\put(400.0,627.0){\rule[-0.200pt]{4.818pt}{0.400pt}}
\put(375,627){\makebox(0,0)[r]{\ \ {$0.015$}}}
\put(1405.0,627.0){\rule[-0.200pt]{4.818pt}{0.400pt}}
\put(400.0,786.0){\rule[-0.200pt]{4.818pt}{0.400pt}}
\put(375,786){\makebox(0,0)[r]{\ \ {$0.02$}}}
\put(1405.0,786.0){\rule[-0.200pt]{4.818pt}{0.400pt}}
\put(503.0,150.0){\rule[-0.200pt]{0.400pt}{4.818pt}}
\put(503,100){\makebox(0,0){\ {$1$}}}
\put(503.0,830.0){\rule[-0.200pt]{0.400pt}{4.818pt}}
\put(605.0,150.0){\rule[-0.200pt]{0.400pt}{4.818pt}}
\put(605,100){\makebox(0,0){\ {$2$}}}
\put(605.0,830.0){\rule[-0.200pt]{0.400pt}{4.818pt}}
\put(708.0,150.0){\rule[-0.200pt]{0.400pt}{4.818pt}}
\put(708,100){\makebox(0,0){\ {$3$}}}
\put(708.0,830.0){\rule[-0.200pt]{0.400pt}{4.818pt}}
\put(810.0,150.0){\rule[-0.200pt]{0.400pt}{4.818pt}}
\put(810,100){\makebox(0,0){\ {$4$}}}
\put(810.0,830.0){\rule[-0.200pt]{0.400pt}{4.818pt}}
\put(913.0,150.0){\rule[-0.200pt]{0.400pt}{4.818pt}}
\put(913,100){\makebox(0,0){\ {$5$}}}
\put(913.0,830.0){\rule[-0.200pt]{0.400pt}{4.818pt}}
\put(1015.0,150.0){\rule[-0.200pt]{0.400pt}{4.818pt}}
\put(1015,100){\makebox(0,0){\ {$6$}}}
\put(1015.0,830.0){\rule[-0.200pt]{0.400pt}{4.818pt}}
\put(1118.0,150.0){\rule[-0.200pt]{0.400pt}{4.818pt}}
\put(1118,100){\makebox(0,0){\ {$7$}}}
\put(1118.0,830.0){\rule[-0.200pt]{0.400pt}{4.818pt}}
\put(1220.0,150.0){\rule[-0.200pt]{0.400pt}{4.818pt}}
\put(1220,100){\makebox(0,0){\ {$8$}}}
\put(1220.0,830.0){\rule[-0.200pt]{0.400pt}{4.818pt}}
\put(1323.0,150.0){\rule[-0.200pt]{0.400pt}{4.818pt}}
\put(1323,100){\makebox(0,0){\ {$9$}}}
\put(1323.0,830.0){\rule[-0.200pt]{0.400pt}{4.818pt}}
\put(400.0,150.0){\rule[-0.200pt]{246.922pt}{0.400pt}}
\put(1425.0,150.0){\rule[-0.200pt]{0.400pt}{168.630pt}}
\put(400.0,850.0){\rule[-0.200pt]{246.922pt}{0.400pt}}
\put(400.0,150.0){\rule[-0.200pt]{0.400pt}{168.630pt}}
\put(150,625){\makebox(0,0){\large{$\frac{error}{signal}$}}}
\put(912,25){\makebox(0,0){{$N$}}}
\put(605.0,263.0){\rule[-0.200pt]{0.400pt}{2.409pt}}
\put(595.0,263.0){\rule[-0.200pt]{4.818pt}{0.400pt}}
\put(595.0,273.0){\rule[-0.200pt]{4.818pt}{0.400pt}}
\put(708.0,279.0){\rule[-0.200pt]{0.400pt}{5.300pt}}
\put(698.0,279.0){\rule[-0.200pt]{4.818pt}{0.400pt}}
\put(698.0,301.0){\rule[-0.200pt]{4.818pt}{0.400pt}}
\put(810.0,292.0){\rule[-0.200pt]{0.400pt}{9.877pt}}
\put(800.0,292.0){\rule[-0.200pt]{4.818pt}{0.400pt}}
\put(800.0,333.0){\rule[-0.200pt]{4.818pt}{0.400pt}}
\put(1015.0,261.0){\rule[-0.200pt]{0.400pt}{4.818pt}}
\put(1005.0,261.0){\rule[-0.200pt]{4.818pt}{0.400pt}}
\put(1005.0,281.0){\rule[-0.200pt]{4.818pt}{0.400pt}}
\put(1220.0,282.0){\rule[-0.200pt]{0.400pt}{5.300pt}}
\put(1210.0,282.0){\rule[-0.200pt]{4.818pt}{0.400pt}}
\put(605,268){\circle*{12}}
\put(708,290){\circle*{12}}
\put(810,312){\circle*{12}}
\put(1015,271){\circle*{12}}
\put(1220,293){\circle*{12}}
\put(1210.0,304.0){\rule[-0.200pt]{4.818pt}{0.400pt}}
\put(605.0,490.0){\rule[-0.200pt]{0.400pt}{19.995pt}}
\put(595.0,490.0){\rule[-0.200pt]{4.818pt}{0.400pt}}
\put(595.0,573.0){\rule[-0.200pt]{4.818pt}{0.400pt}}
\put(708.0,675.0){\rule[-0.200pt]{0.400pt}{22.885pt}}
\put(698.0,675.0){\rule[-0.200pt]{4.818pt}{0.400pt}}
\put(698.0,770.0){\rule[-0.200pt]{4.818pt}{0.400pt}}
\put(810.0,580.0){\rule[-0.200pt]{0.400pt}{22.885pt}}
\put(800.0,580.0){\rule[-0.200pt]{4.818pt}{0.400pt}}
\put(800.0,675.0){\rule[-0.200pt]{4.818pt}{0.400pt}}
\put(1015.0,522.0){\rule[-0.200pt]{0.400pt}{19.995pt}}
\put(1005.0,522.0){\rule[-0.200pt]{4.818pt}{0.400pt}}
\put(1005.0,605.0){\rule[-0.200pt]{4.818pt}{0.400pt}}
\put(1220.0,643.0){\rule[-0.200pt]{0.400pt}{23.126pt}}
\put(1210.0,643.0){\rule[-0.200pt]{4.818pt}{0.400pt}}
\put(605,532){\circle{18}}
\put(708,723){\circle{18}}
\put(810,627){\circle{18}}
\put(1015,564){\circle{18}}
\put(1220,691){\circle{18}}
\put(1210.0,739.0){\rule[-0.200pt]{4.818pt}{0.400pt}}
\put(400.0,150.0){\rule[-0.200pt]{246.922pt}{0.400pt}}
\put(1425.0,150.0){\rule[-0.200pt]{0.400pt}{168.630pt}}
\put(400.0,850.0){\rule[-0.200pt]{246.922pt}{0.400pt}}
\put(400.0,150.0){\rule[-0.200pt]{0.400pt}{168.630pt}}
\end{picture}
\end	{center}
\caption{How the fluctuations on a physically relevant
connected correlator, at $t=0$ ($\bullet$) and $t=a$ ($\circ$),
vary with $N$.}
\label{fig_Nercor}
\end{figure}

To see what happens in practice, I show in Fig.~\ref{fig_Nercor}
how the error to signal ratio on $C(t=0)$ and $C(t=a)$ varies
as a function of $N$, when we perform the same number
of Monte Carlo sweeps,  on the same size lattice,
and for the same lattice spacing
\cite{blmtuwG}.
The correlator is one used 
to calculate a physical quantity, so we infer that the
difficulty of calculating a mass does not grow with $N$ 
beyond the growing difficulty of generating the lattice fields
themselves.

Turning now to the inclusion of quarks fields in the fundamental
representation, the most expensive part 
of current calculations, even for SU(6) and even in the quenched case, 
is the matrix times vector  multiplication (e.g. in propagators)
and this is $\propto N^2$. This may in principle be
partly offset by smaller finite $V$ corrections at larger $N$.

If one now looks at connected correlators (in the quenched case)
and at the higher order correlators that determine their
fluctuations, one finds 
\cite{gbfb}
that the latter generically vanish not at the same rate, but as $O(1/N)$ 
-- which translates into an effective improvement of $\propto N^2$
in statistics. This gain of $N^2$ will compensate the increase in
cost of multiplying matrices by vectors, so that increasing $N$ leads to
no increase in cost. In practice this ideal is not achieved and
the total cost of calculating a mass to a certain accuracy grows
roughly as $\propto N$  
\cite{gbfb}.

Let me emphasise here that
all current calculations have been performed on a small number of
desktops or on a modest cluster. Large $N$ calculations are thus
accessible to all of you!

\section{Large $N$ Physics at high $T$}

The finite $T$ physics of QCD is very topical because of the
dedicated experimental programme at RHIC and the upcoming
experiments at the LHC (where the ALICE detector is dedicated
to this physics). The experiments have confirmed earlier
lattice indications that for quite a large range of $T$ 
above the deconfining temperature, $T_c$, the plasma is
strongly interacting and apparently out of reach of 
straightforward perturbative expansions. 

At the same time, this has become a topical arena for
gauge-gravity duality calculations. Of course, such (top-down) 
approaches are typically applicable to ${\cal{N}}=4$ SUSY, and 
various deformations thereof. And they are only valid in the
limit of $N$ and $g^2$ both large. None of this looks very much
like QCD or SU(3) gauge theory in the low-$T$ confining phase.
However at finite $T$, the adjoint fermions in ${\cal{N}}=4$ SUSY 
acquire $O(T)$ Matsubara masses, from the anti-periodic fermionic 
boundary conditions in the Euclidean time (thermal) direction.
Once SUSY is broken in this way, the adjoint scalars are no
longer protected from acquiring a mass, and will also become
massive. Thus the only remaining light fields are the gauge 
fields -- which begins to look like a gauge theory at $T>T_c$.
Moreover, since the real-world plasma appears to be strongly
coupled, this begins to look like an ideal area for applying
gauge-gravity duality. Of course all the AdS/CFT calculations
are at large $N$ so it is important to check if for the relevant
thermodynamic quantities, $N=3$ is close to $N=\infty$. 
This is what I want to focus upon in this section.

Euclidean lattice calculations of thermal averages are straighforward.
One takes the Euclidean time torus to be of length $l_t=aL_t$,
and imposes (anti)periodic boundary conditions for (fermions) bosons.
The path integral is then just the partition function of the
quantum field theory at a finite temperature $T=1/l_t$, or
$aT=1/L_t$ in lattice units:
\begin{equation}
Z = \int \prod_l dU_l e^{-\beta S} = \sum_n e^{-\frac{E_n}{T}}
\label{eqn_Z}
\end{equation}
Of course we should also make the spatial tori large enough, 
$L_i\gg L_t$, so that we are in the thermodynamic limit, 
where we have a well-defined notion of temperature.

\subsection{Deconfinement}

What do we expect?

If the transition is first order (as indeed it is for $N\geq 3$ in
$D=3+1$ and for $N\geq 4$ in $D=2+1$) it will occur at the value
of $T$ at which the free energies of the confined and  deconfined 
phases become equal, $F_c = F_d$. Now, since the number of gluons 
is $O(N^2)$ we expect $F_d \propto N^2$. On the other hand 
we might naively expect $F_c \propto N^0$, since there are only
colour singlet states in the confined phase. If so, it
immediately follows that $T_c \to 0$ as $N\to \infty$. 

From our numerical results we know that this does not in fact
happen. The reason is easy to see: in the confined phase there is a 
$-O(N^2)$ contribution to $F_c$ that comes from the vacuum energy 
density (the gluon condensate). So in the large $N$ limit, the value 
of $T_c$ is precisely determined by the balance between 
this $O(N^2)$ vacuum contribution to $F_c$ and the $O(N^2)$
piece of  $F_d$. If the plasma had turned out to be weakly coupled,
we could have easily calculated $F_d$ and therefore obtained a direct 
relationship between $T_c$ and the gluon condensate. That would
have been very nice, but as it happens
the plasma is strongly coupled, and so we have no such relation --
but, on the other hand, this opens the door to AdS/CFT calculations.

\begin{figure}[htb]
\begin	{center}
\leavevmode
\input	{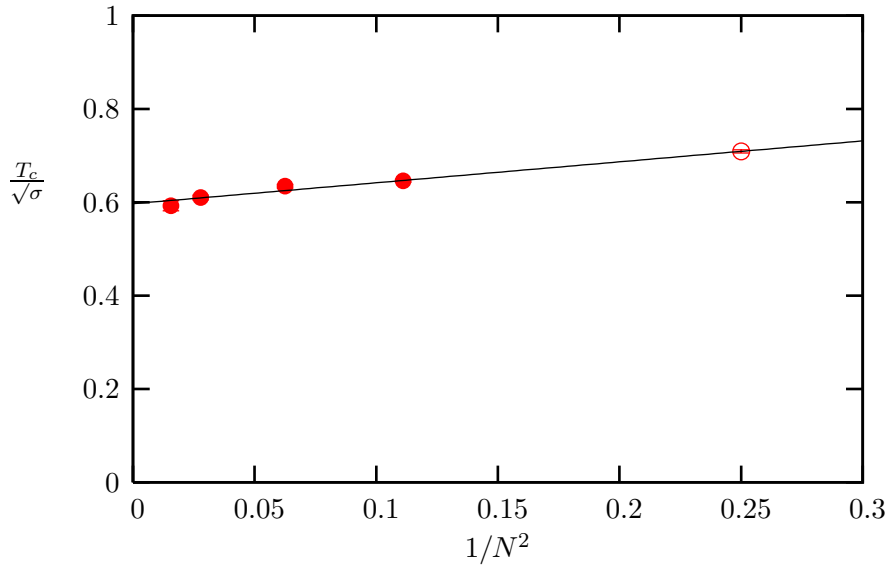}
\end	{center}
\caption{Deconfining temperature in units of the string tension,
for continuum SU($N$) gauge theories in $D=3+1$; with $O(1/N^2)$
extrapolation to $N=\infty$.}
\label{fig_tcN}
\end{figure}

In Fig.~\ref{fig_tcN} I show lattice calculations of $T_c$
in units of the string tension for $N\in [2,8]$ in $D=3+1$. All the
values shown are after extrapolation to the continuum limit 
\cite{blmtuwTc}.
If we fit with an $O(1/N^2)$ correction we obtain
\begin{equation}
\frac{T_c}{\surd\sigma}
\stackrel{N\geq 2}{=}
0.597(4) + \frac{0.45(3)}{N^2} \qquad: \quad D=3+1
\label{eqn_Tcd4}
\end{equation}
which thus provides a prediction $\forall N$. 
It is perhaps surprising that this simple analytic form should
fit all the way down to $N=2$ where the transition has changed
from first to second order. Especially so, given that the errors
for SU(2) and SU(3) are very small, about 0.5\%. We also note
that the coefficient of the correction is $O(1)$ in natural units.

\begin{figure}[htb]
\begin	{center}
\leavevmode
\input	{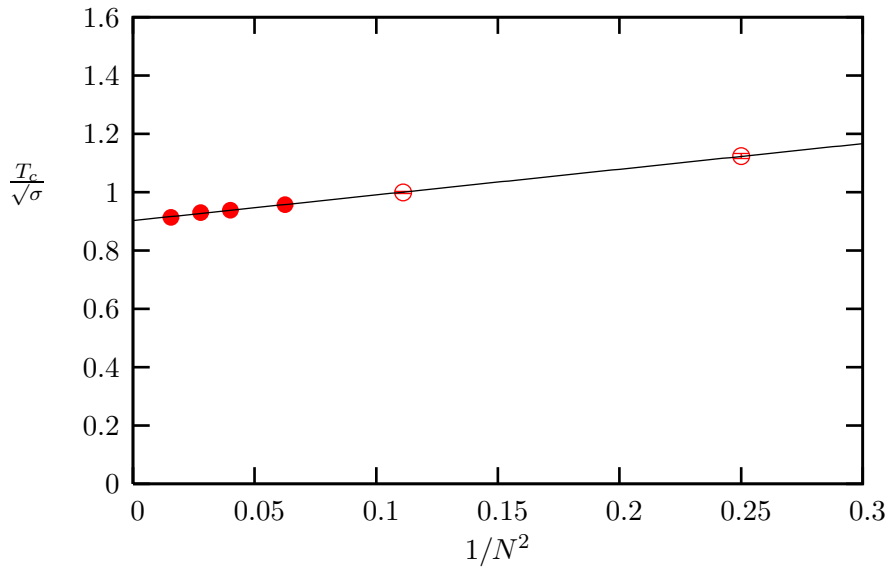}
\end	{center}
\caption{As in Fig.~\ref{fig_tcN} but for $D=2+1$.}
\label{fig_tcNd3}
\end{figure}

It is interesting to see what happens in $D=2+1$. The corresponding
results for $T_c$ 
\cite{Tcd3}
are shown in Fig.~\ref{fig_tcNd3}. Now both $N=2$ and $N=3$ are
second order, but a fit with just the leading correction still
works for all $N$, giving
\begin{equation}
\frac{T_c}{\surd\sigma}
\stackrel{N\geq 2}{=}
0.903(3) + \frac{0.88(5)}{N^2} \qquad: \quad D=2+1
\label{eqn_Tcd3}
\end{equation}
where again the size of the correction is modest in natural units.

Most but not all thermodynamic quantities associated with
the deconfining transition show a modest variation with $N$
\cite{blmtuwTcprop,TQ}.
A striking counterexample is provided by the interface tension,
$\sigma_{cd}$, between the confining and deconfining phases. Although 
this calculation is difficult, one roughly finds 
\cite{blmtuwTcprop}.
\begin{equation}
\frac{\sigma_{cd}}{T^3_c}
\stackrel{N\geq 3}{=}
0.0138(3)N^2 - 0.104(3) \qquad: \quad D=3+1
\end{equation}
Here the coefficient of the subleading term is very large
compared to that of the leading term. Because of this 
the value of $\sigma_{cd}$ is anomalously small for SU(3),
and this is presumably the main reason why the phase transition
appears to be very weakly first order in this case.

\subsection{A strongly coupled gluon plasma?}

We now want to ask whether the gluon plasma continues to 
be strongly coupled at large $N$. One of the measures of this
is the pressure and its deviation from the Stefan-Boltzmann value.
I will focus on this here because the lattice calculation 
is particularly simple. (I reproduce the argument given in
\cite{bbmtP}.)

The pressure is the (infinitesimal) work done when the volume increases
(infinitesimally). So it can be obtained from the change in 
the average energy as we increase the volume, using eqn(\ref{eqn_Z}),
\begin{equation}
p = T \frac{\partial}{\partial V} \log Z(T,V)
= \frac{T}{V}\log Z(T,V) = -f
\end{equation}
where the second equality assumes a sufficiently large and
homogeneous system, and $f=F/V$ is the free energy density.
To calculate the pressure at temperature $T=1/aL_t$ in a volume 
$V=a^3L^3_s$ with lattice cut-off $a(\beta)$, it is convenient
to express $\log Z$ in the integral form:
\begin{equation}
p(T) = \frac{T}{V}\log Z(T,V)
=
 \frac1{a^4(\beta)L_s^3L_t}
\int^\beta_{\beta_0}  d\beta^\prime 
\frac{\partial\log Z }{\partial \beta^\prime}.
\label{eqn_pZintegral}
\end{equation}
There is in general an integration constant, but it will disappear 
when we regularise the pressure in a moment.
This integral form is useful because it is easy to see from 
Eqs.~(\ref{eqn_Z},\ref{eqn_Splaq}) that 
\begin{equation}
\frac{\partial\log Z }{\partial \beta}
=
-\langle S \rangle
=
N_p \langle u_p \rangle
\label{eqn_pZup}
\end{equation}
where $N_p=6L_tL^3_s$ is the total number of plaquettes and
$u_p \equiv {\rm Re}\Tr{U_P}/N$. 

So the pressure can be obtained by simply integrating the average 
plaquette over $\beta$: a very simple calculation. 
This pressure has been defined relative to that of the
unphysical `empty' vacuum and will therefore be ultraviolet 
divergent in the continuum limit. To remove this divergence we 
need to define the pressure relative to that of a more
physical system. We shall follow convention and subtract 
from $p(T)$ its value at $T=0$, calculated with the same 
value of the cut-off $a(\beta)$ (so that the UV divergences
cancel). Thus our pressure will be defined with respect to its
$T=0$ value. Doing so we obtain from eqns(\ref{eqn_pZup},
\ref{eqn_pZintegral})
\begin{equation}
a^4[ p(T) - p(0) ]
=
6 \int^\beta_{\beta_0}  d\beta^\prime 
(\langle u_p \rangle_T - \langle u_p \rangle_0)
\label{eqn_pint1}
\end{equation}
where $\langle u_p \rangle_0$ is calculated on some
$L^4$ lattice which is large enough for it to be 
effectively at $T=0$. 
We replace $ p(T) - p(0) \to p(T)$, where from now on it
is understood that $p(T)$ is defined relative to its value
at $T=0$, and we use  $T=(aL_t)^{-1}$ to rewrite 
eqn(\ref{eqn_pint1}) as
\begin{equation}
\frac{p(T)}{T^4}
=
6 L^4_t \int^\beta_{\beta_0} d\beta^\prime 
(\langle u_p \rangle_T - \langle u_p \rangle_0).
\label{eqn_pint2}
\end{equation}
We remark that when our $L_s^3L_t$ lattice is in the confining
phase, then $\langle u_p \rangle$ is essentially independent 
of $L_t$ and takes the same value as on a $L^4_s$ lattice. This 
should become exact as $N\to\infty$ but is accurate enough even 
for SU(3). Thus as long as we choose $\beta_0$ in eqn(\ref{eqn_pint2})
such that $a(\beta_0)L_t > 1/T_c$ then the integration constant,
referred to earlier, will cancel.

We calculate the pressure using eqn(\ref{eqn_pint2}) on a volume
that is large enough to be effectively infinite. Since the
plasma has a mass gap (the electric and magnetic screening masses) 
this is easy to achieve. We then normalise it to the Stefan-Boltzmann 
value (in an infinite volume). It has long been known that this
ratio is far below unity, even to quite high $T$, for SU(3).
This is now considered to be a reflection of the strong coupling
nature of the gluon plasma. 

\begin{figure}[htb]
\begin	{center}
\leavevmode
\epsfig{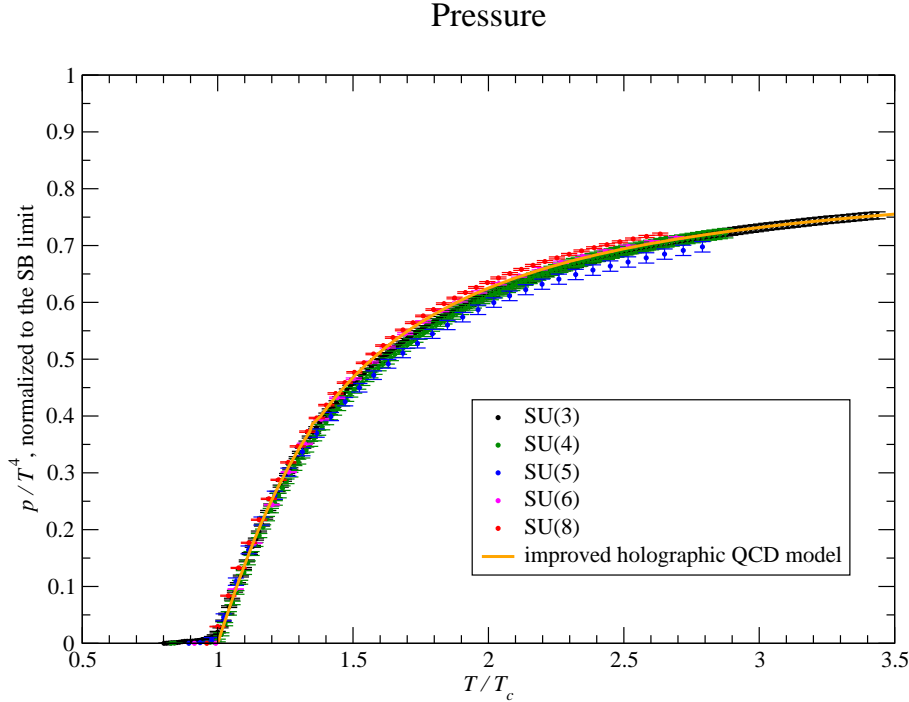} 
\end	{center}
\caption{Normalised pressure above $T_c$ for various SU($N$) gauge theories
\cite{Panero}.}
\label{fig_pressure}
\end{figure}

A simple strategy is to perform similar calculations at larger
$N$ and see whether this ratio continues to remain far from unity 
or not. This was first done in
\cite{bbmtP}
where it was shown that there is essentially no change in
the normalised value of $p/T^4$ as one increases $N$ from
$N=3$ to $N=8$, in the range $1 \leq T/T_c \leq 2$. Recently
\cite{Panero}
there has been a more accurate calculation extending over a larger
range of $T/T_c$, and in Fig.~\ref{fig_pressure} I show the
relevant plot (borrowed from
\cite{Panero}).
We see that any variation is negligible: the $N=\infty$ plasma
is just as strongly coupled as the $N=3$ one.

\vspace*{0.25in}

This is of course good news for the applicability of AdS/CFT
to real world experiments above $T_c$. (And you can see a comparison 
with one such calculation in Fig.~\ref{fig_pressure}.)
In addition, it restricts 
what dynamics we might think is responsible for the strong coupling,
if we make the plausible assumption that this dynamics should
be common to all SU($N$) gauge theories. For example, it 
excludes an important role for topology (since we know that
in the deconfined phase topological fluctuations vanish
roughly exponentially with $N$
\cite{TQ,MTlat08})
or for any colour singlet hadrons that might survive above $T_c$.

\section{And if I had the time ... }

There are many other topics on which there has has been significant 
progress, and which I would have liked to describe to you
if there had been time. Here I will just list some of them:

$\bullet$ space-time reduction

$\bullet$ large-$N$ phase transitions

$\bullet$ topology at large $N$

$\bullet$ interlaced $\theta$-vacua

$\bullet$ chiral symmetry breaking at large $N$

$\bullet$ topology and chiral symmetry breaking

$\bullet$ strong coupling  

$\bullet$ full glueball spectrum and the Pomeron

$\bullet$ $k$-string tensions in $D=2+1$ and $D=3+1$

$\bullet$ Karabali-Nair Hamiltonian approach in $D=2+1$: lattice tests

\section{Flux tubes as strings}
\label{section_tubesstrings}

\subsection{General considerations}

Suppose we place a static fundamental source in our SU($N$) gauge 
theory at $\vec{x} = (0,y,z)$, with a conjugate source at 
$\vec{x} = (r,y,z)$. Suppose our space-time is a 4-torus,
with the Euclidean time extent being $\tau$, and that we
are in the confining phase. As usual we denote by $Z$ the partition 
function of the field theory on the given space-time. We also
define a partition function for the system with these sources by 
\begin{equation}
Z_{s\bar{s}}(r,\tau) = \int dA \  l^\dagger(x=r,y,z)
l(x=0,y,z) \, \exp\{-S[A]\}
\label{eqn_Zss}
\end{equation}
where $l(\vec{x})$ is the traced, path-ordered exponential of the 
gauge potential along a path that encircles the t-torus and is
at $\vec{x}=(x,y,z)$. This is often called a Wilson line, or
a Polyakov loop, and, sometimes, a thermal line. It is the phase factor
that arises from the minimal coupling of the 
static sources to the gauge fields, $j_\mu A_\mu =j_0 A_0$. 
In Section~\ref{section_latticed3} we shall see 
how this translates to the lattice, but for now we shall use a 
continuum language. 

If $r \gg 1/\surd\sigma$ there will be a flux tube between the
sources which, as it evolves in time, sweeps out a sheet bounded
by the periodic sources. (I am making some assumptions e.g. that
the spatial torus is $\gg r$.) This sheet clearly has the topology 
of a cylinder. The partition function can be written as a sum over 
states   
\begin{equation}
\frac{1}{Z} Z_{s\bar{s}}(r,\tau) = \sum_n \omega_n e^{-E_n(r)\tau}
\label{eqn_Zss_open}
\end{equation}
where $E_n(r)$ is an energy level of two sources separated 
by $r$, and $\omega_n$ is its degeneracy. The states
are (excited) flux tubes that begin and end on a source and which
evolves around the $t$-torus. The artifice of static sources,
means that the flux tube states have zero transverse as well as
zero longitudinal momentum.

Now, there is another way to look at this set-up. We are in
Euclidean space time so we are free to think of any of our axes as
being the time direction, with its associated Hamiltonian defined
on the space spanned by the other three coordinates. Taking
$x$ as labelling the `time', $l$ is now a Wilson line that winds 
around what is now a spatial torus of length $\tau$. 
What $Z_{s\bar{s}}(r,\tau)$
represents, in this point of view, is a correlation function whose 
intermediate states consist of flux tubes that wind around this same 
`spatial' torus of length $\tau$ and propagate the distance $r$ 
between the two Wilson lines. The same partition function can 
therefore also be written as
\begin{equation}
\frac{1}{Z} Z_{s\bar{s}}(r,\tau) = 
\sum_{n,p} c_n(p,\tau) e^{-\tilde{E}_n(p,\tau)r}
\label{eqn_Zss_closed}
\end{equation}
where $\tilde{E}_n$ is an energy level of the (excited)
flux tube that winds around a spatial torus of length $\tau$. The
$\tilde{E}_n$ and $E_n$ are different functions in general
because the flux tubes have have different boundary conditions. 
(Often, where the context removes any ambiguity, I will use  $E_n$ 
in place of $\tilde{E}_n$, and I will be casual about distinguishing
energy levels from energy eigenstates.)
Here I have made explicit that the winding flux tubes have to be 
integrated over transverse momentum $p$ since the
operators $l$ are localised at $y,z$. The $c_n$ are the wave-function 
factors for the overlap of a state $|n,p\rangle$ on the operator $l$:
$c_n = \sum_i|\langle vac | l^\dagger | n,i,p \rangle |^2$, where
the sum is over the degenerate eigenstates contributing to the energy 
level $n$. Lorentz invariance enables us to do the integral over $p$
\cite{LW04,HM06}
but I will not pursue that explicitly here.

You may be wondering how one shows that a Polyakov loop correlator 
only involves winding eigenstates (even if this is heuristically 
plausible). I will give the argument in Section~\ref{section_opsd3}.

The above two ways of writing the Polyakov loop correlator,
either as a sum over closed strings or as a sum over open strings,
is a duality that has been well-known since the early '80's, and has 
been used routinely in numerical simulations. However the interesting 
thing for us about this  open-closed string duality is the relatively 
recent realisation 
\cite{LW04}
that it strongly constrains the form of the effective string theory 
describing the dynamics of long flux tubes. 

So suppose that we have an effective string theory, governed by
an effective action $S_{eff}$, which reproduces the long distance
physics of flux tubes. Consider the string partition function
over the $r\times\tau$ cylinder considered above. We will have
\begin{equation}
Z_{cyl}(r,\tau) = \int_{cyl = r\times\tau} dS e^{-S_{eff}[S]}
=
\frac{1}{Z} Z_{s\bar{s}}(r,\tau)
\label{eqn_Zcyl}
\end{equation}
where we integrate over all surfaces $S$ spanning the cylinder.
From eqn(\ref{eqn_Zcyl}) we infer that
$Z_{cyl}(r,\tau)$ can be written as a sum of open or closed
strings as in eqn(\ref{eqn_Zss_open}) and eqn(\ref{eqn_Zss_closed}).
These are nothing but Laplace transforms, in $r$ and $\tau$ 
respectively of $Z_{cyl}(r,\tau)$. So if we have a candidate
string action, $S_{eff}[S]$, we can perform these Laplace transforms
and extract the open and closed string spectra. Conversely,
the particular form of the Laplace transforms in 
eqns(\ref{eqn_Zss_open},\ref{eqn_Zss_closed}), and in particular 
the way Lorentz invariance constrains the energy levels of
different $p$ in eqn(\ref{eqn_Zss_closed}), will constrain
the permitted form of $S_{eff}[S]$ and this may in turn
constrain the possible form of the flux tube energy spectrum.

In the above we have specifically discussed the open-closed duality
\cite{LW04}
associated with a cylinder. One can usefully extend 
\cite{AHEK}
such a discussion to an $r\times\tau$ torus and its associated
closed-closed string duality. Now we would have
\begin{equation}
Z_{torus}(r,\tau) = \int_{T^2=r\times\tau} dS e^{-S_{eff}[S]}
=
\sum_{n,p} \tilde{\omega}_n e^{-\tilde{E}_n(p,\tau)r}
=
\sum_{n,p} \tilde{\omega}_n e^{-\tilde{E}_n(p,r)\tau}
\label{eqn_Ztorus}
\end{equation}
and a useful new constraint on  $S_{eff}[S]$
\cite{AHEK}.
It is clear that we have not exhausted all the possibilities
here and that other boundary conditions may provide further
useful constraints.

Some comments. (For a more detailed discussion, see
\cite{AHEK}.)

As is well-known, string theories are not well-defined
outside their critical dimension. However the resulting anomalies, 
which show up in different ways depending on how one `gauge-fixes' 
the diffeomorphism invariance in one's calculation, typically die
off at long distances, e.g.
\cite{Olesen},
and when one considers a long string
\cite{PS}.
Thus it can make sense, at least technically, to consider a string path 
integral over a single large surface, in an effective string theory 
approach outside the critical dimension
\cite{PS}.
This represents the world sheet swept out by a single long fluctuating 
string.

This effective string theory approach is therefore limited 
to describing the dynamics of a single
long fluctuating flux tube. This is an important physical limitation.
In reality, a sufficiently excited flux tube can decay into a
flux tube of lower energy and a glueball, and such states inevitably
appear in the sum over states in eqn(\ref{eqn_Zss_open}) and
eqn(\ref{eqn_Zss_closed}). In the string picture a 
glueball is a contractible closed loop of string whose length is 
$O(1/\surd\sigma)$ (for light glueballs). 
There is no guarantee that an effective string
theory can consistently describe such extra small surfaces. One
can partially circumvent this by only considering low-lying
string states which are too light to decay:
\begin{equation}
E_n(r) - E_0(r) \ll m_G.
\label{eqn_Elight}
\end{equation}
However even such states will be affected by mixing through virtual
glueball emission, which corresponds to small handles on our large
surface - again something that would be problematic for the string
theory. 

There is of course a limit in which mixings and decays do vanish, and
that is the $N\to\infty$ limit. So it is consistent to use 
eqn(\ref{eqn_Zcyl}) and eqns((\ref{eqn_Zss_open},\ref{eqn_Zss_closed})
for the SU($\infty$) theory. It is then plausible that as we move 
continuously away from that limit, to finite $N$, the corrections
will be under control and small
\cite{AHEK}.
Indeed, we shall see that the low-lying
flux tube spectrum has very little $N$-dependence for $N\geq 3$,
and this increases our confidence in the potential applicability
of the effective string theory approach to SU($N$) gauge theories
in general.

Let us consider a flux tube that winds around a spatial torus 
of length $l$. (We shall often use $l$ in place of $r$ and $\tau$.)
The excited states of this flux tube
are presumably obtained from the ground state, $E_0(l)$, 
by exciting some of the modes living on the  tube. If the 
excited mode is massive we would expect
\begin{equation}
E_n(l)  = E_0(l) + O(\surd\sigma).
\label{eqn_massivemode}
\end{equation}
If the modes are massless, we would expect the extra energy to
be given by their momenta which, for bosons, is quantised to be 
$p=k\pi/l ; k=\pm 1, ...$ on such a periodic flux tube. 
(So to obtain an excited flux tube with zero net longitudinal
momentum, we will need more than one such excitation if, as
is usually the case, a $k=0$ mode is not allowed.) So we expect
\begin{equation}
E_n(l)  = E_0(l) + O\left(\frac{\pi}{l}\right).
\label{eqn_masslessmode}
\end{equation}
So at large $l$, where $\pi/l \ll\surd\sigma$, the low-lying 
flux-tube spectrum is given solely by the excitation of the 
massless modes. 

The first step is therefore to focus on an effective
string action that includes just these massless modes.
In general we expect modes to be massless for symmetry reasons.
In the case of a flux tube there are $D-2$ obvious massless modes.
These are the Goldstone modes that arise from the fact that once
we have specified the location of our flux tube, we have broken
spontaneously the translation invariance in the $D-2$ directions
transverse to the flux tube. Of course it may be that there are other
less obvious massless modes. However it clearly makes sense to 
start with just these Goldstone modes and to calculate from them
properties of the low-lying flux tube spectrum for long flux tubes. 
If these agree with what we find through our direct lattice 
calculations of the spectrum, we can be confident that we have
identified correctly all the massless modes. 

To proceed one needs to fix convenient coordinates to describe
the surface in the path integral. This is a `gauge-fixing' of the
diffeomorphism invariance, and in so doing we risk making the
constraints that follow from this fundamental string symmetry 
less obvious. Here we follow
\cite{LSW,LW04,AHEK}
and do not discuss the details of the important alternative approach
\cite{PS}.
Suppose we are integrating over the surfaces of the cylinder
discussed above. There is a minimal surface which we can
parameterise by $x \in [0,r]$ and $t\in [0,\tau]$. Other
surfaces are specified by a transverse displacement vector
$h(x,t)$ that has two components in the $(y,z)$ directions.
(This is for $D=3+1$; it has only one component for $D=2+1$.) 
This way of parameterising a surface is often  called `static gauge'.
We can now write the effective string action in terms of 
this field $h$; schematically, 
\begin{equation}
S_{eff}[S] \longrightarrow S_{eff}[h]
\label{eqn_Seff_Stoh}
\end{equation}
and the integral over surfaces becomes an integral
over $h(x,t)$ at each value of $(x,t)\in [0,r]\times[0,\tau]$. Since the 
field $h$ is an integration variable in $(0,\infty)$, we can take it
to be dimensionless. Moreover, since the action cannot depend on the 
position of the flux tube (translation invariance), it cannot depend
on $\langle h \rangle$ but only on $\partial_\alpha h$ where
$\alpha = x,t$. That is to say, schematically, 
\begin{equation}
S_{eff}[h] \longrightarrow S_{eff}[\partial h]
\label{eqn_Seff_htodh}
\end{equation}
and we can perform a derivative expansion of  $S_{eff}$ in powers 
of derivatives of $h$; (very) schematically
\begin{equation}
S_{eff} = \sigma r \tau
+ \int^\tau_0 dt \int^r_0 dx \frac{1}{2} \partial h\partial h
+ \sum_{n=2} c_n \int^\tau_0 dt \int^r_0 dx  (\partial h)^{2n}
\label{eqn_Seff_expansion}
\end{equation}
where the derivatives are with respect to $x$ and $t$ and indices
are appropriately contracted.
The coefficients $c_n$ have dimensions [length]$^{(2n-2)}$
to keep the terms dimensionless. So we can expect that for the 
long wavelength fluctuations of a long string, such a higher
order term will make a contribution of $O(1/(\sigma l^2)^{n-1})$
and so the importance of these terms is naturally ordered by
the number of derivatives. All this is entirely analogous to the 
familiar way chiral Lagrangians depend on their Goldstone fields.

Three comments. \\
$\bullet$ The approach just described is typically designed to capture the
physics on energy scales smaller than a dynamical mass scale. Here that
would be $O(\surd\sigma)$. Just as the applicability of chiral 
Lagrangians is typically bounded by the lowest resonances.\\
$\bullet$ For simplicity of presentation we ignore operators that are
located on the boundary of the cylinder.\\
$\bullet$ Such an expansion is unlikely to be better than asymptotic,
and so might well have corrections that are perhaps like
$\exp\{-1/(\partial h)^2\}$ that will lead to corrections
like $\exp\{-c\sigma l^2\}$ in the spectrum. More generally
we need to be cautious about the uniformity of the various limits
being taken in any applications (e.g. large $n$, $r$, $\tau$).\\
$\bullet$ Our chosen `static-gauge' parameterisation does not work 
for general surfaces. To describe a string with an `overhang' or
any kind of `back-tracking', the field $h(x,t)$ would have to multivalued, 
which is something the standard treatments do not allow. That is to say, 
we arbitrarily exclude such rough surfaces from the path integral. For a 
flux tube, its finite width provides a physical lower distance cutoff on such 
fluctuations: any overhang that is within a distance $ \lesssim 1/\surd\sigma$
will in effect be a fluctuation in the intrinsic width of the flux tube 
i.e. a massive mode excitation. Any backtracking/overhang that is larger
will increase the length by $\Delta l >  1/\surd\sigma$ and hence
the energy by  $\Delta E \sim \sigma\Delta l > \surd\sigma$.
In both cases the associated excitation energies will be much
greater than the $O(1/l)$ gap to the stringy modes, once $l$
is large enough.  Thus this should not be a significant issue for the 
long wavelength massless oscillations we have discussed above. But it
needs to be addressed in any analytic treatement that wishes to be more 
ambitious.

\subsection{The Gaussian approximation}

The first non-trivial term in our effective string action
is the Gaussian piece:
\begin{equation}
S_{eff} = \sigma r \tau
+\int^\tau_0 dt \int^r_0 dx \frac{1}{2} 
\partial_\alpha h\partial_\alpha h
\label{eqn_Sgaussian}
\end{equation}
Since it has the fewest derivatives it should provide the leading 
correction to the linear piece of the string energy, at large $l$.
(For the cylinder there is a linear piece, $\propto \tau$,
that comes from the boundary of the cylinder, and represents
a self-energy term for the source. We ignore that in the following.)
Being Gaussian, this can be calculated exactly, and one obtains
\begin{equation}
Z_{cyl}(r,\tau) = e^{-\sigma r\tau} |\eta(q)|^{-(D-2)} 
\quad :\quad q=e^{-\pi\tau/r}
\label{eqn_Zgaussian}
\end{equation}
in terms of the Dedekind eta function
\begin{equation}
\eta(q) = q^\frac{1}{24} \prod^\infty_{n=1} (1-q^n).
\label{eqn_dedekind}
\end{equation}
(See
\cite{LW04}
whose notation I will borrow.) If we expand the product in 
eqn(\ref{eqn_dedekind}) we have a sum of powers of $q$, which,
using  $q=e^{-\pi\tau/r}$, becomes a sum of exponentials in
$\tau$ which is precisely of the form given in eqn(\ref{eqn_Zss_open}).
So matching this result with eqn(\ref{eqn_Zss_open}), we obtain
the prediction
\begin{equation}
E_n(r) = \sigma r + \frac{\pi}{r} 
\left\{n-\frac{1}{24}(D-2)\right\}
\label{eqn_Engaussian}
\end{equation}
for the energy levels. In addition, one also obtains predictions 
for the degeneracies of these levels. This is the exact result, for 
a Gaussian $S_{eff}$, for the energy levels of strings 
with ends fixed to static sources. We note that the excitation
energies display an $O(1/r)$ gap as expected from 
eqn(\ref{eqn_masslessmode}).

The Dedekind eta function possesses a well-known modular invariance:
\begin{equation}
\eta(q) = \left(\frac{2r}{\tau}\right)^{\frac{1}{2}}
\eta(\tilde{q}) \qquad ; \quad \tilde{q}=e^{-4\pi r/\tau}.
\label{eqn_etamodular}
\end{equation}
and so using eqn(\ref{eqn_dedekind}), but now for $\eta(\tilde{q})$,
we can rewrite the expression for $Z_{cyl}$ in 
eqn(\ref{eqn_Zgaussian}) as a sum
of exponentials in $r$ rather than in $\tau$. However this is
not precisely of the form shown in eqn(\ref{eqn_Zss_closed}),
because of the momentum integrations (which can be shown 
\cite{LW04,HM06}
to lead after integration to a sum over Bessel functions rather 
than simple exponentials). Thus a Gaussian $S_{eff}$ does not encode 
the open-closed string duality exactly and cannot be considered
as a candidate for an exact description of strings 
on a cylinder. However if we think of the Gaussian $S_{eff}$ as an 
approximation, possessing higher order terms that we are not 
considering at this stage,
then at large enough $\tau$, where the Bessel functions can
be expanded as exponentials to leading order, we can match
with  eqn(\ref{eqn_Zss_closed}) to obtain the closed
string energies,
\begin{equation}
\tilde{E}_n(\tau) = \sigma \tau + \frac{4\pi}{\tau} 
\left\{n-\frac{1}{24}(D-2)\right\} + O\left(\frac{1}{\tau^2}\right)
\label{eqn_tildeEngaussian}
\end{equation}
together with an expression for the overlaps $c_n$
\cite{LW04}.

The $O(1/r)$ correction to the leading linear term in $E_0(r)$ in 
eqn(\ref{eqn_Engaussian}) is the famous Luscher correction
\cite{LSW} 
for a flux tube with ends fixed on static sources. Physically it
arises from the regularised sum of the zero-point energies of all
the quantised oscillators on the string. It depends only on
the long wavelength massless modes and so is universal: any bosonic 
string theory in which the only massless modes are the transverse
oscillations will have precisely this leading correction. 
The same applies to the $O(1/\tau)$ correction to the leading 
linear term in $\tilde{E}_0(r)$ in eqn(\ref{eqn_tildeEngaussian}).

Although the above results for $E_n(r)$ are obtained in the Gaussian 
approximation to $S_{eff}[h]$, this approximation becomes exact 
as $r\to\infty$, and these predictions for the leading $O(1/r)$
correction are also exact and universal.

\subsection{Nambu-Goto free string theory}
\label{subsection_NG}

There is only one string theory whose spectrum can be calculated
in a closed form (as far as I am aware). That, not surprisingly,
is a free string theory: Nambu-Goto in flat space time (see  e.g.
\cite{NG}),
\begin{equation}
Z = \int dS e^{-\kappa A[S]}
\label{eqn_ZNG}
\end{equation}
where we integrate over all surfaces, with the action proportional
to the invariant area. This is not {\it{\`a priori}} a completely 
unrealistic effective string theory: after all, we recall 
that flux tubes at $N=\infty$ do not interact.

The energy levels of this theory were originally calculated in 
\cite{Arvis}
(and were subsequently extended in various papers). 
Since our numerical calculations
will focus on flux tubes that are closed around a spatial torus
of length $l$, this is the spectrum I will present here. 
 
Consider a string winding once around the $x$-torus. (One can
readily extend this to strings winding $\omega$ times around a torus.) 
Perform the usual Fourier decomposition of $h(x)$. Upon quantisation
the coefficients become creation operators for `phonons'
with momenta $\pm 2\pi k/l$ along the string and energy  
$2\pi k/l$ (since the modes 
are massless). Note that the $k=0$ mode is not included since
it corresponds to a shift to a different transverse position 
of the whole string i.e. to another vacuum of the spontaneously
broken symmetry. We call positive momenta left-moving (L) and the
negative ones right-moving (R). Let $n_{L(R)}(k)$ be the number 
of left(right) moving phonons of momentum $2\pi k/l$. Define the 
total energy (and momentum) of the left(right) moving phonons 
as $2\pi N_{L(R)}/l$, then:
\begin{equation}
N_L = \sum_k \sum_{n_L(k)} n_L(k)k, \qquad
N_R = \sum_k \sum_{n_R(k)} n_R(k)k.
\label{eqn_NLR}
\end{equation}
If $p=2\pi q/l$ is the total longitudinal momentum of the string
then, since the phonons provide that momentum, we must have
\begin{equation}
N_L - N_R = q.
\label{eqn_NLRmom}
\end{equation}

We can now write down the expression for the energy levels of the 
Nambu-Goto string:
\begin{equation}
E^2_{N_L,N_R}(q,l)
=
(\sigma l)^2 
+
8\pi\sigma \left(\frac{N_L+N_R}{2}-\frac{D-2}{24}\right)
+
\left(\frac{2\pi q}{l}\right)^2
\label{eqn_EnNG}
\end{equation}
where the degeneracies corresponding to particular values 
of $N_L$ and $N_R$ will depend on the number of ways these
can be formed from the $n_L$ and $n_R$ in eqn(\ref{eqn_NLR}).
In discussing the states, we shall often write the left and
right moving phonon creation operators of (absolute) momentum 
$2\pi k/l$  as $a_k$ and $a_{-k}$ respectively, and the unexcited
string ground state as $|0\rangle$.

Let us specialise to $q=0$, i.e. $N_L=N_R=n$, and make some general 
comments.\\
$\bullet$ The energy $E_n(l)$  can be expanded for large $l$ in 
inverse powers of $1/\sigma l^2$:
\begin{eqnarray}
E_n(l)
& = &
\sigma l \left(1 + 
\frac{8\pi}{\sigma l^2} 
\left(n -\frac{D-2}{24}\right)\right)^\frac{1}{2} \nonumber \\
& = &
\sigma l + \frac{4\pi}{l} 
\left(n -\frac{D-2}{24}\right) + O\left( \frac{1}{\sigma l^3} \right)
\label{eqn_EnNGexpansion}
\end{eqnarray}
We note that the first correction to the linear piece is exactly
as in eqn(\ref{eqn_tildeEngaussian}). Since we claimed the
latter was `universal', this is as it should be.\\
$\bullet$ The ground state energy becomes tachyonic at small $l$:
\begin{equation}
E^2_0(l)
= (\sigma l)^2 - \frac{\pi\sigma (D-2)}{3}
< 0 \qquad : \quad \sigma l^2 < \frac{\pi (D-2)}{3}.
\label{eqn_E0NG}
\end{equation}
One can regard it as the Hagedorn/deconfining transition
in the Nambu-Goto model, where strings condense
in the vacuum. \\
$\bullet$ The expansion of the square root expression
for the energy $E_n$, in eqn(\ref{eqn_EnNGexpansion}),
only converges for $\sigma l^2 > 8\pi n$ (ignoring the
negligible $D-2$ term). So the higher the excited state,
the larger is the value of $l$ before such an expansion
can be employed. This tells us that the formal expansion of
the action in powers of $1/l$ is not uniform in frequency
-- it is, in fact, only formal. One would expect this to
be the case for any string action, effective or otherwise.
So while the Nambu-Goto action (the invariant area of the
surface) can be expanded as described above, this
expansion is not uniform in energy.\\  
$\bullet$ One can show (see Appendix C of
\cite{LW04})
that the Nambu-Goto model satisfies open-closed duality
exactly. This is in contrast to the Gaussian string action.
Thus, if this is our only constraint, the Nambu-Goto model
is a viable candidate for providing a string action that
simultaneously describes flux tubes attached to static
sources and their dual description as winding flux tubes 
between Polyakov loop operators.

This has at least one important implication. When we use
open-closed (cylinder) string duality to constrain 
terms in the effective action that are higher order in
$\partial h$ (as described earlier), these constraints will 
be satisfied by the Nambu-Goto model. (This is also the case 
\cite{ustobe}
for constraints obtained from the closed-closed duality
\cite{AHEK}
associated with surfaces on a torus.) In particular, where
imposing such constraints
allows us to completely fix the expansion coefficients 
of $E_n(l)$ up to some order in $1/l$, these coefficients will
have to be precisely the same as those obtained by expanding
the Nambu-Goto expression in eqn(\ref{eqn_EnNGexpansion}),
and the corresponding expression for strings with fixed ends,  
to that order.

\subsection{Recent theoretical progress}
\label{subsection_recenttheory}

The seminal work in analysing flux tubes in a string
description in static gauge 
\cite{LSW}
(as described above) and the later more general approach
using conformal gauge
\cite{PS}
(not described here) led to an understanding of the universality
of the leading $O(1/l)$ Luscher correction to the linear growth
of the flux tube energy. Until recently there was, however, very
little further analytic progress along these lines. 

The situation changed in 2004 when major progress took place
independently within both approaches.\\ 
1) In
\cite{LW04}
it was shown that the open-closed duality (discussed above) could 
be used to provide useful constraints on the  higher order terms 
in the expansion of the effective string action. In particular
it was shown that in $D=2+1$ the next, $O(1/l^3)$, term is also
universal and the coefficient is precisely what you get by 
expanding the Nambu-Goto square-root expression to that order.
(As we commented above, the latter has to be the case.) In $D=3+1$
the coefficent is not fixed but there is a relationship predicted
between the coefficients of the two terms in the effective action
that contribute at that order. \\
2) In
\cite{JD}
(and later independently in
\cite{HDPM})
the next order was calculated within the Polchinski-Strominger
framework. The same conclusion was reached as in 
\cite{LW04}
for $D=2+1$, but a stronger conclusion was obtained
in $D=3+1$, where the $O(1/l^3)$ 
term in the action was shown to be universal (and equal to the
value in the Nambu-Goto expansion).

This year there has been further, dramatic progress. In 
\cite{AHEK}
the static gauge approach was used and extended to include the 
constraints that arise from closed-closed (torus) duality.
This enabled 
\cite{AHEK}
to show that the terms up to and including $O(1/l^5)$ are
universal, and of course equal to what you get in the Nambu-Goto 
model. (There are some technical qualifications to this in $D=3+1$ 
that I am omitting.) This work demonstrated how one can extend one's
predictions for the effective string action, by finding new physical 
conditions that it must satisfy. One may speculate that further progress
could be made by going beyond the cylinder and torus, to consider
other boundary conditions for the surfaces that one is integrating
over, so as to create new, more powerful constraints on $S_{eff}$.
 
In addition the authors of
\cite{AHEK})
calculated the effective string action in some confining gauge 
theories with a gauge-gravity dual, and showed  explicitly that 
in these cases the coefficients up to and including $O(1/l^5)$ 
are indeed as predicted by their general arguments.

Finally, as I have been writing this section, some papers
have appeared
\cite{Dass09}
extending the Polchinski-Strominger approach
\cite{PS,JD}
and also  showing that the terms up to  $O(1/l^5)$ are universal.
The most recent of these papers 
\cite{Dass09b}
makes the dramatic claim that (with certain constraints) the 
energy spectrum of any effective string theory 
is, to all orders, the same as that of Nambu-Goto.

While I have not had the time to digest the most recent papers,
it is clear that this is an exciting area in which a great deal of 
progress is being made at this moment.

\subsection{Lattice calculations - a potted history}

Having spent quite a lot of time describing the analytic
work, I do not have the time to do more than point briefly
to some of the numerical work that has been carried out over
this same period. This is particularly inappropriate because
there has been a huge amount of this work, and its increasing 
range, sophistication and precision has provided strong motivation 
for the recent analytic work that I have been describing. I will
just point to some work in various directions, and leave it
to you to follow their references and citations to build a
more detailed and balanced picture for yourselves.

In the early to mid-80's there was already a great
deal of work testing string model 
predictions with numerical lattice calculations, and with
open-closed string duality in mind, by for example the Copenhagen 
group, e.g
\cite{AOP}.
Some of the earliest numerical work that produced reliable 
ground state energies for closed strings and for the Luscher
coefficient was in that same period 
\cite{deFSST}.
The development of blocking
\cite{block}
and smearing algorithms
\cite{smear}
in the mid-80's finally made the accurate calculation of 
energies and string tensions routine.

The potential between static sources was, of course, of continued
interest, but the pioneering calculations for excited string states
date to the early-90's, e.g.
\cite{CM}.
The interest here was both theoretical and phenomenological: the
excited string states could be used in a Schrodinger equation
to get predictions for the masses of hybrid mesons where some
of the quantum numbers are carried by excited glue.   
In the 90's there was alot of progress by the Torino group, e.g.
\cite{Caselle},
investigating numerically the match between string theory
predictions for Wilson loop expectation values and what one
obtains in various gauge and spin models. It is in this body of work
that one first sees a prolonged and serious focus on matching
the Nambu-Goto model to numerical results, a focus which became
commonplace in later work. This work has continued into
the 00's with, for example, calculations in more `exotic'
theories
\cite{Caselle2}.

In the late 90's and early-00's there was the first sequence of 
calculations
\cite{kuti}
that was dedicated to calculating the full string spectrum 
(open and closed) in SU(2) and SU(3) gauge theories.
In the same period,  powerful new Monte Carlo techniques were developed
\cite{multihit}
that made possible the numerical calculation of the Luscher coefficient
with greater accuracy, thus confirming the earlier work that had pointed
to it being in the same universality class as the simplest bosonic
string model. The last decade has also seen, for the first time,
reliable numerical calculations at large $N$ 
\cite{MT98d3,BLMTd3d4,HMMT04},
in some cases for a very large $N$ indeed
\cite{jkrn,rn}.
There have simultaneously been a number of very accurate
calculations of open strings, both of the ground state 
(i.e. the heavy quark potential)
\cite{pot}
and with some work on excited states
\cite{expot}
and work on sources in other representations than the fundamental
\cite{balideldar}.
And since the mid-00's there has been the dedicated calculation
of the spectrum of closed flux tubes, in both $D=2+1$ 
\cite{BBMTk1d3,AABBMTk1d3,BBMTk2d3,AABBMTk2d3}
and $D=3+1$
\cite{AABBMTd4}
that I am going to devote the rest of my lectures to.

Before moving on let me say a few comparative words about calculations 
of Wilson loops, open strings (potentials) and closed strings 
(torelons). Expectations of Wilson loops are transforms
of eigenspectra (although care has to be taken with the
self-energies associated with the boundaries) and are an
alternative way to test string models. With open strings
there is a transition between short-distance perturbative
physics (the Coulomb potential) 
and the long-distance confining physics. This is potentially
of great interest (especially at large $N$) but it can
introduce extra ambiguities in the extraction of the flux tube
spectrum, if that is what one is primarily interested in
(as we are here).
The spectrum of closed flux tubes, stabilised by closing
them around a spatial torus, is a particularly clean way to
investigate the properties of flux tubes, and that is why
we have focused on that approach in the work below.

\section{The spectrum of closed flux tubes: $D=2+1$}
\label{section_latticed3}

In this Section I will discuss the energy spectrum of closed
flux tubes in SU($N$) gauge theories in $2+1$ dimensions.
The numerical results that I shall show are all from
\cite{BBMTk1d3,AABBMTk1d3,BBMTk2d3,AABBMTk2d3} 
or from unpublished work within that collaboration.

I am not going to say much about $D=2+1$ except to remind you that the
gauge coupling $g^2$ has dimensions of mass, so the dimensionless
expansion parameter for physics on a scale $l$ will be $lg^2$.
Thus the theory is free in the UV and strongly coupled in the IF:
much like $D=3+1$ in other words. (At least in this respect.)
Lattice calculations
\cite{MT98d3}  
have shown that the theory is linearly confining at low $T$,
just like SU($N$) gauge theories in $D=3+1$. Our understanding
of the dynamics of confinement, and our analytic control over the 
long-distance physics, is not much better in $2+1$ than in $3+1$ 
dimensions. (But see
\cite{nair}.)  
In many ways, the fundamental dynamical questions
are similar in both dimensions and it is useful to pose them
in both contexts simultaneously, as I will be doing in these lectures.

\subsection{Quantum numbers and operators}
\label{section_opsd3}

Consider a confining flux tube, with the flux in the fundamental 
representation. Let it wind once around the $x$-torus, which we take
to be of length $l$. There are a number of symmetries some of which
are interesting and some of which are not. 

$\bullet$ Let $p$ be the longitudinal momentum of the flux tube.
By periodicity this is quantised, $p=2\pi q/l$ where $q$ is an integer.
(When we are on a lattice we will express $p$ and $l$ in lattice units.)
It is plausible that the ground state, with energy $E_0(l)$, is invariant
under longitudinal translations, and so must have longitudinal momentum
$q=0$. To have $q\neq 0$ a flux tube must have a deformation so that
it is not invariant under longitudinal translations. That is to say,
it must be excited in some non-trivial way. Thus 
$E^2(p) \neq E^2_0 + p^2$ and the calculated value of $E(p)$ carries
non-trivial dynamical information: $p$ is an interesting quantum number.

$\bullet$ By contrast, if $p_\perp$ is the transverse momentum then
we simply expect $E^2(p)= E^2_0 + p^2_\perp$, so  $p_\perp$ is not,
for us, an interesting quantum number.

$\bullet$ Under charge conjugation, $C$, the direction of the flux is 
reversed. Thus a flux tube has zero overlap onto its charge-conjugated 
homologue (see below) and so linear combinations of definite 
$C$ are trivially degenerate. (Except for SU(2) where $C$ is trivial.)
Thus $C$ is not interesting and we shall consider flux tubes whose
flux is in the +ve $x$ direction.

$\bullet$ Consider the 2 dimensional parity operation 
$(x,y) \stackrel{P}{\to} (x,-y)$.
It is plausible that the absolute ground state, with energy $E_0(l)$, 
is invariant under reflections, and so will have $P=+$, with the
$P=-$ linear combination being null. Thus the lightest non-null 
$P=-$ state involves a flux-tube with a non-trivial deformation,
and so $P$ is certainly an interesting quantum number.
More specifically, in a string model $h(x)  \stackrel{P}{\to} -h(x)$
so that the its Fourier coefficients also satisfy
$a_k  \stackrel{P}{\to} - a_k$
and hence the parity of a string excitation is simply
\begin{equation}
P = (-1)^{number \ of \ phonons}.
\label{eqn_Pd3}
\end{equation}
So the lightest $P=-$ state will have one excited phonon with $k=1$.

$\bullet$ Suppose our (x,y) torus is symmetric. Consider rotations of 
the flux tube by $\pi/2$, so that it winds instead around the $y$-torus,  
or by $\pi$ so that the flux is reversed. Both these flux tubes
will have zero overlap onto the original flux tube. Thus rotations
are uninteresting. (In 2 space dimensions there are, of course,
no rotations around the axis of the flux tube.)

For the above reasons we choose to calculate the flux tube energy
as a function of its length $l$, its longitudinal momentum 
$p=2\pi q/l$, and its parity $P$.

\begin{figure}[htb]
\begin	{center}
\leavevmode
\includegraphics[width=2.5cm]{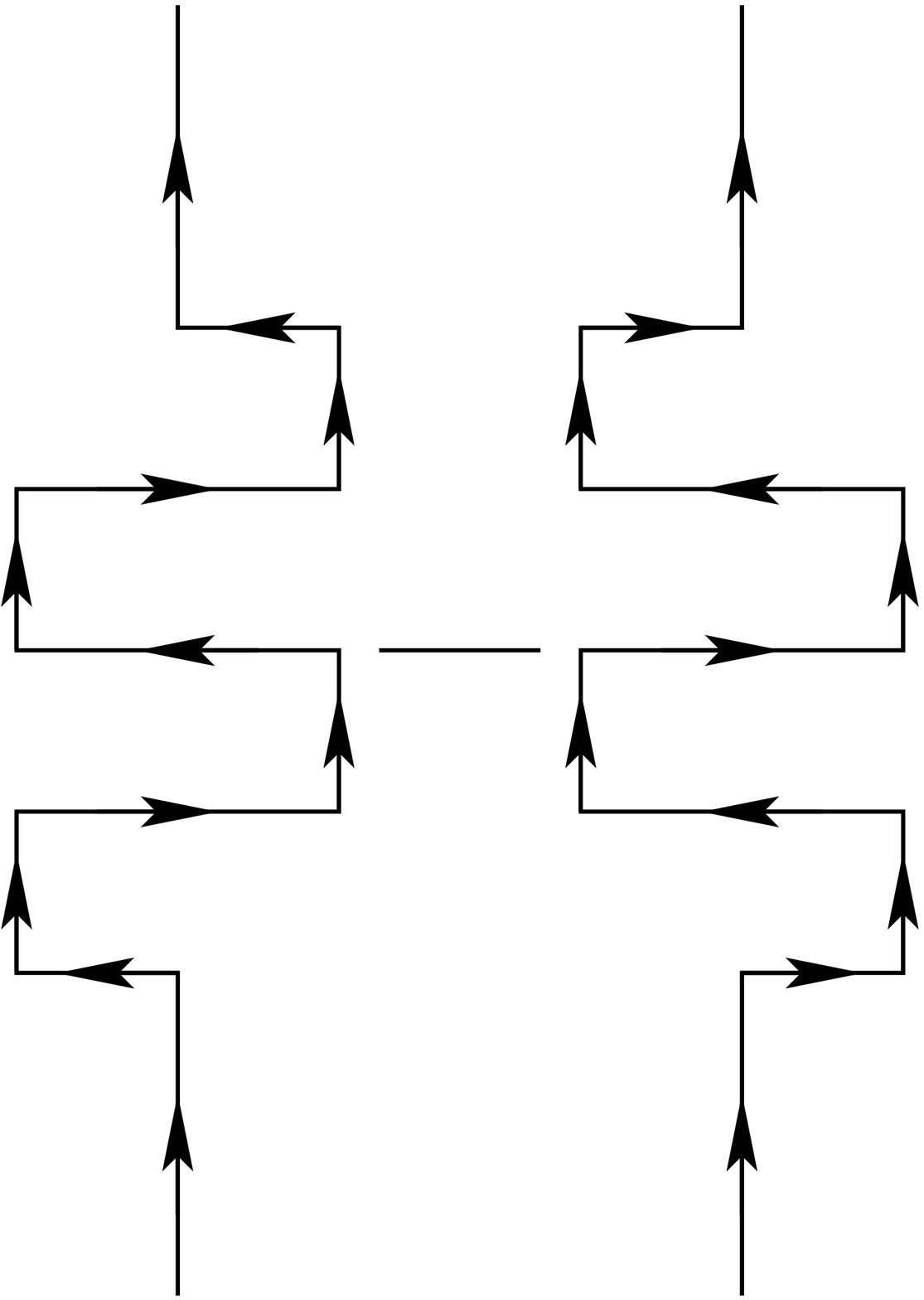} $\quad$
\includegraphics[width=2.5cm,height=3.0cm]{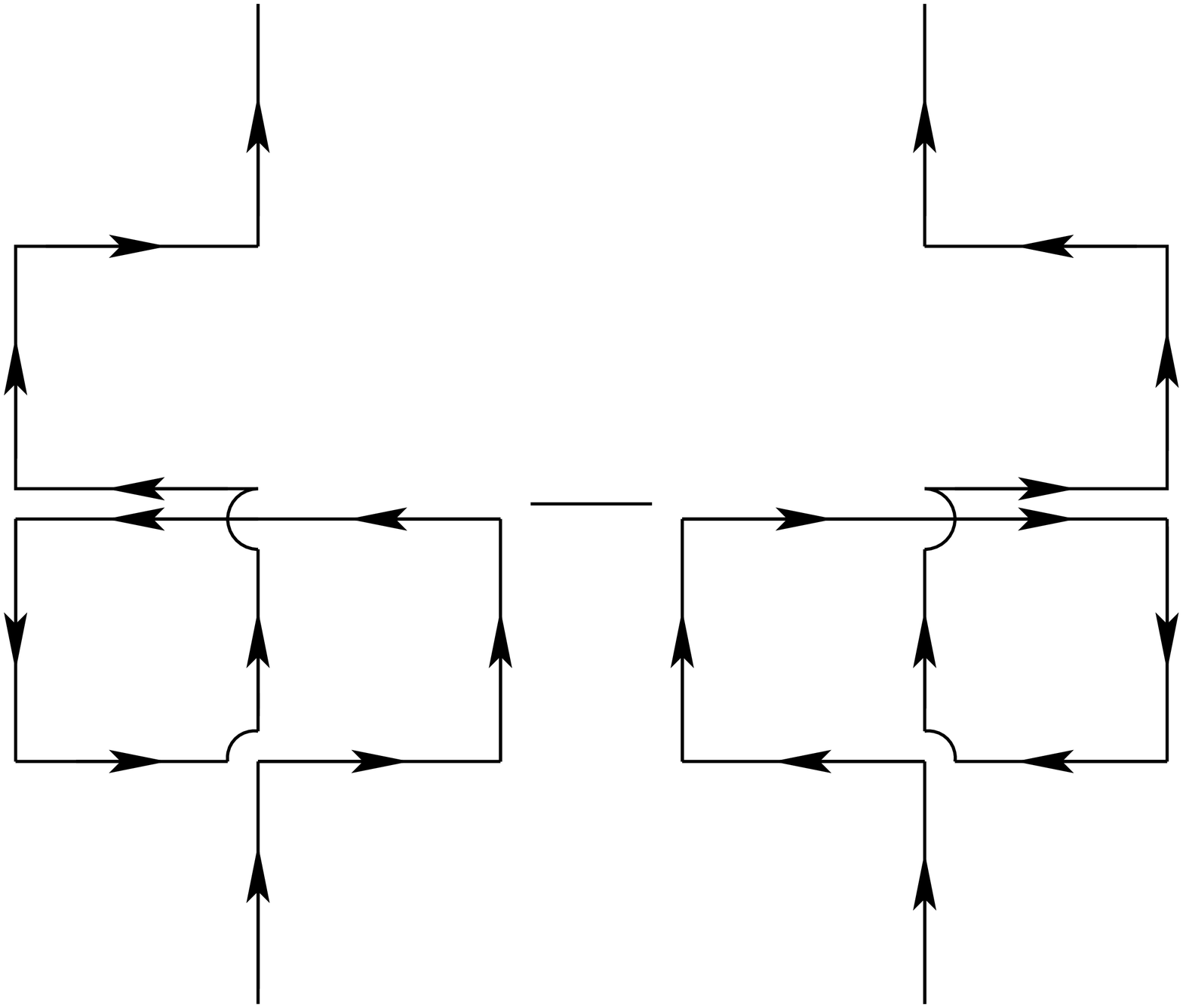} $\quad$
\includegraphics[width=2.5cm]{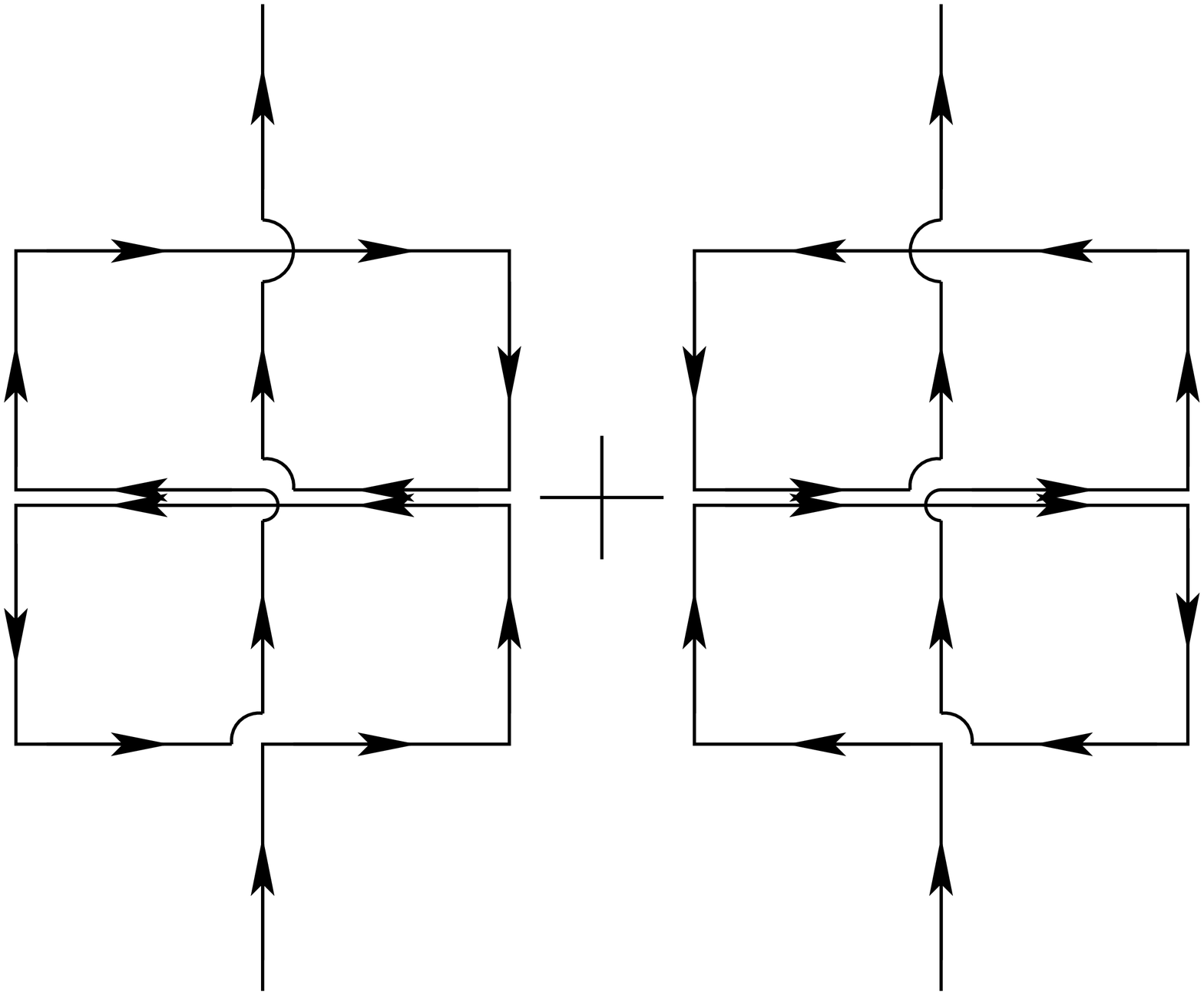} $\quad$
\includegraphics[width=2.5cm,height=3.0cm]{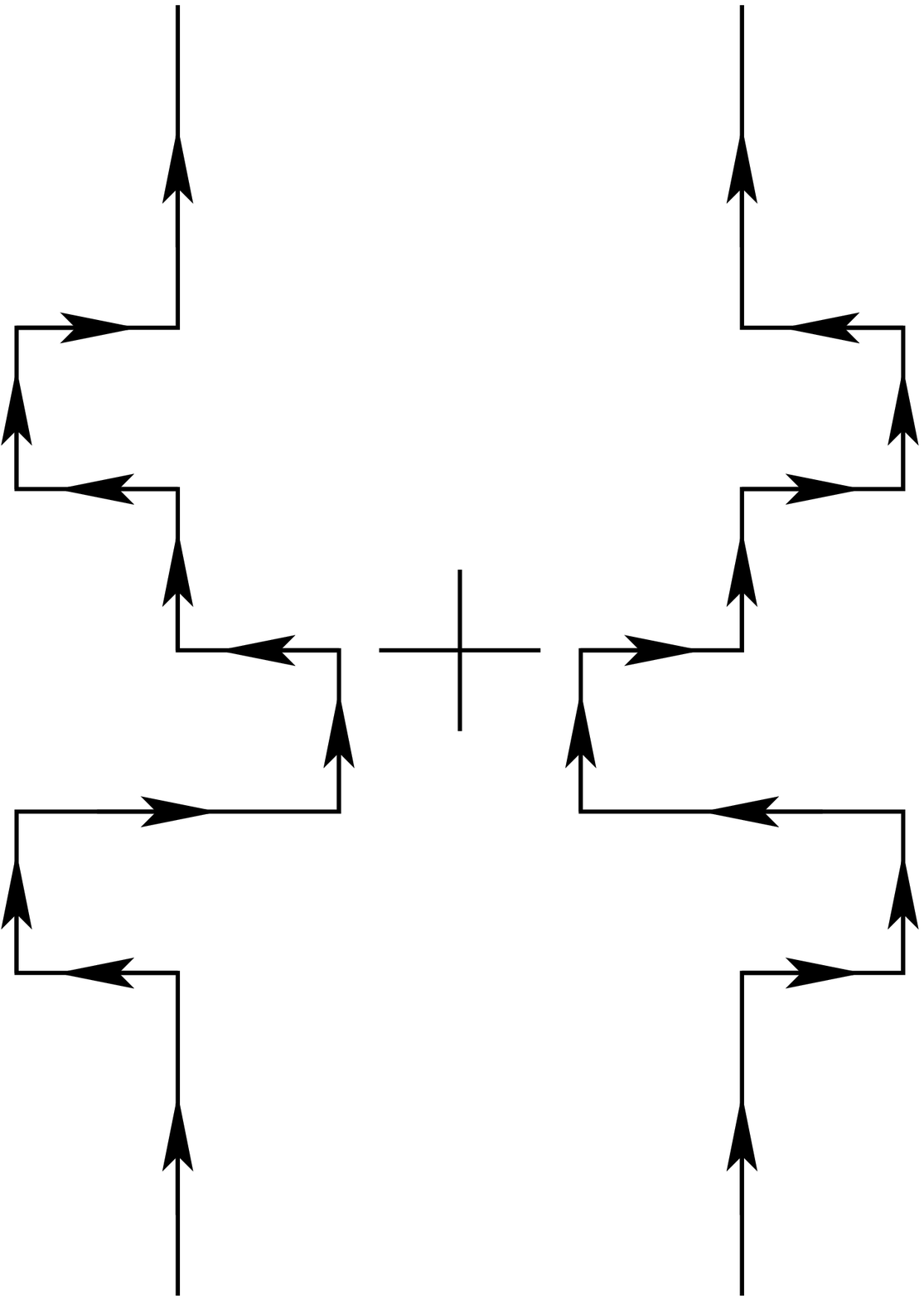}
\end	{center}
\caption{Examples of operators with non-trivial parity: $P=-,-,+,+$
starting from the left.}
\label{fig_ops}
\end{figure}

We now need some lattice operators whose correlators will give us
the desired spectrum. The most elementary such operator is the
simple Polyakov loop:
\begin{equation}
l_p(n_y,n_t) = \mathrm{Tr} 
\left\{\prod^L_{n_x=1} U_x(n_x,n_y,n_t)\right\} 
\label{eqn_poly}
\end{equation}
where $l=La$ and we take a product of the link matrices in 
the $x$-direction once around the $x$-torus. 

That such a winding operator should couple to winding flux tubes 
is heuristically plausible, but we can argue a little better than that.
Consider a transformation 
\begin{equation}
U_x(n_x=L,n_y,n_t) \longrightarrow zU_x(n_x=L,n_y,n_t),
\qquad z\in Z_N, \,\, \forall n_x,n_y
\label{eqn_zgauge}
\end{equation}
where all links in the $x$-direction at, say, the boundary $n_x=L$
are multiplied by an element of the centre. This does not affect
the value of any plaquette, since a $z$ will always be accompanied by
a $z^\star$ from the conjugate link at the opposite side of the
plaquette, which gives unity, and so the action is not changed.
And neither is the invariant Haar group measure. Thus the weighting
of this transformed field is identical to that of the original field.
Moreover, by a similar argument to that for the plaquette, the 
product of link matrices around any contractible closed loop $l_C$ 
is invariant, since the number of $z$ and $z^\star$ will be the same. 
However a non-contractible closed loop like the Polyakov
loop is not invariant: clearly $l_p \to z l_p$. We therefore
see that
\begin{equation}
\langle l^\dagger_p  l_C \rangle
=
z \langle l^\dagger_p  l_C \rangle, \ \ 
\forall z\in Z_N
\quad \Longrightarrow \quad 
\langle l^\dagger_p  l_C \rangle 
=
0.
\label{eqn_zlG}
\end{equation}
(This assumes that the center symmetry is not spontaneously broken.)
Eqn(\ref{eqn_zlG}) is true for any winding loop $l_p$ and any
closed contractible loop $l_C$. So the corresponding states are
completely orthogonal. The contractible loops clearly generate
localised states like glueballs, while the winding loops generate
non-local winding states which include winding flux tubes. 
Note that a similar argument to that in eqn(\ref{eqn_zlG}) tells
us that $l_p$ and its charge conjugation, $l^\dagger_p$, are
orthogonal, except in SU(2). In the continuum the field transformation
in eqn(\ref{eqn_zgauge}) arises when we note that periodicity for 
adjoint fields implies periodicity up to a gauge transformation
that is an element of the center. 

The operator in  eqn(\ref{eqn_poly}) is localised in $n_y$ and
so has $p_\perp \neq 0$. If we sum over $n_y$, to get
$l_p(n_t) = \sum_{n_y} l_p(n_y,n_t)$ we get an operator with 
$p_\perp = 0$, and from now on we assume this has been done.  
This operator is manifestly invariant under longitudinal translations 
as well, so $p=0$. It is also invariant under parity $P$. So to have 
$p\neq 0$ or $P\neq +$ we must introduce a deformation in the operator 
in eqn(\ref{eqn_poly}). Some examples of $P=-$ operators are shown in 
Fig.{\ref{fig_ops}}. By translating such an operator by $\Delta x$ in the
$x$ direction, multiplying it by the phase factor 
$\exp\{i 2 \pi q \Delta x/L\}$,
and then adding all such translations, one obtains an operator with
longitudinal momentum $p=2\pi q/L$. Doing so for the operator in
eqn(\ref{eqn_poly}) would have yielded a null operator for $p\neq 0$,
but for the operators in Fig.{\ref{fig_ops}}, that will not in general
be the case.

In practice, to obtain good overlaps onto any states at all, one needs 
to smear
\cite{smear}
and block 
\cite{block}
the `link matrices' that appear in operators such as those 
in  Fig.{\ref{fig_ops}}. And one also needs many operators 
in addition to those shown here in order to have adequate overlaps 
onto a good number of excited states. So, as described in more detail in
\cite{AABBMTk1d3},
we typically have 150-200 operators in each of our $D=2+1$ calculations.

\subsection{Ground state energy}

\begin{figure}[htb]
\begin	{center}
\leavevmode
\input	{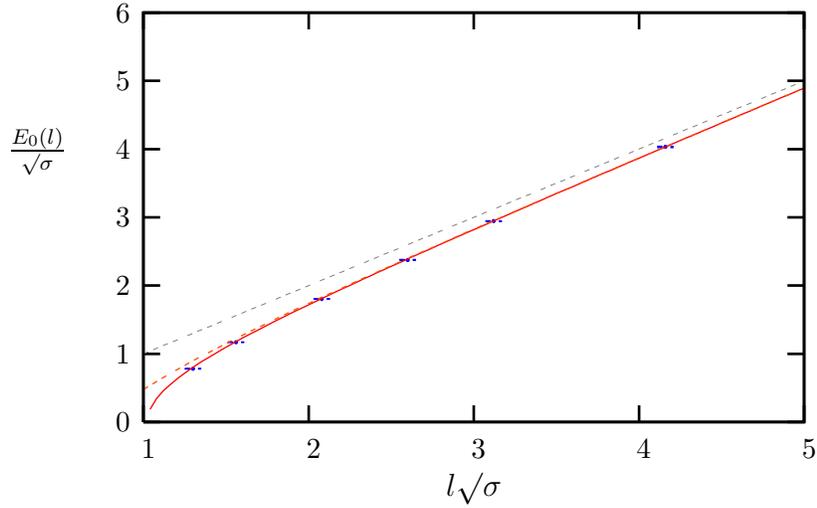}
\end	{center}
\caption{Ground state flux tube energy in SU(5) for lattice spacing
$a\surd\sigma \simeq 0.130$. Shown are fits with a Luscher correction
(dashed line) and with Nambu-Goto (solid line). The linear piece of 
such a fit, $\sigma l$, is also shown (dots) for orientation.}
\label{fig_E0n5b80}
\end{figure}

In Fig.{\ref{fig_E0n5b80}} I display how the energy, $E_0$, of the
lightest flux tube in the $q=0, \, P=+$ sector of states, varies
as a function of its length $l$. This is a calculation in SU(5) 
and at a value of the bare inverse coupling, $\beta = 80$,
which corresponds to a lattice spacing in physical units
of $a \simeq 0.130/\surd\sigma $. (Of course the latter is something
we only know after our calculation of the string tension, $a^2\sigma$.)
This is a small value of $a$ at which $O(a^2)$ lattice spacing 
corrections are known to be negligible
\cite{MT98d3,BBMTk1d3}.
It therefore makes sense to present the values of $E_0$ and $l$ in 
physical units, as we have done in Fig.{\ref{fig_E0n5b80}}, using
the value of $a\surd\sigma$ we obtain by fitting the $L$-dependence
of $aE_0(L)$.

Also shown  in  Fig.{\ref{fig_E0n5b80}} are the best fits with either a 
simple Luscher correction, $E_0(l)=\sigma l - \pi/6l$, or with the
full Nambu-Goto result,  $E_0(l)=\sigma l (1 - \pi/3\sigma l^2)^{1/2}$.
Fitting the lattice values of $aE_0(L)$ as a function of $L=l/a$ gives
us a value for the string tension $a^2\sigma$ which we then use to
produce the rescaled coordinates plotted in Fig.{\ref{fig_E0n5b80}}.
We see that although the Luscher correction captures much of the
deviation from linearity at small $l$, Nambu-Goto clearly works
even better. I also show the linear piece of the fit, so that you
can see the deviation from linearity more explicitly.

As remarked earlier, in the context of the glueball calculation
in Section~\ref{subsection_lattice}, the numerical calculation 
gets less reliable
as the energy becomes large -- and here that happens as $l$ becomes
large. That limits how far we can go in $l$. To see what this means
in practice, recall that we are looking for an exponential decay of our
correlation function at large enough $t$: $C(t=an_t) \propto e^{- aEn_t}$.
One way to see how well this is being determined is to fit neighbouring
values of $t=an_t$ with a simple exponential,
\begin{equation}
\frac{C(n_t)}{C(n_t-1)} =  \exp\{-aE_{eff}(n_t)n_t\}. 
\label{eqn_Eeff}
\end{equation}
If $aE_{eff}(n_t)$ is independent of $n_t$ within errors 
for $n_t \geq n_0$, then
we can fit the data with a single exponential for $n_t \geq n_0$.
Clearly, the smaller the errors on $C(n_t \geq n_0)$ the more reliable 
will be the estimate of the energy.

\begin{figure}[htb]
\begin	{center}
\leavevmode
\input	{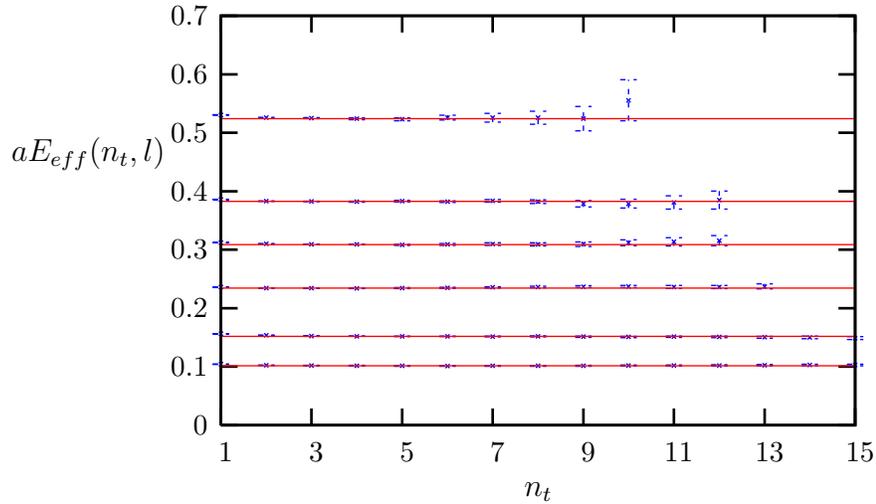}
\end	{center}
\caption{Effective energies, as in eqn(\ref{eqn_Eeff}),
which lead to the energies in Fig.~\ref{fig_E0n5b80}.}
\label{fig_E0n5b80eff}
\end{figure}

In Fig.{\ref{fig_E0n5b80eff}} I plot the values of a $aE_{eff}(n_t;l)$ 
corresponding to the various values of $l$ in Fig.{\ref{fig_E0n5b80}}.
At the smallest values of $n_t$ the errors are invisible on this plot,
but it is clear that we have precise control of the asymptotic exponential 
decay for all our values of $l$. It is also clear, however, that 
the range of $n_t \geq n_0$ where  $aE_{eff}(n_t;l)$ is determined
with useful accuracy shrinks rapidly as $l \, \uparrow$ and that if we
were to attempt a calculation at significantly larger $l$ we would
soon lose control of the asymptotic exponential decay.

\begin{figure}[htb]
\begin	{center}
\leavevmode
\input	{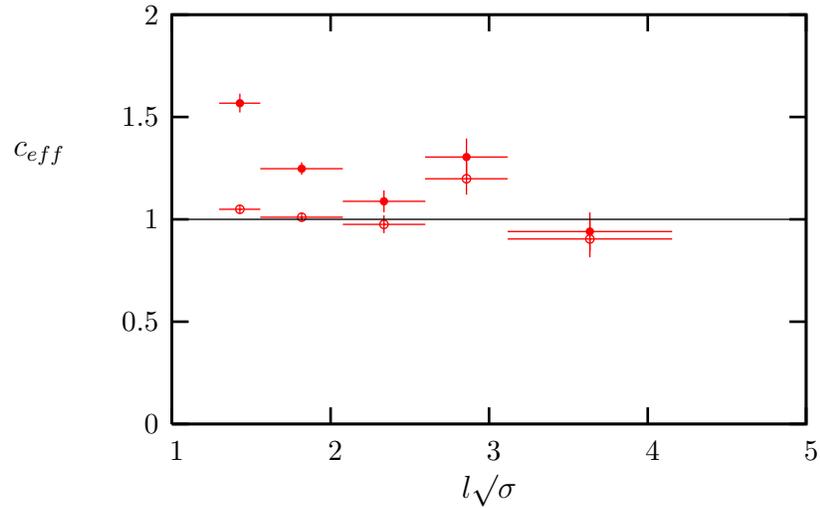}
\end	{center}
\caption{SU(5) effective central charge: from the Luscher ($\bullet$)
and Nambu-Goto ($\circ$) fits in eqn(\ref{eqn_MLceff}) and
eqn(\ref{eqn_NGceff}) respectively.}
\label{fig_ceffn5}
\end{figure}

Let us now turn to a more precise analysis of how well the fits shown 
in  Fig.{\ref{fig_E0n5b80}} actually work. There are many ways to 
do this, and here we do the following, in analogy with the effective 
energy plots just shown. We define effective coefficients (and
string tensions) by
\begin{equation}
E_0(l) = \sigma_{eff} l - c_{eff}\frac{\pi(D-2)}{6l}
\label{eqn_MLceff}
\end{equation}
and
\begin{equation}
E_0(l) = \sigma_{eff} l 
\left( 1 - c_{eff}\frac{\pi(D-2)}{3\sigma_{eff} l^2} \right)^{\frac{1}{2}}
\label{eqn_NGceff}
\end{equation}
for the two kind of fits. We determine $a^2\sigma_{eff}$ and
$c_{eff}$ for each pair of values $l=l_i,l_{i+1}$ where 
$l_i \leq l_{i+1} \, \forall i $. Plotting the result in 
Fig.{\ref{fig_ceffn5}} - the horizontal `error bars' indicate the distance
between the values of $l_i\surd\sigma$ and $l_{i+1}\surd\sigma$ -
we observe that we appear to have
\begin{equation}
c_{eff} \stackrel{l\to\infty}{\longrightarrow} 1
\label{eqn_asymceff}
\end{equation}
in both cases. That is to say, the central charge corresponds
to a bosonic string theory where the only massless modes on the
flux tube are the transverse oscillations.

You may be disturbed by the `peak' in $c_{eff}$ around
$l\surd\sigma = 3$. This is a nice example of the kind of
statistical fluctuation (here $\sim 2.5$ standard deviations) 
which is large enough to cause some hesitation, while
being small enough to arise quite regularly 
in numerical calculations. To check that it is 
indeed a statistical fluctuation, I show in Fig.{\ref{fig_ceffn4}
a similar calculation, but this time in SU(4) and with greater 
statistical accuracy. There are other examples that I could
show as well.

\begin{figure}[htb]
\begin	{center}
\leavevmode
\input	{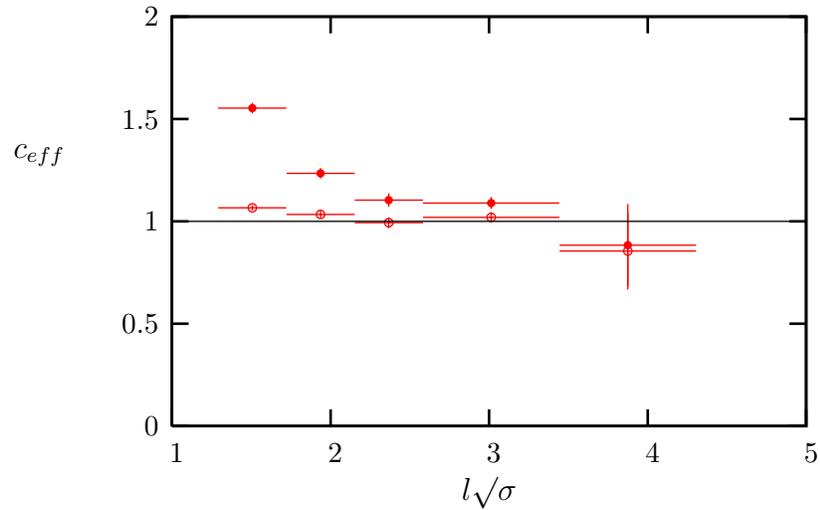}
\end	{center}
\caption{As in Fig.~\ref{fig_ceffn5} but for SU(4).}
\label{fig_ceffn4}
\end{figure}

Since the Luscher correction is only the leading correction
in an expansion in powers of $1/\sigma l^2$, it is entirely expected
that $c_{eff}(l)$ should deviate significantly from unity as $l$
decreases. What is much more of a surprise is that for the
Nambu-Goto fit, $c_{eff}\simeq 1$ for all values of $l$.
That is to say, the deviations from Nambu-Goto are very small at 
all our values of $l$, even down to $l\surd\sigma \sim 1.5$ 
where the flux tube is about as wide as it is long and where
it no longer `looks' anything like an ideal thin string. 

One might imagine that any such free string-like behaviour will
arise only at large $N$, if it is to arise anywhere. 
It is therefore interesting 
to go to the smallest possible $N$ and see what happens there.
So in  Fig.{\ref{fig_ceffn2} I show the corresponding plot for
SU(2). (These calculations are fast and it is easy to achieve
very high statistical accuracy.) We observe a very similar pattern
to what we have seen at larger $N$. It appears that the free
bosonic string theory provides a very good approximation for all $N$.

\begin{figure}[htb]
\begin	{center}
\leavevmode
\input	{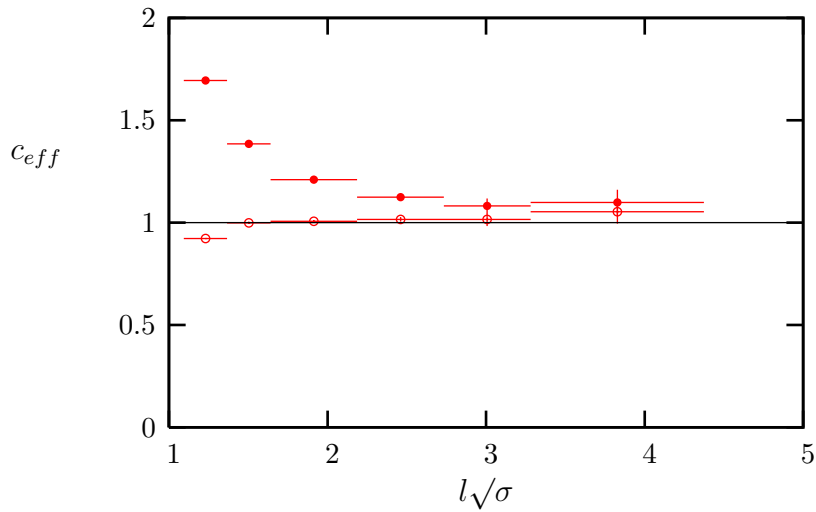}
\end	{center}
\caption{As in Fig.~\ref{fig_ceffn5} but for SU(2).}
\label{fig_ceffn2}
\end{figure}

Now, the Nambu-Goto expression for $E_0(l)$ becomes tachyonic for
$l\surd\sigma < \pi/3$, as we see from eqn(\ref{eqn_EnNG}). 
Given that we find such good agreement
with Nambu-Goto, one might ask what happens as we approach this
stringy `Hagedorn' transition. For $N>3$ the critical deconfining 
length $l_c\surd\sigma > \pi/3$, so we no longer have a flux tube when
$l$ is reduced to $\pi/3$ and the question cannot be addressed. 
However for SU(2) we have a second order phase transition at
$l_c\surd\sigma \sim 0.9$. What we find there, as shown in 
Fig.{\ref{fig_E0n2crit}}, is what one would expect: as we approach 
$l_c$ the energy dependence is governed by the critical exponents 
appropriate to the universality class of the transition (the $D=2$ 
Ising model) and so vanishes linearly in $(l-l_c)$. As $l$ increases
this linear behaviour matches smoothly onto the square root behaviour
of the Nambu-Goto prediction. As we see from Fig.~\ref{fig_E0n2crit}
this happens before the turn-over to the would-be stringy transition
at $l\surd\sigma=\pi/3$.

\begin{figure}[htb]
\begin	{center}
\leavevmode
\input	{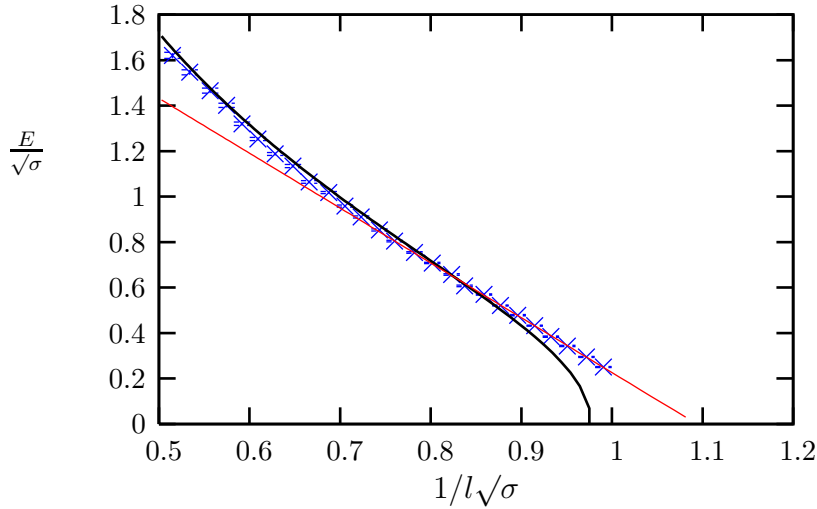}
\end	{center}
\caption{Ground state energy of closed flux tube in SU(2) when the
length $l$ approaches the deconfining transition.}
\label{fig_E0n2crit}
\end{figure}

\subsection{Excited state energies}

In Nambu-Goto, the deviation of the ground state energy, $E_0(l)$, 
from $\sigma l$ is, to lowest 
order in $1/l^2$, simply the (regularised) sum of the zero point energies
arising from the quantisation of all the oscillation modes of all 
possible wavelengths, $\lambda_n = l/n$. The fact that the
flux tube is well-described by this at small $l$ is surprising because 
one would not expect that the excitations of a short fat flux tube are
the same as those of a thin string. Clearly the next step is to calculate 
explictly the low-lying energy spectrum and see how it varies with $l$.

In Fig.~\ref{fig_Eq0n3} I show the low-lying spectrum we obtain 
in SU(3) at a lattice spacing $a\surd\sigma \simeq 0.174$ and for
zero longitudinal momentum, $q=0$. The ground state is well fitted
by Nambu-Goto as shown. This fit determines $a^2\sigma$ and so the
predictions shown for the excited states are completely parameter-free.
We find that the first excited state of the flux tube has positive parity, 
$P=+$, just as in the Nambu-Goto model. Moreover, as one can see from 
Fig.~\ref{fig_Eq0n3}, it agrees well with the predicted Nambu-Goto energy,
corresponding to $N_L=N_R=1$ in eqn(\ref{eqn_EnNG}), for 
$l\surd\sigma \geq 2$. At this value of $l$
the flux tube is almost as wide as it is long, so it is quite remarkable
that the excitation energy should be so close to what one obtains 
from a pair of left and right moving phonons, of the longest possible 
wavelengths, on an ideal string.

The next four states become degenerate for larger $l$ and are
already nearly so for $l\surd\sigma \simeq 3$. We find that two have
$P=+$ and two have $P=-$. This degeneracy pattern is precisely
as predicted by the Nambu-Goto model, where the states are
$a_1a_1 a_{-1} a_{-1}|0\rangle,\, a_2a_{-2}|0\rangle,\,
a_2 a_{-1} a_{-1}|0\rangle,\,a_1a_1 a_{-2}|0\rangle$ in the
notation of Section~\ref{subsection_NG}. Moreover we see that
the average energy of these four states is very close to the 
Nambu-Goto prediction, even at quite small $l$.

\begin{figure}[htb]
\begin	{center}
\leavevmode
\input	{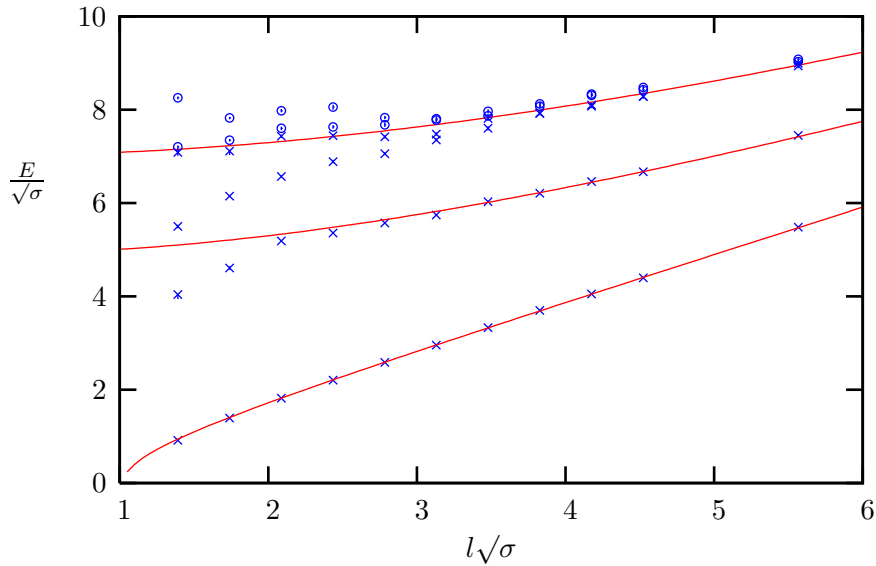}
\end	{center}
\caption{Low-lying spectrum of $p=0$ flux tubes in SU(3), with
$P=+$ ($\times$) and $P=-$ ($\circ$) states. Lines are Nambu-Goto
predictions.}
\label{fig_Eq0n3}
\end{figure}

Of course, all this is at a particular finite lattice spacing. To see 
whether these results survive the continuum limit, we repeat the 
calculations at a lattice spacing  $a\surd\sigma \simeq 0.087$,
which is smaller by about a factor of two. (We also increase all our
lattice sizes, in lattice units, by a factor of 2 so that the volume
is the same in physical units.) In 
Fig.~\ref{fig_Eq0n3cont} I compare the energies obtained with the 
coarser and finer lattice spacings. There is no visible variation,
and we can confidently assume that our results also hold
in the continuum limit of SU(3).

\begin{figure}[htb]
\begin	{center}
\leavevmode
\input	{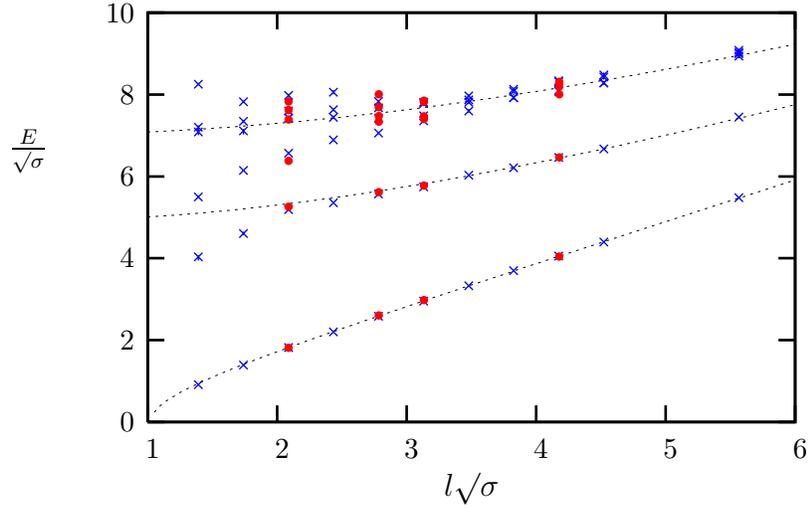}
\end	{center}
\caption{Spectrum from Fig.~\ref{fig_Eq0n3}, $\times$, compared to
spectrum for $a\to a/2$, $\bullet$.}
\label{fig_Eq0n3cont}
\end{figure}

It is clearly interesting to see what happens as we go to larger $N$
since that is the limit in which the effective string description
has the most compelling motivation. In Fig.{\ref{fig_Eq0n3n6}} I compare 
what one obtains in 
SU(6) and SU(3) at a (nearly) common value of $a\surd\sigma \sim 0.17$.
There is clearly very little $N$-dependence. So what we are finding
is also representative of SU($\infty$) in the continuum limit.

\begin{figure}[htb]
\begin	{center}
\leavevmode
\input	{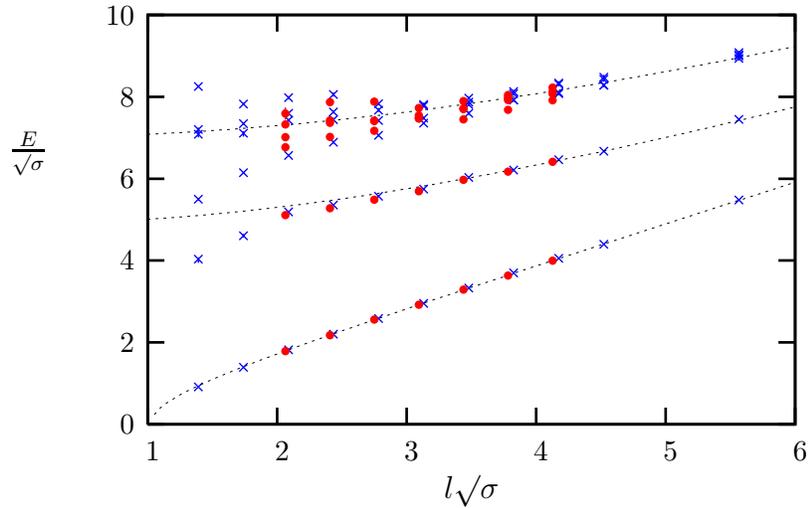}
\end	{center}
\caption{Spectrum from Fig.~\ref{fig_Eq0n3}, $\times$, compared to
SU(6) spectrum, $\bullet$, at the same $a$.}
\label{fig_Eq0n3n6}
\end{figure}

We now turn to flux tubes that have a non-zero longitudinal momentum.
As we remarked earlier, a flux tube with $q\neq 0$ must possess a 
deformation if it is not to be null state, so this will provide us with 
further information about the excitation spectrum. Moreover, since
the energy of the ground state in any sector with given quantum numbers 
is usually the one that has the smallest systematic errors in our
variational calculation, this
will enable us to obtain the corresponding flux tube excitation energies 
with no ambiguity. 

In Figs.~\ref{fig_Eq1n3} and ~\ref{fig_Eq2n3} I show the results for 
the lowest non-zero  momenta, $q=1$ and $q=2$. We see that
the ground states in each
case fall very precisely onto the parameter-free Nambu-Goto prediction
and that they have the same quantum numbers as in the Nambu-Goto spectrum,
i.e. $P=-$ with $q=1$ (corresponding to $a_1|0\rangle$) and a pair 
of degenerate $P=+$ and $P=-$ states with $q=2$ (corresponding to
$a_1a_1|0\rangle$ and $a_2|0\rangle$). We also note that the 
energies of the excited states are, in each case, very 
close to the Nambu-Goto prediction, although their large energies
(especially for $q=2$) means that the systematic errors are likely
to be significant.

\begin{figure}[htb]
\begin	{center}
\leavevmode
\input	{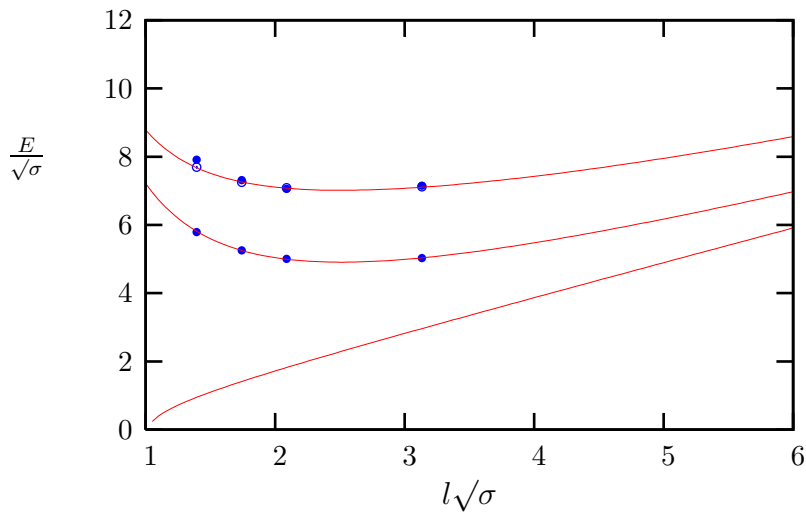}
\end	{center}
\caption{Spectrum of closed flux tubes with $p=2\pi/l$ in SU(3).
Nambu-Goto predictions shown, including $p=0$ ground state
from which we extract $a\surd\sigma$.}
\label{fig_Eq1n3}
\end{figure}

\begin{figure}[htb]
\begin	{center}
\leavevmode
\input	{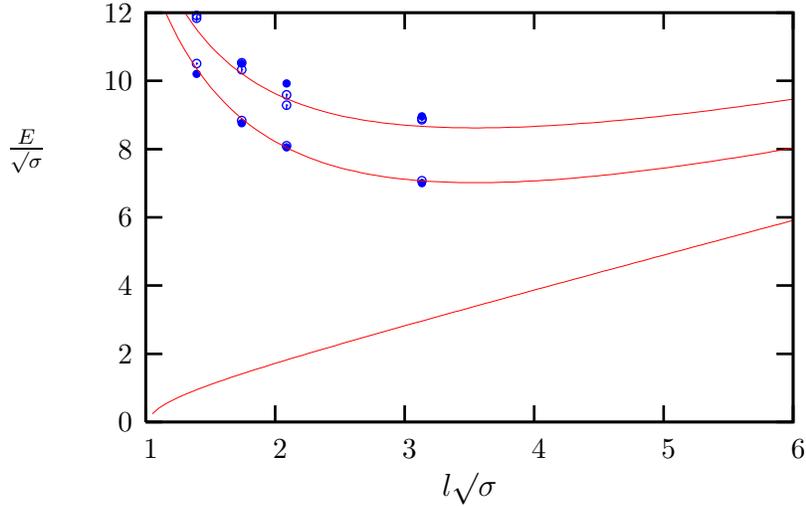}
\end	{center}
\caption{As in Fig.~\ref{fig_Eq1n3} but for $p=4\pi/l$.}
\label{fig_Eq2n3}
\end{figure}

\subsection{Comparison with theory}

The fact that the physical flux tube has an excitation
spectrum so similar to that of a free string theory is
striking and, especially at small $l$, counter-intuitive.
However, as we described in Section~\ref{subsection_recenttheory}, 
recent analytic studies have shown that the effective string
action, when expanded in powers of $1/l$, is the
same as the Nambu-Goto action up to $O(1/l^5)$. It is
natural to ask whether what we have seen is no more than
a reflection of these results.

To answer this question we show again the $q=0$ SU(3)
spectrum in  Fig.{\ref{fig_Eq0n3theory}}, but this time
accompanied by the theoretical prediction to $O(1/l)$
\cite{LSW,PS}
that one obtains from the leading Gaussian approximation
(the `Luscher correction'), the one to $O(1/l^3)$
obtained by Luscher and Weisz, and by Drummond in 2004
\cite{LW04,JD,HDPM}
and finally the one to $O(1/l^5)$ obtained this year by
Aharony and collaborators
\cite{AHEK}.
(To decrypt the figure, use the rule that the curves that
are higher order in $1/l$ are closer to the Nambu-Goto
curves at larger $l$.) 

\begin{figure}[htb]
\begin	{center}
\leavevmode
\input	{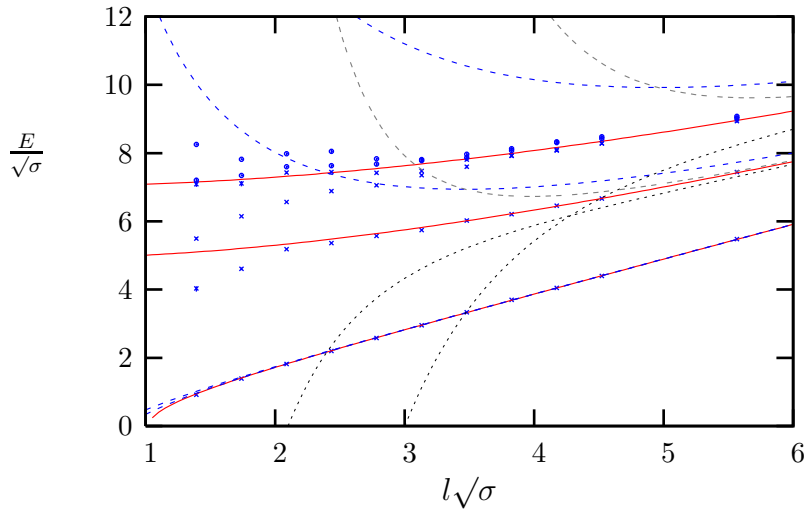}
\end	{center}
\caption{Spectrum as in Fig~\ref{fig_Eq0n3} with full
NG predictions (solid) and the derivative
expansion (dashed) to $O(1/l)$, $O(1/l^3)$, and $O(1/l^5)$.}
\label{fig_Eq0n3theory}
\end{figure}

We see that for the ground state these theoretical predictions 
do indeed approach the numerically determined energy as we 
include terms of higher order in $1/l$. However this is not the 
case for the excited states, where there is no tendency to
approach the data -- at least for $l\surd\sigma < 4$ for the
first excited level and  $l\surd\sigma < 6$ for the
second excited level. Indeed the oscillating behaviour of the
fits as we add extra terms is instantly reminiscent of the behaviour
of a series outside its radius of convergence. That this is indeed
what must be happening is clear from the radius of convergence of the
series expansion of the Nambu-Goto expansion:
\begin{eqnarray}
E_n(l) 
& = &
\sigma l\left(1 + \frac{8\pi}{\sigma l^2}\left(n-\frac{D-2}{24}\right)
\right)^{\frac{1}{2}} \nonumber \\
& \stackrel{n \geq 1}{\simeq} &
\sigma l\left(1 + \frac{8\pi}{\sigma l^2}n
\right)^{\frac{1}{2}} \nonumber \\
& = &
\sigma l \sum_i c_i \left(\frac{8\pi n}{\sigma l^2}\right)^i
\qquad \quad : \qquad  \quad l \surd{\sigma} \gtrsim \sqrt{8\pi n}
\end{eqnarray}

Thus the fact that we find near-NG behaviour for $E_{n}(l)$ for 
$l \surd{\sigma} < \sqrt{8\pi n}$ is something that does not follow 
from the recent analytic developments. The latter address the $l\to\infty$
limit where $E_n(l)-E_0(l) \ll \surd\sigma$. By contrast, we find 
that the close agreement persists down to values of $l$ where, as we 
see from  Fig.{\ref{fig_Eq0n3theory}}, we have 
$E_n(l)-E_0(l) > \surd\sigma$. Thus the analytic and numerical 
results are largely complementary, although the fact that they
both point towards the relevance of Nambu-Goto is surely no co-incidence.

\subsection{Where are the massive modes?}

The confining flux tube has a finite width $\sim 1/\surd\sigma$.
This implies that the full description of the flux tube requires
massive degrees of freedom in addition to the massless transverse
oscillations encoded in the free bosonic string theory. These
should make their presence felt in the observed excitation spectrum.
For example, the excitations of such massive modes should lead
to extra excited states with an energy gap $\Delta E \sim O(\surd\sigma)$.
(At least this should be so in the simplest case of weak coupling between 
the different modes.)

This is a quite general expectation, whether one starts from a
semi-classical intuition where one thinks of the confining flux
tube as some kind of dual non-Abelian Nielsen-Olesen vortex, 
or whether one relies on a gauge-gravity duality intuition,
where the massive modes arise from the highly non-trivial metric 
in the vicinity of the horizon where the string, hanging deep into the 
$AdS_5$ space, acquires its linear energy.
 
What is the scale of these massive modes? Obviously $O(\surd\sigma)$,
but can we be more precise? Presumably this same scale appears in
the spectrum of glueballs -- either in the mass of the lightest glueball,
or in the typical excitation energy of the lightest glueballs. 
Extracting the values of these from
\cite{MT98d3}
we can make a plausible estimate
\begin{equation}
\Delta E = E - E_0 \simeq 2\surd\sigma - 4 \surd\sigma .
\label{eqn_gap}
\end{equation}

Returning to the spectrum shown in  Fig.~\ref{fig_Eq0n3} we find
no sign of such an extra state at the smaller and intermediate 
values of $l$ where we have confidence in our identification of all 
the states with $E \lesssim 8  \surd\sigma$ as being string-like.
This raises two possibilities:\\
$\bullet$ the excitation energy of the massive modes is much larger
than expected so that they effectively play no role in the spectrum
at any value of $l$;\\
$\bullet$ our basis of lattice operators, although apparently large,
has in fact a small overlap onto these massive excitations so that they
do not appear in the numerically determined spectrum. \\
Both of these possibilities would have interesting consequences. The first 
would imply that we might well be able to describe most of the physics
of the gauge theory within a simple bosonic string model, with other
modes being so massive as to be largely decoupled. The second would suggest 
that there is very little mixing between these massive modes and
the stringy modes, since we know that our operator basis has a good
overlap onto all the light stringy modes. This again should lead to
a simplification in the dynamical description of the theory.
My suspicion is that the second possibility is the correct one and that
with a suitably extended basis of operators we will find some
massive excitations with the expected energy gap shown in
eqn(\ref{eqn_gap}). 

The above discussion suggests that it would be useful to look at 
other kinds of flux tubes where we know that there exist extra massive 
modes. This leads us smoothly onto the subject of $k$-strings.

\subsection{$k$-strings}

So far we have considered confining flux tubes carrying
flux in the fundamental representation i.e. the kind of flux tube
that forms between distant sources that are in the fundamental 
representation. One can also choose to consider sources in higher 
representations of SU($N$), and the corresponding confining tubes
carrying this flux. There are however gluons in the vacuum and they
can screen the sources down to other representations. This means
that a flux tube will be unstable if it can be screened by gluons
to a flux tube with a smaller string tension. For example the
adjoint flux tube can become a state with no flux tube. (Each
adjoint source being totally screened by an adjoint gluon.)
As $N\to\infty$ any given unstable flux tube will become stable,
and even in SU(3) it appears to make sense to discuss the 
qualitative  properties of such flux tubes
\cite{gbfb}.
However, since what we want to obtain are quite precise calculations
of energy eigenstates, an unstable flux tube, with a finite energy 
width, is clearly not ideal. 

Fortunately some of these flux tubes are absolutely stable at larger 
but finite $N$. A simple way to see this is to recall that because 
gluons are adjoint, they do not feel the centre $Z_N$. So if we have a
source that transforms non-trivially under the centre as
\begin{equation}
\psi \longrightarrow z^k \psi  \qquad : \quad z \in Z_N 
\label{eqn_kstring}
\end{equation}
this transformation cannot change under screening. So we can categorise
such sources by the value of $k$ in eqn(\ref{eqn_kstring}). 
When such a source and its conjugate are far apart, the flux tube 
between them is often referred to, generically, as a
$k$-string. If the distance is large enough it will be
energetically favourable to screen the sources with gluons so that
the resulting flux tube is the one with the smallest string tension
in the sector of given $k$: call it $\sigma_k$. This  flux tube is
completely stable and when we speak of a $k$-string this is what we 
are usually referring to. (Which usage is being employed, the generic 
or the specific, will usually be clear from the context.) 
We note that $\sigma_{k=1}=\sigma_f =\sigma$ 
is just the fundamental string tension.
The different values of $k$ for a given SU($N$) group are constrained by
\begin{equation}
z^N=1  \qquad z^k=z^{N-k} \qquad : \quad z \in Z_N 
\end{equation}
which immediately implies that the non trivial values of $k$ are
$k=1,2, ..., [N/2]$. Thus we have to go to at least SU(4) to find 
a new stable flux tube of this kind.

We can think of a local source for a generic $k$ string as a 
localised collection of $k$ fundamental
sources (heavy `quarks') with any number of gluons. So
the simplest example of a $k$-string is simply $k$ separate
fundamental flux tubes between the $k$ fundamental sources and 
their conjugates. Since one finds 
\cite{BLMTd3d4,BBMTk2d3}
that $\sigma_k < k \sigma$, we can think of the ground-state
$k$-string as a bound state of the $k$ fundamental flux tubes.

As $N\to\infty$ we have  $\sigma_k \to k\sigma$ and so the binding
energy vanishes. Associated with the binding there will presumably
be some massive excitation of the string. For example, if one imagines 
that the $k=2$ binding is due to the exchange of scalar glueballs between
the two fundamental flux tubes, then these would provide the 
scale for the massive excitation. Note that even though the binding
energy $\to$ $0$ as $N\to \infty$ this does not mean that the mass
scale also vanishes; the loss of binding may simply be because 
the relevant coupling vanishes in this limit (as would be the case 
for our glueball exchange example). Nonetheless the fact that
a $k$-string does not survive at $N=\infty$ does weaken the
theoretical argument for a clean effective string theory
description of the kind discussed earlier in these lectures.
So, one way or another, $k>1$ flux tubes promise to provide an
interesting contrast to the fundamental $k=1$ flux tubes that
we have studied so far, and that is our main motivation
in this Section.

\begin{figure}[htb]
\begin	{center}
\leavevmode
\input	{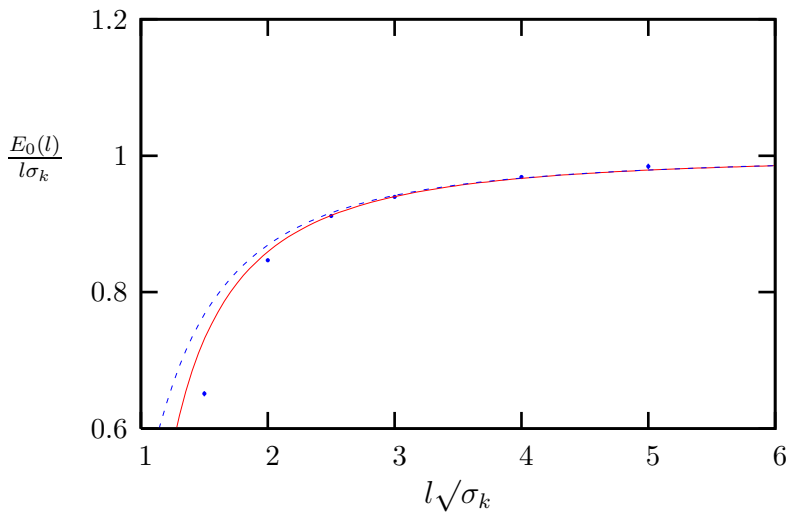}
\end	{center}
\caption{Ground state energy of the $k=2$ flux tube, $E_0(l)$,
normalised by its linear piece, plotted versus $l$. For SU(4) at 
$\beta=32$. Fits with the Luscher correction 
(dashed) and with full Nambu-Goto (solid) are shown.}
\label{fig_Ek2n4}
\end{figure}

So let us begin with the specific case of SU(4) which is the smallest
group with stable $k=2$ flux tubes. It is known 
\cite{BLMTd3d4,AABBMTk2d3}
that $\sigma_{k=2} \simeq 1.35 \sigma_f$ which implies that the
two fundamental flux tubes are quite strongly bound. Associated
with this binding must be some massive excitation, and we
would hope to see its presence clearly encoded in the spectrum
of $k=2$ flux tubes that wind around a spatial torus.

In Fig.~\ref{fig_Ek2n4} I plot the value of the ground state
energy, $E_0(l)$, divided by its asymptotic linear component, 
$l \sigma_k $, versus the length in units of $ \surd\sigma_k$.
(For purposes of comparison with earlier figures, note that 
$\surd\sigma_{k=2} \simeq 1.16 \surd\sigma $.)
I also show a fit with a Luscher correction and a fit with
Nambu-Goto. While both of these fits clearly capture a large
part of the deviation from linearity, and the latter does 
somewhat better than the former, the Nambu-Goto fit now 
has significant corrections, in contrast to the case of a
$k=1$ flux tube.

\begin{figure}[htb]
\begin	{center}
\leavevmode
\input	{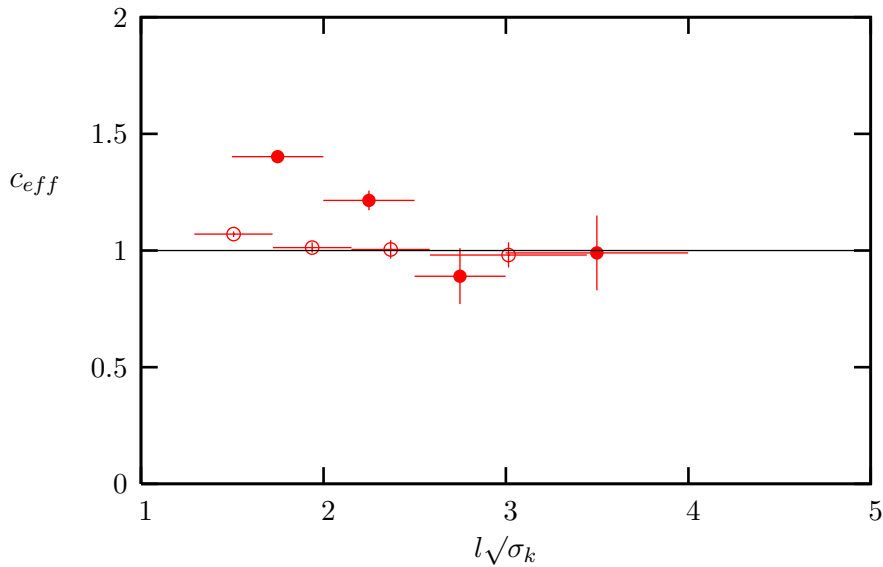}
\end	{center}
\caption{The effective central charge for $k=2$ ($\bullet$)
and $k=1$ ($\circ$) flux tubes.} 
\label{fig_ceffk1k2n4}
\end{figure}

An alternative way to see this is to do a fit with 
an effective central charge, using eqn(\ref{eqn_NGceff}).
We plot the result in Fig~\ref{fig_ceffk1k2n4} together
with the results obtained for a fundamental flux tube
for the same lattice. We see that, while the corrections
are very much larger for $k=2$ than for $k=1$, we still
appear to have $c_{eff}(l) \to 1$ as $l\to \infty$, i.e. the
$k=2$ flux tube also appears to belong to the universality class
of the simple bosonic string theory.

\begin{figure}[htb]
\begin	{center}
\leavevmode
\input {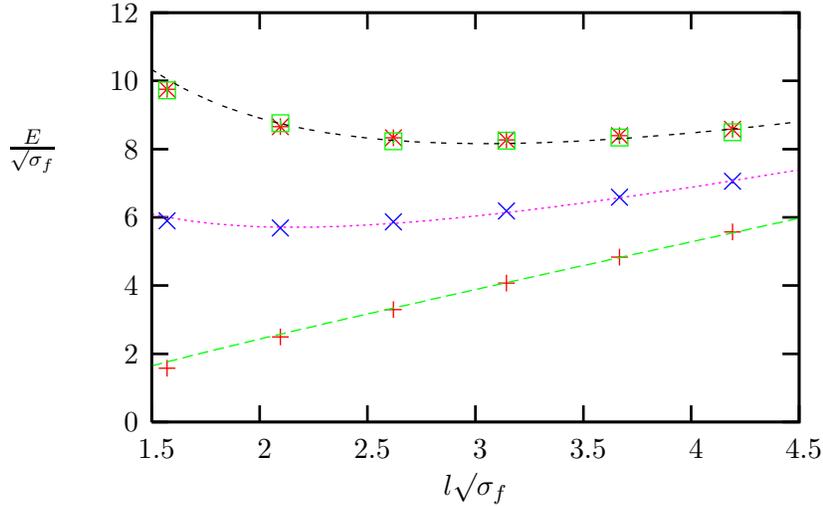}
\end	{center}
\caption{Spectrum of lightest $k=2$ flux tubes with momenta
$q=0,1,2$, in SU(4).} 
\label{fig_Ek2q012n4}
\end{figure}

We now turn to the lightest states with $q\neq 0$. 
Our results for these are shown in Fig.~\ref{fig_Ek2q012n4}
and are obtained in SU(4) at $\beta=50$ (so at a lattice spacing that 
is about 2/3 of the one considered above). For comparison we also show 
the $q=0$ ground state which serves to fix the value of $a^2\sigma_k$
for the  $q\neq 0$ Nambu-Goto predictions. (Note we use a different 
normalisation than in Fig.~\ref{fig_Ek2n4}, both for $E$ and for $l$.) 
In Nambu-Goto
the $q=1$ state has one phonon and so has $P=-$, while for the $q=2$ 
level there are two degenerate possibilities: two phonons each with unit 
momentum ($P=+$) and one phonon with two units of momentum ($P=-$).
We note that these are precisely the quantum numbers of the states 
we find, whose energies are displayed in  Fig.~\ref{fig_Ek2q012n4}. 
We also note that the observed energies are very close to the 
Nambu-Goto predictions for all values of $l$ -- just as we saw for
the fundamental flux tube in Figs.~\ref{fig_Eq1n3},~\ref{fig_Eq2n3}.

\begin{figure}[htb]
\begin	{center}
\leavevmode
\input{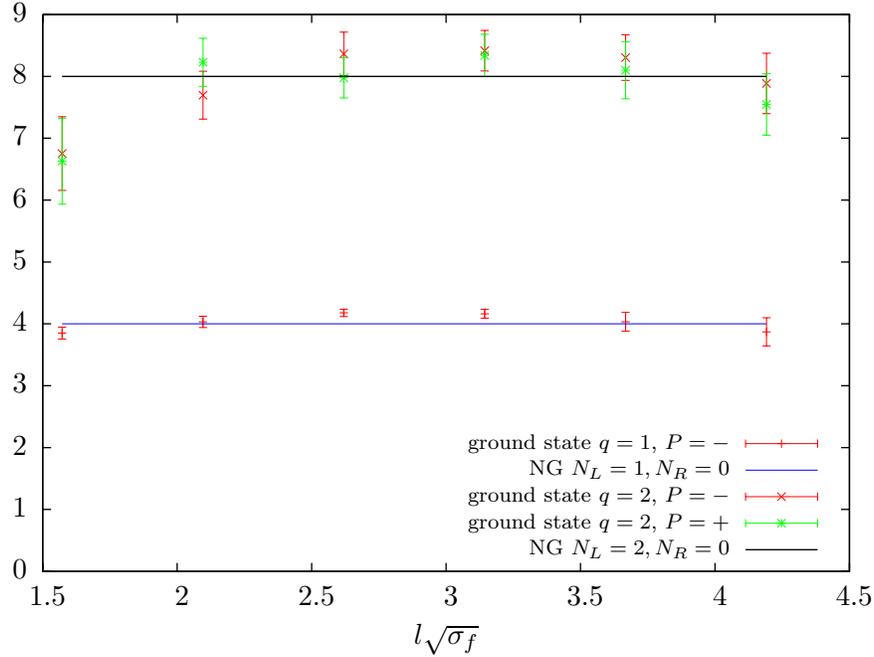} 
\end	{center}
\caption{The $q=1,2$ excitation energies, $\Delta E$ in
eqn(\ref{eqn_DelE}), plotted versus $l$.}
\label{fig_DEk2q012n4}
\end{figure}

Of course some of the extra energy of the $q\neq 0$ states comes
from the $p^2$ contribution to $E^2(p)$ and that is not peculiar 
to the Nambu-Goto model. So to better expose the level of 
(dis)agreement with the model in eqn(\ref{eqn_EnNG}), we form 
the combination
\begin{equation}
\Delta E = 
\frac{1}{\pi\sigma_k}
\left( E^2_n(l) - E^2_0(l) - \left(\frac{2\pi q}{l}\right)^2 \right)
\stackrel{NG}{=} 4\left(N_L + N_R\right)
\label{eqn_DelE}
\end{equation}
and plot the results in Fig.~\ref{fig_DEk2q012n4}. We see quite
striking evidence for the remarkably early onset, in $l$, of agreement 
with this simple free string model.

\begin{figure}[htb]
\begin	{center}
\leavevmode
\input{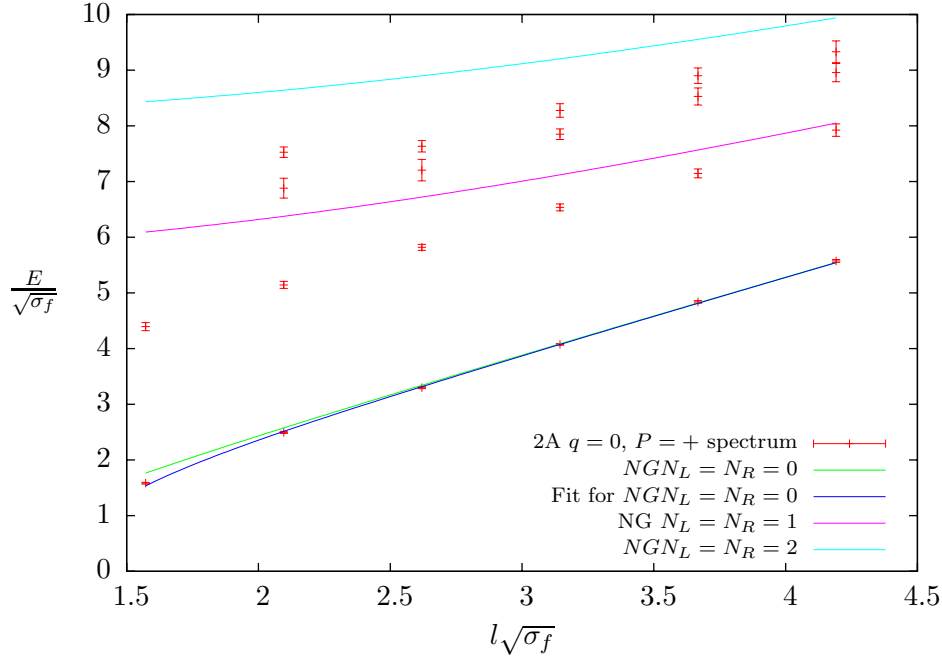}
\end	{center}
\caption{The low-lying spectrum of closed $k=2$ flux tubes with
$q=0$ and $P=+$, in SU(4).}
\label{fig_Enk2q0n4}
\end{figure}

We turn now to the first few excited $k=2$ states in the $q=0$ and 
$P=+$ sector, which is the analogue of the $P=+$ sector of the $k=1$
plot in Fig.~\ref{fig_Eq0n3}. We plot the resulting energies in
Fig.~\ref{fig_Enk2q0n4}. We simultaneously show the predictions
of the Nambu-Goto model with, as usual, the ground state fit
providing the value of $a^2\sigma_k$. The contrast with 
Fig.~\ref{fig_Eq0n3} is quite striking. Here we have large deviations
from the free string predictions, even for our longest flux tubes.
In particular, we observe that the first excited state is roughly 
consistent with our expectation for a massive mode excitation of the 
flux tube, as in eqn(\ref{eqn_massivemode}). However it is also
consistent with being a stringy excitation, with large corrections, 
which is approaching the Nambu-Goto prediction at very large $l$. 
One way to resolve this ambiguity is to compare the wave-function of
this first excited $k=2$ excitation with that of the first 
excited $k=1$ excitation, since we are confident that the latter is 
a massless stringy mode. This was done in
\cite{AABBMTk2d3}
where we showed that the wave-functions are in fact nearly identical.
This makes it very likely that what we are seeing here is not
the excitation of some quite different massive mode, but a 
conventional massless mode, albeit with large corrections to the
free string result.

\begin{figure}[htb]
\begin	{center}
\leavevmode
\input{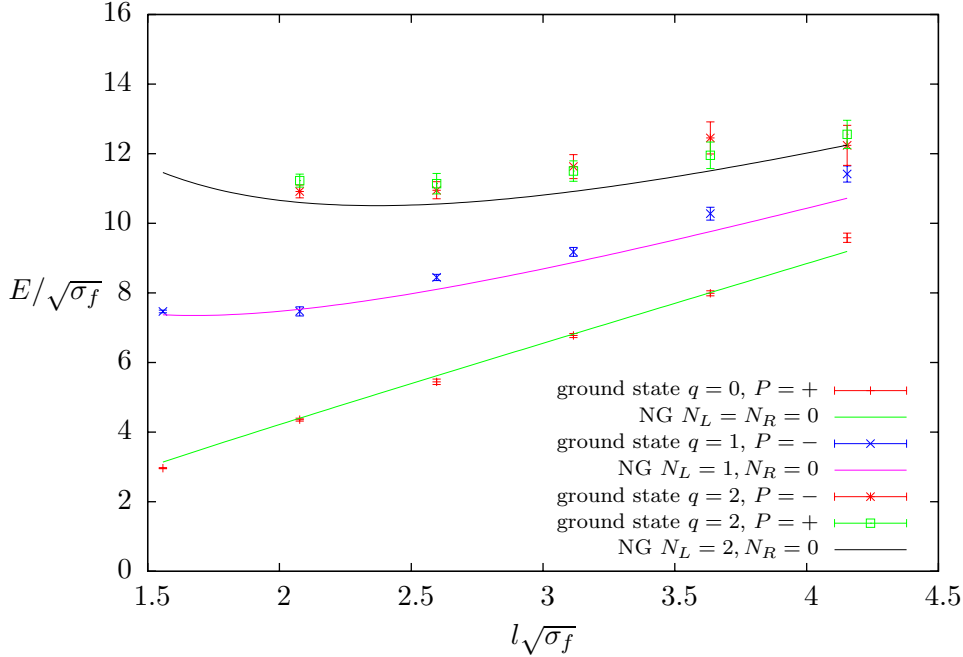}
\end	{center}
\caption{Spectrum of lightest $k=2$ flux tubes, with momenta
$q=0,1,2$, in the totally symmetric representation, for SU(4).}
\label{fig_Ek2sq012n4}
\end{figure}

We now briefly turn to the question of flux tubes that carry flux
in different representations of SU($N$) for a given $k$. Our basis 
of $k=2$ operators contains two representations, the totally
anti-symmetric, $k=2A$, and the totally symmetric, $k=2S$.
(If we had had the foresight to use a larger basis, we
would have been able to discuss other representations!) 
When we perform the variational calculation in this basis
we find that the low-lying ground and excited states fall
either into the $k=2A$ or the $k=2S$ sectors with very little
mixing between the two sectors
\cite{BBMTk2d3,AABBMTk2d3}.
In particular the lightest states with $q=0,1,2$ are $k=2A$.
The energies of the corresponding $k=2S$ states are displayed
in Fig.~\ref{fig_Ek2sq012n4}. These states are much heavier
and consequently our calculations suffer from much greater 
systematic (and statistical) errors. Nonetheless it is clear
that these states are also well-described by the Nambu-Goto
model, albeit with a considerably larger string tension, 
$\sigma_{2S}$. (Aside: the $k=2S$ $q=0$ ground state would have
been lighter, for some $l$, than some of the excited states
shown in Fig.~\ref{fig_Enk2q0n4}, but to avoid a messy plot
I did not include it therein.) I do not show the $q=0$
excited states of $k=2S$ flux tubes: these have very large
deviations from the Nambu-Goto predictions
\cite{AABBMTk2d3}.

While our above foray into non-fundamental flux tubes has not
uncovered any unambiguous signal for the excitation of massive modes,
we have seen that at least some excitations of $k=2$ flux tubes 
have much larger deviations from the Nambu-Goto predictions than
the $k=1$ fundamental flux tube. However the overall picture is still
one of a surprising level of agreement with the free string theory,
down to quite small $l$. When we consider flux in higher representations
the deviations become even larger. But what appears to be the case is 
that such unstable `resonance'-like flux tubes can be labelled by the
flux representation, to a good approximation, as can the corresponding
tower of string-like excitations.

\section{The spectrum of closed flux tubes: $D=3+1$}

We are, of course, ultimately more interested in $D=3+1$ than 
in $D=2+1$. So what do we find if we study the closed flux tube 
spectrum in that case?

One difference is that one loses confinement on a slightly longer 
length scale: $l_c\surd\sigma = \surd\sigma/T_c \sim 1.6$ 
from eqn(\ref{eqn_Tcd4}). The large-$l$ theoretical analysis 
in Section~\ref{section_tubesstrings}
goes through, with some qualifications that I will not dwell upon. 
And there are more quantum numbers, which is where I shall start.

The numerical results are all taken from
\cite{AABBMTd4}.
We have performed calculations in SU(3) at two values of $a$.
This enables us to confirm that what we are seeing is, to a
very good approximation, the physics of the continuum theory.
We also perform a calculation of the ground state in SU(6) and
of the excited state spectrum in SU(5) (at the coarser lattice
spacing) which allows us to confirm that there is very little $N$
dependence. We will not dwell on these points any further in the
following brief overview of our current, still incomplete results
and analysis.

\subsection{Quantum numbers and operators}

There are now two transverse directions, and so we have rotations
and corresponding angular momenta around the `symmetry axis' of 
the flux tube. The massless `phonons' living on the string now 
carry not only momentum and energy but also unit angular momentum, 
with positive or negative helicity. 

The quantum numbers of the flux tube can be conveniently encoded 
as follows. \\
$\bullet$ There is the length $l$ of the $x$-torus around which
the flux tube winds. \\
$\bullet$ There is also the number of times, $w$, that 
the flux tube winds around this torus, but we shall only consider
$w=1$ from now on. \\
$\bullet$ Then there is the momentum along the flux tube, $p=2\pi q/l$. 
(Transverse momentum is not interesting, for the same reason as 
in $D=2+1$.) \\
$\bullet$ There is the projection of angular momentum onto the symmetry 
axis of the flux tube, $J=0, \pm 1, ...$. \\
$\bullet$ There is also a transverse parity, $P_\rho$,  in the plane 
that is transverse to the axis, and which is analogous to the $D=2+1$
parity. Under this parity $J\to -J$. Since we choose to use this 
parity to label our states, we must use $|J|$ rather than $J$ as the
spin label, and so when we refer to  $J$, it is to be usually 
understood as $|J|$ from now on. \\ 
$\bullet$ For $p=0$ there is a reflection symmetry 
$x \to -x$ which defines a corresponding parity we call $P_r$;
it reverses the momenta of the individual phonons. 

To have a chance of a good overlap onto the ground state and some 
excited states for each of these quantum numbers, we need a very 
large basis of operators. If we imagine taking the $O(200)$ deformations
we used in $D=2+1$ and using them independently in the two transverse 
directions, we have $\sim 40000$ operators which is much too large
in practice. So instead we choose only $\sim 700$ operators.

In practice we find that our overlaps for the excited states are 
not nearly as good as they were in $D=2+1$, and our calculations
have, inevitably, larger statistical and systematic errors. 
Moreover, for some
quantum numbers the ground states themselves appear to have too large
a mass for us to be able to extract an energy with any confidence.
However even though the spectrum we obtain is incomplete, it
possesses, as we shall see, some striking regularities.

\subsection{Ground state energy}

Our first calculation is in SU(3) at $\beta=6.0625$, which corresponds
to a lattice spacing of $a\surd\sigma \simeq 0.195$. We focus on
the absolute gound state, with no phonon excitations, and with
quantum numbers $p=0, J=0, P_r=+, P_\rho=+$.

\begin{figure}[htb]
\begin	{center}
\leavevmode
\input	{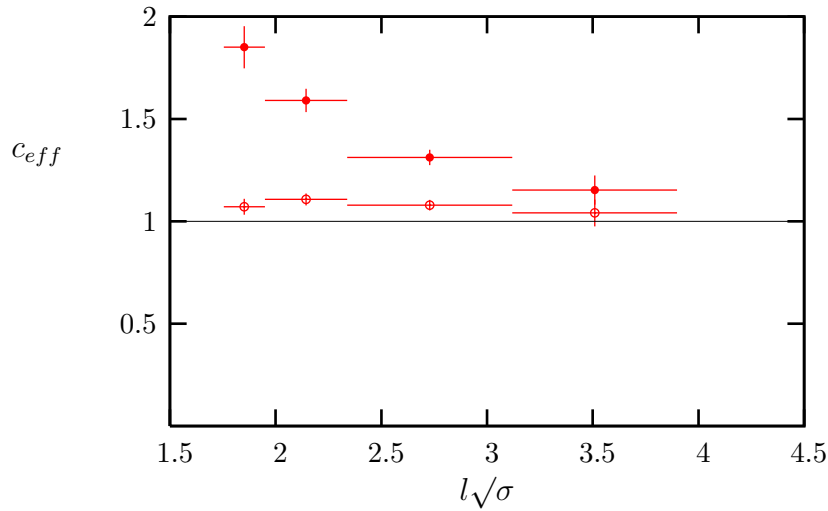}
\end	{center}
\caption{Effective central charge: from Luscher ($\bullet$) and 
Nambu-Goto ($\circ$) using eqns(\ref{eqn_MLceff},\ref{eqn_NGceff}).}
\label{fig_ceffd4n3}
\end{figure}

We perform calculations for several values of $l$, starting
close to the critical value $l_c\surd\sigma \sim 1.55$.
We then extract an effective central charge, $c_{eff}(l)$ using
eqns(\ref{eqn_MLceff},\ref{eqn_NGceff}). The resulting plot,
in Fig.~\ref{fig_ceffd4n3}, can be compared to the similar
plot for $D=2+1$ in, say,  Fig.~\ref{fig_ceffn4}. At first sight
they look very similar. Just as in $D=2+1$, we clearly have 
$c_{eff}\to 1$ as $l\to \infty$.
So here too the effective string theory is in the universality class
of the bosonic string theory where the only massless modes are the
goldstone modes associated with the spontaneous breaking of transverse 
translation symmetry. Moreover we find $c_{eff} \simeq 1 \,\, \forall\, l$ 
when using the Nambu-Goto expression in eqns(\ref{eqn_NGceff}),
showing that this expression captures most of the $l$-dependence.
However, looking more closely, it is clear that the deviations from
$c_{eff}=1$ while small are nonetheless significantly larger in 
$D=3+1$ than they were in $D=2+1$.

\subsection{Excited state energies}

We now turn to the lightest states with momenta $q=0,1,2$. 
We show our results for the ground states in these channels in
Fig.~\ref{fig_En3b6338q012}, together with the Nambu-Goto predictions.
(These have been obtained in the SU(3) calculation with the 
finer lattice spacing.)
As usual the string tension comes from the fit to the $q=0$ ground
state, so that the Nambu-Goto predictions for $q=1,2$ 
are completely parameter-free.

\begin{figure}[htb]
\begin	{center}
\leavevmode
\input	{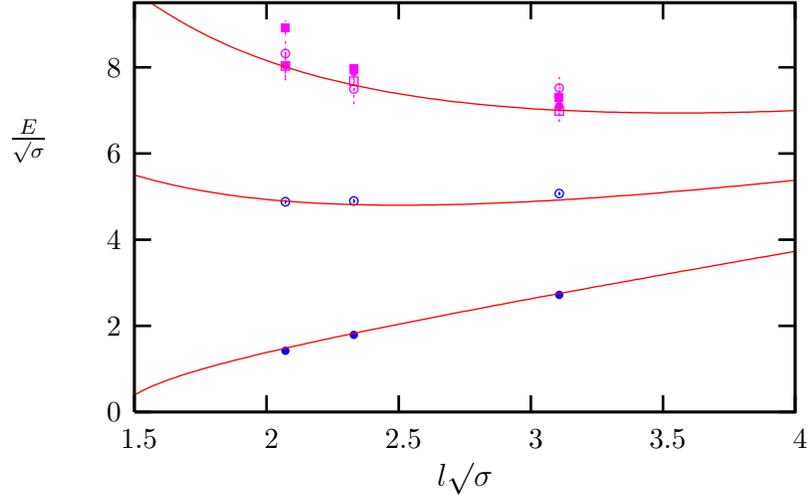}
\end	{center}
\caption{Lightest states with momenta $q=0,1,2$, and the Nambu-Goto
predictions.}
\label{fig_En3b6338q012}
\end{figure}

We observe that the states are in remarkable agreement with
the  Nambu-Goto predictions. Moreover the quantum numbers are
as expected from the latter: \\
$\bullet$ The $q=0$ ground state has no phonon excitations and
so has $J=0, P_\rho = P_r = +$. \\
$\bullet$ The $q=1$ ground state has one phonon of minimum 
momentum. So this has $J=1$. On a cubic lattice, this state 
has an exact $P_\rho = \pm$ degeneracy, so we only calculate 
and show the $P_\rho = +$ state. \\
$\bullet$ The $q=2$ ground states can be formed from two phonons 
of minimum momentum each, or one phonon with double this momentum.
(These are all degenerate in the Nambu-Goto model.) The
latter can have unit positive or negative helicity, or equivalently,
in our preferred basis, $J=1$ and $P_\rho = \pm$. In the former 
case each phonon has positive or negative helicity, which means
there are three states with $J=0,\pm 2$, or, again in our preferred 
basis, $J=0$ and $J=2, P_\rho = \pm$. These are precisely the
quantum numbers of the nearly degenerate states shown in   
Fig.~\ref{fig_En3b6338q012}. (There are four in our plot rather than 
five because we do not show both of the exactly degenerate $J=1$ states.)

While the energies of the $q=2$ ground states are nearly
degenerate, there is a visible splitting between them. 
While this may be real, we caution the reader that for such
massive states, for which some of our operator overlaps are
modest, it is possible that the systematic errors are comparable
to these splittings. Better calculations are needed.
 
\begin{figure}[htb]
\begin	{center}
\leavevmode
\input	{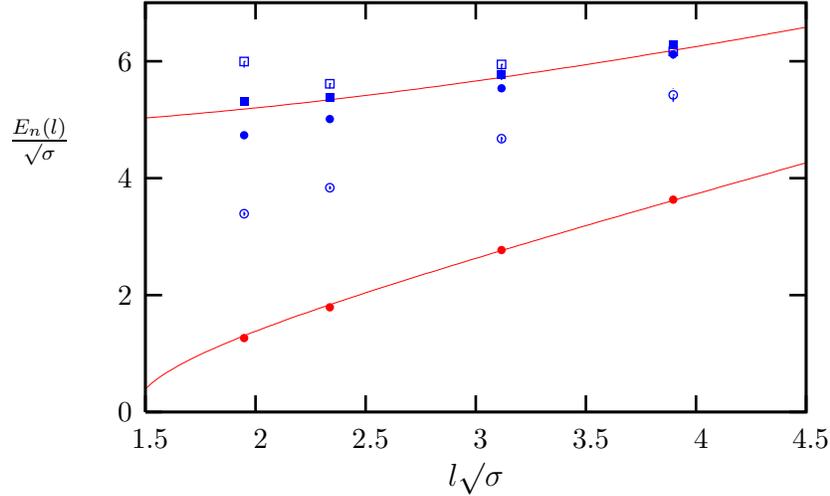}
\end	{center}
\caption{The ground and first excited level in the $q=0$
sector.}
\label{fig_AAn3d4q0}
\end{figure}

We turn now to a comparison of the lightest and first excited levels
in the $q=0$ sector. These are shown in Fig.~\ref{fig_AAn3d4q0}
and come from the SU(3) calculation on the coarser lattice,
at $\beta =6.0625$. The Nambu-Goto predictions for the ground and
first excited energy levels are also shown. The fit to the ground 
state fixes the value of $a^2\sigma$ so that the  prediction for the
first excited level is parameter free. 
The first excited level will have two phonons of equal and opposite
momentum. Since each of these can have positive or negative helicity
there will be four states, two with $J=0$ and two with $J=2$. These
four states of various $P_\rho$ and $P_r$, are degenerate in the 
Nambu-Goto model.

What we see in  Fig.~\ref{fig_AAn3d4q0} is that while the lightest
four states above the ground state do indeed have precisely the
Nambu-Goto quantum numbers, only 3 of them are nearly degenerate
and close to the Nambu-Goto prediction, while the 
fourth is far from that prediction, even for $l\surd\sigma \sim 4$. 
The quantum numbers of this anomalous state 
are $J=0, P_\rho = P_r = -$. While one might
be reassured that at least it appears to be approaching the other 
states as $l\uparrow$, we also observe that its gap from
the ground state is roughly independent of $l$. This might
suggest that it is a massive rather than a stringy excitation and 
that it will `cross' the first excited stringy energy level rather 
than asymptoting towards it as $l\to\infty$. So at this stage 
we are left not
knowing whether this state is a stringy excitation with an anomalously
large interaction energy or something quite different: for
example an excitation of the massive
modes associated with the non-trivial structure of the confining
flux tube. We remark that because this state is relatively light, 
and because it has a good overlap onto our operator basis, the
energy calculation is particularly reliable.
We also note that we obtain essentially the same $q=0$
spectrum at the smaller lattice spacing, and in the SU(5)
calculation. This tells us that the anomalous behaviour of
this $J=0, P_\rho = P_r = -$ state is neither a lattice artifact
nor some finite-$N$ correction. It is indeed a feature of the 
large-$N$ continuum theory.

\begin{figure}[htb]
\begin	{center}
\leavevmode
\input	{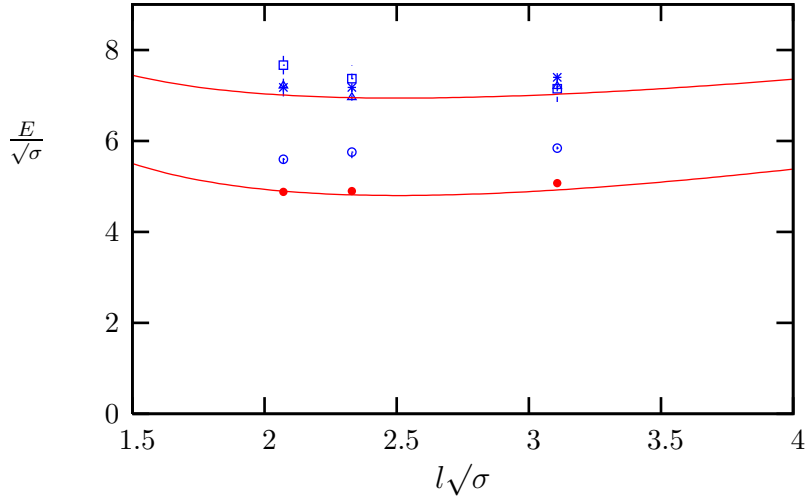}
\end	{center}
\caption{The ground state and first excited level in the $q=1$ sector.}
\label{fig_En3b6338q1}
\end{figure}

We turn now to a comparison of the ground and first excited levels 
in the $q=1$ sector. Here the ground state has one phonon
and so can have spin $\pm 1$, or $J=1$ and $P_\rho = \pm$ in
our preferred basis. As usual we will only show one of these exactly
degenerate states. The first excited level can be produced by
2 left and 1 right moving phonons, all with unit momentum,
or by 1 left moving phonon with momentum two and and 1 right mover
with unit momentum. Since all these phonons have spin $\pm 1$
one has states with $J=0,2,3$ each with $P_\rho = \pm$, and 2 sets
of states with $J=1,P_\rho = \pm$, i.e. 10 states in all, and they 
are all degenerate in the Nambu-Goto model. (In counting states
there are fewer than naive combinatorics would suggest because 
some are the same state because the phonon operators commute.) 

We show in Fig.~\ref{fig_En3b6338q1} how our calculated $q=1$ energy 
levels, obtained in SU(3) at the finer lattice spacing, compare 
to those of the Nambu-Goto model. Since $a^2\sigma$
has been fixed by the fit to the $q=0$ ground state, these are
all parameter-free predictions. The $q=1$ ground state agrees 
very well with the prediction, but we have already seen that in 
Fig.~\ref {fig_En3b6338q012}. As for the first excited level,
our calculation is incomplete because the combination of large 
energies and, in some cases, poorer overlaps means that we are
only able to extract 4 states with any reliability. These have 
quantum numbers that are amongst those expected within the
Nambu-Goto model, and three of the states have energies that are 
close to those predicted by the model. However the fourth state,
with quantum numbers $J=0, P_\rho=-$ is much lower, and
indeed much closer to the $q=1$ ground state than
to the first excited level, and shows no sign of
approaching the NG prediction as $l$ increases. This discrepancy
is slightly less on the coarser lattice spacing, indicating that
it will probably be even larger in the continuum limit. It is also
larger in the SU(5) calculation, suggesting that this state will
be even closer to the ground state at $N=\infty$. Again, this
strikingly anomalous state is clearly a feature of the continuum
large-$N$ physics.

So our overall conclusion is that, just as in $D=2+1$, typical
states are very close to the Nambu-Goto prediction, even down
to very small values of $l\surd\sigma$ where the flux tube is
far from being a `thin string' and where an expansion of $E_n(l)$ in 
inverse powers of $1/\sigma l^2$ has long ceased to converge. However 
at the same time there are some states which deviate so far
from the free string theory prediction that it is entirely
plausible that they involve massive rather than just stringy 
excitations.

\section{Conclusions}

My lectures have covered two loosely linked topics.

\vspace*{0.05in}

In the first part of my lectures I described in some detail how
the lattice has answered some important questions about SU($N$) 
gauge theories in the 't Hooft large $N$ limit. For example,
we now have quite convincing evidence that the SU($\infty$) theory
is linearly confining for temperatures $T < T_c$, and that the
deconfining temperature remains finite and non-zero as $N\to\infty$.
Importantly - from the phenomenological point of view - it turns
out that many quantities show small corrections as we go from
SU(3) to SU($\infty$). That is to say, $N=3$ is `close to' $N=\infty$.
And where we have looked, we find that 't Hooft's large $N$ counting,
derived by looking at diagrams to all orders, is corroborated by
the non-perturbative lattice calculations: e.g. $g^2 \propto 1/N$
as $N\to\infty$. (This comment does not apply to the way
$\sigma_k\to k\sigma$ as $N\to\infty$
\cite{BBMTk2d3},
but that is another story.)
We also spent some time on physics at finite $T$,
in particular in the region above but close to $T_c$ where RHIC
experiments have been very active and calculations based on 
gauge-gravity duality have been widely employed. We showed that
the strong coupling plasma in this region is very similar 
in SU(3) and SU($\infty$), as needs to be the case if the AdS/CFT
calculations are to have any relevance. All this is `good news' for 
the application of large $N$ arguments to the real world. 

The situation is currently less clear as far as the meson spectrum of
QCD is concerned. Here there have been three calculations of the
$\rho$ and $\pi$ mesons. Two have shown that $m_\rho$ has a 
weak dependence on $N$, while one shows a strong dependence. While
the methods used in the two sets of calculations are very different, 
the overlap in what is actually being calculated is large enough that
it is clear that (at least) one set of calculations must be incorrect. 
This needs to be resolved and urgently.

The computational cost of such calculations is modest and, as discused
in my lectures, grows with $N$ quite weakly for calculations with 
quarks. Assuming that when things are clarified we find that 
$QCD_{N=3} \simeq QCD_{N=\infty}$, for the prominent low-lying
mesons there are some very interesting questions to address here.
The point is that at large $N$ mixings vanish so we will have
a clean separation between $q\bar{q}$ mesons and glueballs.
And all the states become stable so we should get well-defined
excited states, including $q\bar{q}$ radial excitations.
Interactions between colour singlet states vanish, so `molecular'
states should disappear from the spectrum. One can therefore obtain
a clean and precise spectrum to as high a mass as one's computing
resources allow. Such a spectrum, beyond its elegance and the dynamical 
insights it might provide, would be of great value in interpreting
hadron spectroscopy in the real world.   

\vspace*{0.05in}

In the second part of these lectures I focussed on a very
specific question: what is the effective string theory that
describes confining flux tubes? Even a partial answer to this question
will clearly be important in the search for an answer to the broader
question of what is the string theory that describes SU($N$) 
gauge theories in the $N\to\infty$ limit. Gauge-gravity
duality has brought a new and more precise focus to this
long-standing issue.

At the theoretical level it is simplest to restrict oneself
to the massless modes of the string that are associated with the 
spontaneous breaking of the transverse translation invariance. 
For a long string, $l \gg 1/\surd\sigma$, the lightest such modes lead to 
energy levels that are within $\Delta E \sim O(\pi/l) \ll \surd\sigma$ 
of the ground state, and this is where one would expect such an 
effective theory to be valid. The derivative expansion of the effective
action leads to contributions of higher order in $1/l^2$ to the
string partition function evaluated on a cylinder or on a torus, and 
so to the corresponding partition functions for the open and/or closed
flux tubes that sweep out the surfaces with the corresponding 
boundaries. The relationship between these string and field theoretic
partition functions can only become exact in the $N=\infty$ limit.
I sketched, in some detail, the recent, quite dramatic analytic 
progress on using this relationship to constrain the form of the
corrections to the energy spectrum of long strings. 
As a result we now know that not only is the Luscher
$O(1/l)$ correction to the linear $\sigma l$ piece universal,
but so are the terms of $O(1/l^3)$ and of $O(1/l^5)$. (Up to some 
dimension-dependent qualifications.) Moreover these terms are exactly 
the same as one finds in the free string theory described by the 
Nambu-Goto action in flat space-time. (This follows automatically 
since Nambu-Goto theory satisfies the same theoretical constraints.)
Thus the effective string theory describing the low-energy 
excitations of a long flux-tube is, to this quite high order, 
precisely a free string theory. 

The numerical calculations are largely complementary to the analytic 
analysis in that they provide information on the energy spectrum of 
confining flux tubes that are of short or intermediate length, i.e.
$l \sim O(1/\surd\sigma)$. While there is a plausible  
overlap in $l$ between the range of validity of the 
analytic and numerical calculations for the ground
state, this quickly disappears as we go to more highly excited states.
This is not by choice, but rather because the energy of a very long 
flux tube is large, and the corresponding correlation function
disappears very rapidly, $\propto \exp\{-aE_n(l)n_t\}$, into the 
statistical noise of the Monte Carlo calculation. 
For $l\surd\sigma \in (1.5,4.5)$ however, we are able to obtain very 
accurate results for the lightest energy levels, particularly in $D=2+1$. 
In this case we find that the ground and excited states
can be accurately described by the Nambu-Goto model, almost down to
the critical length $l_c\simeq 1/\surd\sigma$ where one loses 
confinement. For such values of $l$, 
for example for $l\surd\sigma \simeq 2$, the flux tube is
almost as wide as it is long -- naively it is more like a fat blob
than a thin string. It is remarkable that the lowest excitations of 
this `fat blob' remain almost precisely those of a thin string. The
more so that at such low values of $l$ the Nambu-Goto expression 
for the energy of the excited states can no longer be expanded in a 
convergent power series in powers of $1/l$. This suggests that the
massive and stringy modes must be weakly coupled to each other.

So in $D=2+1$, the theoretical analysis tells us that the 
effective string action is Nambu-Goto-like to some high order
in $1/l$. Numerical results tell us that this continues to
be the case even when $l$ is so small that an expansion in
$1/l^2$ has long ceased to converge. We take this to be strong
evidence that there is a useful effective string action
for confining flux tubes not only at very large $l$, but at almost
all values of $l$, and that an accurate first approximation to
such an action is the Nambu-Goto free string theory.

We can be a little more specific. Let the actual energy level be 
$E_n(l)$, and let $E^{NG}_n(l)$ be the corresponding energy level
in the Nambu-Goto model, as given by the square root expression
in eqn(\ref{eqn_EnNGexpansion}) arising from eqn(\ref{eqn_EnNG}).
Then if at large $l$ the leading correction arises at $O(1/l^\alpha)$,
we can write for the low-lying excited states
\begin{equation}
E_n(l) = E^{NG}_n(l) + 
\frac{ c \surd\sigma}{(l\surd\sigma)^\alpha} F_n(l\surd\sigma)
\qquad , \quad F_n(l\surd\sigma) \stackrel{l\to\infty}{\longrightarrow} 1.
\label{eqn_Ecorrn}
\end{equation}
where we know, from the theoretical analyses that
\begin{equation}
\alpha \geq 7  
\label{eqn_alpha}
\end{equation}
and from the numerical calculations that the correction
\begin{equation}
\Delta E_n(l)
=
\frac{ c \surd\sigma}{(l\surd\sigma)^\alpha)} F_n(l\surd\sigma)
\ll E_n(l), 
\qquad l\in [l_0,l_1]
\label{eqn_bound}
\end{equation}
where the range $[l_0,l_1]$ extends from well below the value 
of $l$ where the expansion of $E^{NG}_n(l)$ no longer converges. 
(Recall that $E^{NG}_n(l)$ is completely
fixed once we have determined the string tension from the calculation
of the ground state, $E_0(l)$).

An important puzzle is that we observe no excitations of massive modes, 
with say $E(l) - E_0(l) \sim O(\surd\sigma)$, even at the smaller 
values of $l$ where they should be clearly visible (unless the energy 
gap is unexpectedly large). Presumably our basis of operators has a small
overlap with such states. This again suggests a weak coupling
between stringy and massive modes, and needs clarification.

We also studied $k=2$ flux tubes which can be thought of as bound 
states of two fundamental flux tubes. Here the binding, if nothing
else, tells us that there must be associated massive modes. While 
we do not see any modes that are clearly massive excitations, we
do find quite large corrections to the Nambu-Goto eigenspectrum
for some eigenstates.
Since there is no large-$N$ limit to such flux tubes, in the sense
that the binding vanishes at $N=\infty$, the theoretical basis for 
this analysis is less rigorous. However the comparison does confirm that
there is something very special and simple about the elementary flux 
tubes that carry flux in the fundamental representation.

Our results for $D=3+1$ are incomplete and not yet published.
There is a richer spectrum of states, because there are more
relevant quantum numbers once we have two transverse directions.
By the same token our calculations are less accurate because
our basis of operators is significantly less complete than it
was in the lower dimension. Despite these caveats, we have
established some striking regularities. As in the lower dimension,
many states have energies close to the Nambu-Goto prediction
even for $l$ close to the deconfining length scale, 
$l_c\surd\sigma \sim 1.5$. However, now there is a new feature, and that
is that there are a few states that are very far from Nambu-Goto
and show no sign of approaching the predicted energy levels as
$l$ increases. Are these anomalous states related to the massive
excitations of the flux tube that have eluded us, so far, in two 
spatial dimensions?  This remains to be understood. The fact that the
overall picture has this clear binary character, gives us some
confidence that it can be simply understood. 

\vspace*{0.05in}
 
The calculations that I have summarised or merely referred to in these
lectures, are mostly a first attempt to get a rough idea of the physics of
SU($\infty$) gauge theories. Perhaps their main virtue is to point 
to the huge amount of interesting physics that is waiting to be done,
and to provide a demonstration that lattice calculations can
address such questions with readily available resources.
I hope that this will encourage you to get actively involved.

\section*{Acknowledgements}

I would like to thank the organisers
for providing me with the opportunity to lecture
at this School and for creating such an excellent 
environment for useful and critical discussions
between all the participants. I would also like 
to thank my collaborators Andreas Athenodorou and
Barak Bringoltz: the second part of my lectures is based 
on our joint work and what I understand about this
subject is largely due to the extensive discussions 
we have enjoyed over the last few years. 

\vfil\eject


\begin{thebibliography}{99}
%
%

\bibitem{MT98d3}
M. Teper, Phys. Rev. D59 (1999) 014512 (hep-lat/9804008).\\
M. Teper, Phys. Lett. B397 (1997) 223 (hep-lat/9701003).

\bibitem{maldacena}
J. Maldacena, Adv. Theor. Math. Phys. 2 (1998) 231 (hep-th/9711200).\\
O. Aharony, S. Gubser, J. Maldacena, H. Ooguri and Y. Oz,
Phys. Rept. 323 (2000) 183 (hep-th/9905111).

\bibitem{GtH74}
G. 't Hooft, Nucl. Phys. B72 (1974) 461.

\bibitem{Nreviews}
E. Witten, Nucl. Phys. B160 (1979) 57.\\
S. Coleman, 1979 Erice School Lectures.\\
A. Manohar, 1997 Les Houches Lectures, hep-ph/9802419.

\bibitem{GVNcNf}
G. Veneziano, Nucl. Phys. B117 (1976) 519.

\bibitem{MSGV}
A. Armoni, M. Shifman and G. Veneziano,
Nucl. Phys. B667 (2003) 170.


\bibitem{DMsu6}
R. Dashen, E. Jenkins and A. Manohar,
Phys. Rev. D51 (1995) 3697 (hep-ph/9411234).

\bibitem{lat_books}
M. Creutz, {\it{Quarks, gluons and lattices}} (CUP, 1983). \\
I. Montvay and G. Munster, {\it{Qauntum Fields on a Lattice}}
(CUP, 1994).\\
H. Rothe, {\it{Lattice Gauge Theories}} (World Scientific, 1997).\\
J. Smit, {\it{Introduction to Qauntum Fields on a Lattice}}
(CUP, 2002).\\
T. Degrand and C. DeTar, {\it{Lattice Methods for Quantum Chromodynamics}}
(World Scientific, 2006).\\
C. Gattringer and C. Lang, {\it{Quantum Chromodynamics on the Lattice}}
(Springer, 2010).


\bibitem{smear}
APE Collaboration,  
Phys. Lett. B192 (1987) 163; B197 (1987) 400.

\bibitem{block}
M. Teper, Phys. Lett. B183 (1987) 345; B185 (1987) 121.

\bibitem{variational}
K. Wilson, closing remarks at Cosener's House Lattice Gauge Theory meeting
(Abingdon, UK, March 1981).

\bibitem{veryoldG}
K. Ishikawa, G. Schierholz and M.Teper. Phys. Lett. B110 (1982) 399.\\
M. Falcioni, E. Marinari, M. Paciello, G. Parisi, F. Rapuano, B. Taglienti
and Y. Zhang, Phys. Lett. B110 (1982) 295.\\
B. Berg and A. Billoire, Phys. Lett. B113 (1982) 65.\\
B. Berg, A. Billoire and C. Rebbi, Ann. Phys. 142 (1982) 185.

\bibitem{uwmlvar}
B. Berg and A. Billoire, Nucl. Phys. B221 (1983) 109.\\
M. Luscher and U. Wolff, Nucl. Phys. B339 (1990) 222.


\bibitem{Symanzik}
K. Symanzik,
Nucl. Phys. B226 (1983) 187; Nucl. Phys. B226 (1983) 205.

\bibitem{mGold}
C. Michael and M. Teper,
Nucl. Phys. B314 (1989) 347.

\bibitem{CloseG}
F. Close and Q. Zhao,
Int. J. Mod. Phys. A21 (2006) 821 (hep-ph/0509305).

\bibitem{HMMT04}
H. Meyer and M. Teper,
JHEP 0412 (2004) 031 (hep-lat/0411039).

\bibitem{blmtuwG}
B. Lucini, M. Teper and U. Wenger,
JHEP 0406 (2004) 012 (hep-lat/0404008).

\bibitem{gbfb}
G. Bali and F. Bursa,
PoSLAT2007:050 (2007) (arXiv:0708:3427) ; 
JHEP 0809 (2008) 110 (arXiv:086:2278).

\bibitem{ldd}
L. Del Debbio, B. Lucini, A. Patella and C. Pica,
JHEP 0803 (2008) 062 (arXiv:0712:3036).

\bibitem{ahrn}
A. Hietanen, R. Narayanan, R. Patel and C. Prays,
arXiv:0901:3752.

\bibitem{blmtuwTc}
B. Lucini, M. Teper and U. Wenger,
Phys. Lett. B545 (2002) 197 (hep-lat/0206029);
JHEP 0401 (2004) 061 (hep-lat/0307017).

\bibitem{BBMTk1d3}
B. Bringoltz and M. Teper,
Phys. Lett. B645 (2007) 383 (hep-th/0611286). 

\bibitem{Parisi_MFI}
G. Parisi in {\it{High Energy Physics}} - 1980 (AIP 1981).

\bibitem{Lepage_TI}
G. Lepage, Schladming lectures, hep-lat/9607076.

\bibitem{BLMTgkd4}
B. Lucini and M. Teper, 
JHEP 0106 (2001) 050 (hep-lat/0103027).

\bibitem{MTlat08}
M.Teper, PoS (Lattice 2008) 022 (arXiv:0812.0085).


\bibitem{CAMTAT}
C. Allton, M. Teper, A. Trivini, 
JHEP 0807 (2008) 021 (arXiv:0803.1092).

\bibitem{SFsu3}
S. Capitani, M. Luscher, R. Sommer and H. Wittig,
NUcl. Phys. B544 (1999) 669  (hep-lat/9810063).

\bibitem{bethke} 
S. Bethke, 
Prog. Part. Nucl. Phys. 58 (2007) 351 (hep-ex/060603).

\bibitem{SFsu4} 
B. Lucini, G. Moraitis, 
Phys. Lett. B668 (2008) 226 (arXiv:0805.2913); 
PoSLAT2007:058 (2007) (arXiv:0710.1533).

\bibitem{blmtuwTcprop}
B. Lucini, M. Teper and U. Wenger,
JHEP 0502 (2005) 033 (hep-lat/0502003)

\bibitem{jkrn}
J. Kiskis and R. Narayanan,
JHEP 0809 (2008) 180 (arXiv:0807.1315)

\bibitem{Tcd3}
J. Liddle and M. Teper, PoSLAT2005:188 (2005) (hep-lat/0509082); 
arXiv:0803.2128.\\
K. Holland, 
JHEP 0601 (2006) 023 (hep-lat/0509041).\\
K. Holland, M. Pepe and U-J Wiese, 
JHEP 0802 (2008) 041 (arXiv:0712.1216).

\bibitem{TQ}
B. Lucini, M. Teper and U. Wenger,
Nucl. Phys. B715 (2005) 461 (hep-lat/0401028). \\
L. Del Debbio, H. Panagopoulos and E. Vicari,
JHEP 0409 (2004) 028 (hep-th/0407068).

\bibitem{bbmtP}
B. Bringoltz and M. Teper,
Phys. Lett. B628 (2005) 113 (hep-lat/0506034).

\bibitem{Panero}
M. Panero, Phys. Rev. Lett. 103:232001 (2009) (arXiv:0907.3719).

\bibitem{LW04}
M. Luscher and P. Weisz,
JHEP 0407 (2004) 014 (hep-th/0406205).


\bibitem{HM06}
H. Meyer,
JHEP 0605 (2006) 066 (hep-th/0602281).


\bibitem{AHEK}
O. Aharony and E. Karzbrun, 
JHEP 0906 (2009) 012 (arXiv:0903.1927).


\bibitem{Olesen}
P. Olesen, Phys. Lett. 160B (1985) 144.


\bibitem{PS}
J. Polchinski and A. Strominger, 
Phys. Rev. Lett. 67 (1991) 1681.


\bibitem{LSW}
M. Luscher, K. Symanzik and P. Weisz,
Nucl. Phys. B173 (1980) 365.\\
M. Luscher, Nucl. Phys. B180 (1981) 317.

\bibitem{NG}
B. Zwiebach, {\it{A First Course in String Theory}}
(CUP, 2004).

\bibitem{Arvis}
J. Arvis, Phys. Lett. 127B (1983) 106.

\bibitem{ustobe}
A. Athenodorou, B. Bringoltz and M. Teper,
in preparation.

\bibitem{JD}
J. Drummond, hep-th/0411017;  hep-th/0608109.

\bibitem{HDPM}
N. Hari Dass and P. Matlock, hep-th/0606265; hep-th/0612291;
arXiv:0709.1765.

\bibitem{Dass09}
N. Hari Dass, P. Matlock and Y. Bharadwaj, arXiv:0910.5615.\\
N. Hari Dass and Y. Bharadwaj, arXiv:0910.5620.

\bibitem{Dass09b}
N. Hari Dass, arXiv:0911.3236.

\bibitem{AOP}
J. Ambjorn, P. Olesen and C. Peterson,
Nucl. Phys. B244 (1984) 262; B240 (1984) 189.

\bibitem{deFSST}
Ph. de Forcrand, G. Schierholz, H. Schneider and M. Teper,
Phys. Lett. 160B (1985) 137.


\bibitem{CM}
S. Perantonis and C. Michael, Nucl. Phys. B347 (1990) 854.
C. Michael, hep-ph/9809211.

\bibitem{Caselle}
M. Caselle, R. Fiore, F. Gliozzi, M. Hasenbusch and P. Provero,
Nucl. Phys. B486 (1997) 245 (hep-lat/9609041).\\
M. Caselle, F. Gliozzi, U. Magnea and S. Vinti,
Nucl. Phys. B460 (1996) 397 (hep-lat/9510019).

\bibitem{Caselle2}
M. Caselle, M. Hasenbusch and M. Panero,
JHEP 0603 (2006) 084 (hep-lat/0601023); 
JHEP 0601 (2006) 076 (hep-lat/0510107);
JHEP 0503 (2005) 026 (hep-lat/0501027).\\
F. Gliozzi, M. Panero and A. Rago,
hep-lat/0309061.

\bibitem{kuti}
K. Juge, J. Kuti and C. Morningstar,
Phys. Rev. Lett.90 (2003) 161601 (hep-lat/0207004).\\
K. Juge, J. Kuti, F. Maresca, C. Morningstar and M. Peardon,
Nucl. Phys. Proc. Suppl. 129 (2004) 703 (hep-lat/0309180).\\
J. Kuti, 
PoS LAT2005:001 (2006)  (hep-lat/0511023).

\bibitem{multihit}
M. Luscher and P. Weisz,
JHEP 0109 (2001) 010 (hep-lat/0108014);
JHEP 0207 (2002) 049 (hep-lat/0207003).

\bibitem{BLMTd3d4}
B. Lucini and M. Teper,
Phys. Rev. D64 (2001) 105019 (hep-lat/0107007).

\bibitem{rn}
J. Kiskis and R. Narayanan,
Phys. Lett. B681 (2009) 372 (arXiv:0908.1451).

\bibitem{pot}
N. Hari Dass and P. Majumdar,
PoS Lat2007: (2007) 316 (arXiv:0709.4170);
JHEP 0610 (2006) 020 (hep-lat/0608024).\\
S. Necco and R. Sommer,
Nucl. Phys. B622 (2002) 328 (hep-lat/0108008).\\
S. Necco, hep-lat/0306005.\\
H. Meyer, 
Nucl. Phys. B758 (2006) 204 (hep-lat/0607015).

\bibitem{expot}
B. Brandt and P. Majumdar,
arXiv:0905.4195; PoS Lat2007: (2007) 027 (arXiv:0709.3379).

\bibitem{balideldar}
S. Deldar,
Eur. Phys. J. C47 (2006) 163 (hep-lat/0607025);
Phys. Rev. D62 (2000) 034509 (hep-lat/9911008).\\
G. Bali, 
Phys. Rev. D62 (2000) 114503 (hep-lat/0006022)

\bibitem{AABBMTk1d3}
A. Athenodorou, B. Bringoltz and M. Teper,
Phys. Lett. B656 (2007) 132 (arXiv:0709.0693).

\bibitem{BBMTk2d3}
B. Bringoltz and M. Teper,
Phys. Lett. B663 (2008) 429 (arXiv:0802.1490).

\bibitem{AABBMTk2d3}
A. Athenodorou, B. Bringoltz and M. Teper,
JHEP 0905 (2009) 019 (arXiv:0812.0334).

\bibitem{AABBMTd4}
A. Athenodorou, B. Bringoltz and M. Teper, arXiv:0912.3238;
in preparation.

\bibitem{nair}
D. Karabali, V. P. Nair and A. Yelnikov,
Nucl. Phys. B824 (2010) 387 (arXiv:0906.0783). 

\end{thebibliography}
\end{document}